\documentclass[12pt]{article}
\usepackage{graphicx,amssymb,amsmath,tensor,empheq,color}

\usepackage{mathtools}

\usepackage{hyperref}
\hypersetup{colorlinks = true, linkcolor=black, citecolor=black, urlcolor=blue}
\usepackage{url}

\makeatletter

\usepackage{esint}

\newlength{\fighskip} \fighskip=2pt
\newlength{\figvskip} \figvskip=3pt

\newcommand*{\figbox}[2]{{
  \def\figscale{#1}
  \def\arraystretch{0.8}
  \arraycolsep=0pt
  \begin{array}{c}
    \vbox{\vskip\figscale\figvskip
      \hbox{\hskip\figscale\fighskip
        \includegraphics[scale=\figscale]{#2}}}
  \end{array}}}
  
\newcommand{\be}{\begin{equation}}
\newcommand{\ee}{\end{equation}}
\newcommand{\bea}{\begin{eqnarray}}
\newcommand{\eea}{\end{eqnarray}}
\newcommand{\nn}{\nonumber\\}

\newcommand*{\widebox}[1]{\setlength{\fboxsep}{1ex}%
  \fbox{#1}}
\newcommand*{\wideboxed}[1]{\setlength{\fboxsep}{1ex}%
  \fbox{\m@th$\displaystyle#1$}}

\def\ubrace#1_#2{%
  \underbrace{#1}_{\hb@xt@\z@{\hss$\scriptstyle#2$\hss}}}

\newcommand*{\V}[1]{\boldsymbol{#1}}

\newcommand\bdot{\mathbin{\mathpalette\bdot@{0.5}}}
\newcommand*\bdot@[2]{\vcenter{\hbox{\scalebox{#2}{$\m@th#1\bullet$}}}}

\newcommand{\hgf}{%
\,\tensor[_{2\kern-1.2pt}]{F}{_{\kern-0.8pt 1}}\kern-1.2pt}
\newcommand{\hgfs}{\mathbf{F}}

\newcommand{\lt}{\left}
\newcommand{\rt}{\right}

\newcommand{\blangle}{\bigl\langle}
\newcommand{\brangle}{\bigr\rangle}
\newcommand{\dlangle}{\langle\kern-1.5pt\langle}
\newcommand{\drangle}{\rangle\kern-1.5pt\rangle}
\newcommand{\bdlangle}{\blangle\kern-3pt\blangle}
\newcommand{\bdrangle}{\brangle\kern-3pt\brangle}

\newcommand*{\bra}[1]{\langle{#1}|}
\newcommand*{\ket}[1]{|{#1}\rangle}
\newcommand*{\braket}[2]{\langle{#1}|{#2}\rangle}

\newcommand*{\bbra}[1]{\blangle{#1}\big|}
\newcommand*{\bket}[1]{\big|{#1}\brangle}
\newcommand*{\bbraket}[2]{\blangle{#1}\big|{#2}\brangle}

\newcommand*{\corr}[1]{\langle{#1}\rangle}
\newcommand*{\bcorr}[1]{\blangle{#1}\brangle}

\renewcommand{\ge}{\geqslant}

\newcommand{\vp}{\varphi}

\newcommand{\vt}{\vartheta}

\newcommand{\calC}{\mathcal{C}}
\newcommand{\calD}{\mathcal{D}}

\newcommand{\calF}{\mathcal{F}}
\newcommand{\calG}{\mathcal{G}}
\newcommand{\calH}{\mathcal{H}}

\newcommand{\calL}{\mathcal{L}}
\newcommand{\calM}{\mathcal{M}}

\newcommand{\calO}{\mathcal{O}}

\newcommand{\calX}{\mathcal{X}}
\newcommand{\calY}{\mathcal{Y}}

\newcommand{\ZZ}{\mathbb{Z}}

\newcommand{\RR}{\mathbb{R}}
\newcommand{\CC}{\mathbb{C}}

\DeclareMathOperator{\tr}{tr}
\DeclareMathOperator{\Tr}{Tr}
\DeclareMathOperator{\sgn}{sgn}

\newcommand{\const}{\mathrm{const}}
\newcommand{\Tt}{\mathrm{T}}

\let\Re\relax\DeclareMathOperator{\Re}{Re}

\DeclareMathOperator{\SO}{SO}

\DeclareMathOperator{\SL}{SL}
\DeclareMathOperator{\PSL}{PSL}
\DeclareMathOperator{\sL}{\mathfrak{sl}}
\DeclareMathOperator{\tSL}{\widetilde{\mathrm{SL}}}
\DeclareMathOperator{\Diff}{Diff}

\DeclareMathOperator{\Sch}{Sch}

\DeclareMathOperator{\AdS}{AdS}
\DeclareMathOperator{\tAdS}{\widetilde{AdS}}
\DeclareMathOperator{\HH}{H}
\DeclareMathOperator{\area}{area}
\DeclareMathOperator{\len}{length}

\newcommand{\lar}{{\leftarrow}}
\newcommand{\rar}{{\rightarrow}}
\newcommand{\IN}{\text{in}}
\newcommand{\OUT}{\text{out}}

\newcommand{\la}{\text{L}}
\newcommand{\ra}{\text{R}}

\newcommand{\GG}{\mathfrak{G}}
\newcommand{\tGG}{\widetilde{\GG}}

\newcommand{\TFD}{\mathrm{TFD}}

\newcommand{\SYK}{\text{SYK}}
\newcommand{\ch}{\text{ch}}
\newcommand{\g}{\text{g}}

\newcommand{\tu}{\tilde{u}}
\newcommand{\tz}{\tilde{z}}

\newcommand{\tpsi}{\tilde{\psi}}

\newcommand{\bA}{\breve{A}}
\newcommand{\bC}{\breve{C}}
\newcommand{\ww}{\mathbf{w}}
\newcommand{\Euc}{\mathrm{E}}
\newcommand{\unit}{\mathbf{1}}
\newcommand{\rcont}{\rho_{\text{cont}}}
\newcommand{\rdisc}{\rho_{\text{disc}}}
\newcommand{\rPl}{\rho_{\text{Pl}}}

\newcommand{\al}{\alpha}
\newcommand{\bt}{\beta}

\newcommand{\tht}{\theta}

\newcommand{\rpsi}{\mathring{\psi}}
\newcommand{\rPsi}{\mathring{\Psi}}
\newcommand{\rG}{\mathring{G}}
\newcommand{\rQ}{\mathring{Q}}

\newcommand{\Rho}{\mathrm{P}}
\newcommand{\Iota}{\mathrm{I}}

\newcommand{\lam}{\lambda}

\newcommand{\om}{\omega}

\newcommand{\ga}{\gamma}

\newcommand{\de}{\delta}
\newcommand{\De}{\Delta}

\def\eg{e.g.\ }
\def\ie{i.e.\ }

\newcommand{\ov}{\over}
\newcommand{\p}{\partial}

\makeatother

\title {Statistical mechanics of a two-dimensional black hole}

\author{Alexei Kitaev\footnote{kitaev@caltech.edu}\; and
S.\ Josephine Suh\footnote{suh@caltech.edu}\\
\normalsize\it California Institute of Technology, Pasadena, CA 91125, U.S.A.\vspace{0.5cm}}

\date{January 23, 2019}

\begin{document}

\setcounter{tocdepth}{2}

\maketitle
\begin{abstract}
The dynamics of a nearly-$\AdS_2$ spacetime with boundaries is reduced to that of two particles in the anti-de Sitter space. We determine the class of physically meaningful wavefunctions, and prescribe the statistical mechanics of a black hole. We demonstrate how wavefunctions for a two-sided black hole and a regularized notion of trace can be used to construct thermal partition functions, and more generally, arbitrary density matrices. We also obtain correlation functions of external operators.
\end{abstract}

\tableofcontents
\newpage

\begin{figure}
\centerline{\begin{tabular}{c@{\hspace{3cm}}c}
\includegraphics{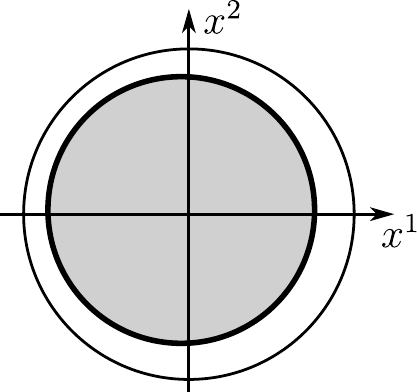} & \includegraphics{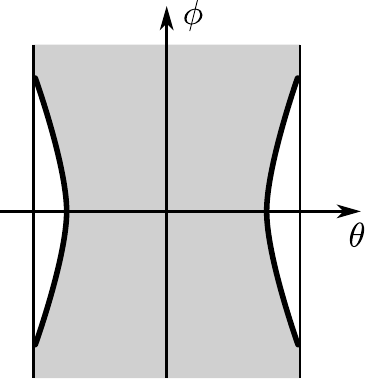}
\vspace{8pt}\\
a) & b)
\end{tabular}}
\caption{The Euclidean (a) and Lorentzian (b) geometries in the Jackiw-Teitelboim theory. The physical spacetime  (shaded) is embedded in the Poincare disk or the global anti-de Sitter space.}
\label{fig_geometry}
\end{figure}

\section{Introduction}

Dilaton gravity in $1+1$ dimensions is free of UV divergences and therefore should allow a fully quantum treatment. A particularly simple model, due to Jackiw~\cite{Ja85}, Teitelboim~\cite{Te83}, and Almheiri and Polchinski~\cite{AlPo14}, is well-studied semiclassically and represents a whole universality class. Its vacuum solution describes an eternal black hole. The spacetime is rigid with constant negative curvature, and thus can be embedded in $\widetilde{\AdS}_2$. The entire dynamics is associated with two time-like boundaries that are close to the spatial infinities. They may be regarded as particles moving in the anti-de Sitter space, see Figure~\ref{fig_geometry}. However, the quantization of this system and the construction of a canonical ensemble pose a challenge because the phase volume is infinite. This issue is also pertinent to higher-dimensional black holes and to the early Universe~\cite{HaHa83}. Completely resolving it in the simplest case might help to make progress in the more realistic settings.

There are in fact several related problems that are reasonable to consider together. The Sachdev-Ye-Kitaev (SYK) model~\cite{SaYe93,Kit.KITP,SoftMode} is a well-defined quantum system with a finite-dimensional Hilbert space. At low temperatures, it exhibits a collective soft mode with gravity-like behavior, whose effective action involves the Schwarzian derivative, $\Sch\bigl(f(x),x\bigr)
=\frac{f'''}{f'}-\frac{3}{2}\bigl(\frac{f''}{f'}\bigr)^2$\,~\cite{Kit.KITP,MS16,SoftMode}. Specifically, the Euclidean action is
\begin{equation}\label{ISch}
I_{\Sch}[\vp]=-\gamma\int_{0}^{L}\Sch(e^{i\vp},\ell)\,d\ell,
\end{equation}
where $\ell=J\tau_{\SYK}$ and $L=J\beta_{\SYK}$ are the imaginary time and inverse temperature in natural units, and $\gamma=\alpha_{S}N$ with $N$ being the system size and $\alpha_{S}$ some numerical coefficient. The dynamical degree of freedom is a smooth orientation-preserving map $\vp$ from a circle of length $L$ (representing the imaginary time) to the standard circle of length $2\pi$. The effective action \eqref{ISch} is applicable when $J\beta_{\SYK}\gg 1$ and $N\gg 1$. Under these assumptions, the SYK partition function is given by the formula
\begin{equation}\label{FSYK}
\ln Z_{\SYK}\approx -\beta_{\SYK}E_0+S_0+\ln Z_{\Sch},
\end{equation}
where $S_0=Ns_0+\const$ is a so-called ``zero-temperature entropy'', and $Z_{\Sch}$ is defined as the integral of $\exp(-I_{\Sch}[\vp])$ with a suitable measure.

While the SYK problem has two large parameters, $Z_{\Sch}$ depends only on their ratio, $\beta=L/\gamma$. Indeed, the effective action can be written as $-(2\pi/\beta)\int_{0}^{2\pi}\Sch(e^{i\vp},\theta)\,d\theta$. If $\beta\ll 1$, the problem is classical. The minimum action is achieved at the function $\vp(\theta)=\theta$; hence
\begin{equation}
Z_{\Sch}(\beta) \sim \exp\Bigl(-\min_{\vp} I_{\Sch}[\vp]\Bigr)
=e^{2\pi^2/\beta}\qquad \text{for }\, \beta=L/\gamma\ll 1.
\end{equation}
In general, the Schwarzian partition function and density of states are as follows:
\begin{equation}\label{ZSch}
Z_{\Sch}(\beta) =\int_{0}^{\infty}e^{-\beta E_{\Sch}}\rho_{\Sch}(E_{\Sch})\,dE_{\Sch}
\propto \beta^{-3/2} e^{2\pi^2/\beta},\qquad
\rho_{\Sch}(E_{\Sch}) \propto\sinh\biggl(2\pi\sqrt{2E_{\Sch}}\biggr).
\end{equation}
(The unspecified coefficients of proportionality depend on the normalization of the integration measure.) This result was derived in several ways, in particular, by solving the SYK model in the double-scaling limit~\cite{randmat}, by evaluating the Schwarzian path integral exactly~\cite{StWi17}, and by reducing the problem to Liouville quantum mechanics~\cite{BaAlKa16,BaAlKa17}. The last method is the most powerful one as it can also be used for calculation of matrix elements. Two different reductions of the Schwarzian theory to a 2D CFT with a large central charge were proposed in~\cite{MeTuVe17}. Our approach will be similar to that of~\cite{BaAlKa16,BaAlKa17}, but we consider a more general problem, one that has two parameters but fewer infinities to worry about. As a consequence, the wavefunction, including the overall factor, is defined unambiguously. 

The Schwarzian action also arises from two-dimensional Jackiw-Teitelboim theory, which involves the metric tensor $g$ and a dilaton field $\Phi$\,~\cite{Jen16,MSY16,EMV16}. The Euclidean action is
\begin{equation}\label{IJT}
I_{\text{JT}}[g,\Phi]
=-\frac{1}{4\pi}\int_{D}\Phi(R+2)\sqrt{g}\,d^2x
-\frac{1}{2\pi}\int_{\partial D}\Phi K\,d\ell,
\end{equation}
where $D$ is a disk, $d\ell$ is the boundary length element, and $K$ is the extrinsic curvature. The boundary term is such that the variation of the action depends only on $\delta g$ and $\delta\Phi$ but not their derivatives; this is necessary to define boundary conditions. The condition $\Phi|_{\partial D}=\Phi_*$ (for some constant $\Phi_*$) is imposed and the total boundary length $L$ is fixed.

The bulk term in \eqref{IJT} gives the constraint $R=-2$ but vanishes on-shell. Thus one can isometrically embed (or more generally, immerse) $D$ in the Poincare disk so that the action becomes $-\frac{\Phi_*}{2\pi}\int_{\partial D}K\,d\ell$. It is convenient to also add a trivial term proportional to $L$:
\begin{equation}\label{IJT1}
I_{\g}=I_{\text{JT}}+\gamma L=-\gamma\int_{\partial D}(K-1)\,d\ell,\qquad\quad
\gamma=\frac{\Phi_*}{2\pi}.
\end{equation}
Now, consider polar coordinates $r,\vp$ on the Poincare disk as functions on the curve $\partial D$, which is parametrized by the proper length $\ell$. If $L\gg 1$, it is reasonable to assume that $r(\ell)$ is close to~$1$ and that the curve is roughly parallel to the unit circle. Then
\begin{equation}\label{K-1}
K-1 \approx\Sch(e^{i\vp(\ell)},\ell).
\end{equation}
(For the reader's convenience, this equation is derived in the beginning of the next section.) We conclude that action \eqref{IJT1} is approximately equal to the Schwarzian action.

Not making any approximations, one can still simplify action \eqref{IJT1}. By the Gauss-Bonnet theorem, $\int_{\partial D}K\,d\ell$ equals $2\pi$ plus the area enclosed by the curve. Then we arrive at the following geometric action and global constraint for a closed curve $X$ in the Poincare disk:
\begin{equation}\label{I0}
\wideboxed{
I_{\g}[X]=-\gamma\bigl(\area[X]-L+2\pi\bigr),\qquad \len[X]=L.
}
\end{equation}
We assume that $\gamma>0$ and take $\,\area[X]$ to be positive if $X$ goes counterclockwise. As has just been explained, this model is classically equivalent to Jackiw-Teitelboim theory. However, the functional integrals appear to be different. Indeed, each of the integrals should include all curves (even self-intersecting ones) for which the corresponding action makes sense. The area is defined for all closed curves, whereas in the dilaton problem, a curve should bound an immersed disk. On the other hand, both models are quantum mechanically equivalent to the Schwarzian model if $\gamma$ and $L$ are large. The rough argument is that under this assumption, typical curves have $K\approx 1$ and do not wiggle too much, so that one can use equation \eqref{K-1}. We will refer to the condition $\gamma,L\gg 1$ as the \emph{Schwarzian limit}.

There are several ways to think about problem \eqref{I0}. One is that it describes a particle with an imaginary charge in a constant magnetic field. We prefer a slightly different interpretation: that there is a particle with spin $\nu=-i\gamma$ on the hyperbolic plane. One may also view the region enclosed by a curve $X$ as a balloon whose wall is flexible but cannot be stretched; the air pressure inside tries to maximize the two-dimensional volume, that is, $\area[X]$. 

To elaborate on the previous statement regarding the fully quantum geometric model, we need to define the functional integral. This involves regularization, whereby $I_{\g}[X]$ is replaced by another action $I[X]$ that is quadratic in derivatives, see Section~\ref{sec_Euc}. We choose not to include the term $-2\pi\gamma$ in the regularized action, which results in the multiplication of the partition function by $e^{-2\pi\gamma}$. This partition function will be expressed as $Z=\int e^{-\beta E}\rho(E)\,dE$, with $\rho(E)$ calculated explicitly. In general, the renormalized parameters $\beta$ and $E$ depend on $L$, $E_{\g}$, and the UV cutoff. But in the Schwarzian limit, there is a cutoff-independent renormalization scheme,
\begin{equation}
\beta=L/\gamma,\qquad\quad
E=\gamma E_{\g}-\frac{\gamma^2}{2}+\frac{1}{8},\qquad \gamma E_{\g}=E_{\Sch},
\end{equation}
under which
\begin{equation}
\ln Z(\beta)\approx -\beta\,\bigl(-\tfrac{\gamma^2}{2}+\tfrac{1}{8}\bigr)
-2\pi\gamma + \ln Z_{\Sch}(\beta),\qquad
\rho(E)\approx e^{-2\pi\gamma}\rho(E_{\Sch}).
\end{equation}

Let us stress some unusual features of the geometric model. To define the partition function, we divide an infinite Euclidean path integral by the volume of the hyperbolic plane (and also by $2\pi$, so that we are actually dividing by the volume of the Euclidean symmetry group $\PSL(2,\RR)$). This makes for a reasonable statistical mechanics problem, but does not guarantee that it can be formulated in terms of a Hilbert space and a Hamiltonian. In fact, although $Z=\int e^{-\beta E}\rho(E)\,dE$ with $\rho(E)\geq 0$, we cannot write $Z=\Tr\lt(e^{-\beta H}\rt)$. Or rather, a formula like that exists, but the trace is not the conventional one. We will see that
\begin{equation}
Z(\beta)=\frac{1}{2}\tr\bigl(e^{-\beta H}\Rho\bigr),
\end{equation}
where $\tr$ is the usual trace divided by the volume of the Lorentzian symmetry group $\tSL(2,\RR)$, and $\Rho$ commutes with the Hamiltonian. Furthermore, the thermofield double state is given by $Z^{-1/2}e^{-\beta H/2}\Phi$ for a certain $\Phi$ that is anti-Hermitian, squares to $-\Rho$, and commutes with $H$.

Some existing work related to our subject matter is as follows: the semi-classical wavefunction for the Hartle-Hawking state in Jackiw-Teitelboim gravity was studied in \cite{HaJa18}, and the quantum entropy of the Hartle-Hawking state in the same theory was studied in \cite{Lin18}.

\section{Geometry and classical trajectories}

The metric on the hyperbolic plane $\HH^2$ (with unit curvature radius) is described by the Poincare disk model:
\begin{equation} \label{Pdisk}
ds^2=4\,\frac{(dx^1)^2+(dx^2)^2}{(1-r^2)^2},\qquad\quad
r^2=(x^1)^2+(x^2)^2.
\end{equation}
Depending on the situation, it may be convenient to use polar coordinates $(r,\vp)$ or complex variables $z=x^1+ix^2$ and $\bar{z}=x^1-ix^2$. The metric has a symmetry group $\GG$ that is isomorphic to $\PSL(2,\RR)=\SL(2,\RR)/\{\pm1\}$. It consists of all linear fractional maps $z\mapsto\frac{az+b}{cz+d}$ preserving the unit disk, where the matrix $\left(\kern-1pt\begin{smallmatrix}a&b\\ c&d\end{smallmatrix}\kern-0.5pt\right)$ has unit determinant and is defined up to sign. To work with spinors, we need to fix a gauge, \ie a cross section of the principal $\widetilde{\SO}(2)$ bundle over $\HH^2$ (where the tilde indicates the universal cover). This is essentially equivalent to choosing an orthonormal frame $(v_1,v_2)$ at each point. The spin connection is given by the set of coefficients
\begin{equation}
\tensor{\omega}{_\mu^a_b}=\omega_{\mu}\tensor{\Xi}{^a_b},
\end{equation}
where the matrix $\begin{pmatrix} \tensor{\Xi}{^1_1} & \tensor{\Xi}{^1_2}\\[2pt] \tensor{\Xi}{^2_1} & \tensor{\Xi}{^2_2} \end{pmatrix} =\begin{pmatrix} 0 & -1\\ 1 & 0 \end{pmatrix}$ is the rotation generator. For example, in the \emph{disk gauge} $(\mathring{v}_1,\mathring{v}_2)$,
\begin{equation}
\label{dgauge}
\begin{pmatrix} \mathring{v}_1^1 & \mathring{v}_2^1\\[2pt]
\mathring{v}_1^2 & \mathring{v}_2^2 \end{pmatrix}
=\frac{1-r^2}{2} \begin{pmatrix} 1 & 0\\ 0 & 1 \end{pmatrix},\qquad\qquad
\figbox{1.0}{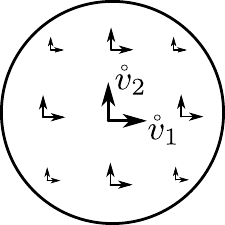}
\end{equation}
the spin connection is
\begin{equation}
(\mathring{\omega}_r,\mathring{\omega}_\vp)
=\left(0,\,\frac{2r^2}{1-r^2}\right).
\end{equation}

Let us consider a closed, counterclockwise curve $X$ parametrized by proper length $\ell$, and let $\alpha$ be the angle between the tangent to the curve and circumferential direction. Then
\begin{empheq}[left=\figbox{1.0}{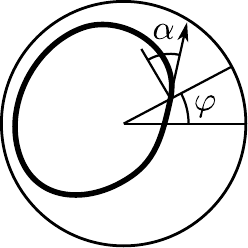}\hspace{2cm}]{gather}
\label{rpa1}
r\vp'=\frac{1-r^2}{2}\,\cos\alpha,\qquad\quad
r'=\frac{1-r^2}{2}\,\sin\alpha,\\[8pt]
\label{rpa2}
K=\vp'-\alpha'+\mathring{\omega}_\mu(X^\mu)'
=\frac{1+r^2}{2r}\cos\alpha-\alpha'.
\end{empheq}
(In the last equation, $\vp'-\alpha'$ is the rotation rate of the tangent vector relative to the local frame.) Knowing $\vp$ as a function of $\ell$, one can try to solve for $r$ and $\alpha$. The task is simplified if $1-r$ and $\alpha$ are small. In the first approximation, $1-r\approx\vp'$ and $\alpha\approx-\vp''/\vp'$. Hence
\begin{equation}
K-1\approx \frac{1}{2}\,\vp'^2
-\frac{1}{2}\biggl(\frac{\vp''}{\vp'}\biggr)^2
+\biggl(\frac{\vp''}{\vp'}\biggr)'
=\Sch(e^{i\vp},\vp)\,\vp'^2+\Sch(\vp,\ell)=\Sch(e^{i\vp},\ell),
\end{equation}
as was stated in the introduction.

We now discuss the variational problem \eqref{I0}. Since the hyperbolic plane has scalar curvature $R=-2$, the area inside a closed curve $X$ is equal to  $\int(-R/2)\sqrt{g}\,d^2x=\int\omega_{\mu}dX^{\mu}$. The last expression represents the holonomy of a local frame; it serves as a (gauge-dependent) analogue of the area for open curves. Imposing the constraint\, $\len[X]=L$ using a Lagrange multiplier $E_{\g}$, we obtain the modified action $I_{\g}-E_{\g}L$ which is expressed in detail as 
\begin{equation}\label{Imod}
I_{\g}[X]-E_{\g}\kern1pt\len[X]
= \int\bigl(M\,d\ell-\gamma\omega_{\mu}\,dX^{\mu}\bigr)-2\pi\gamma,\qquad\quad
M=\gamma-E_{\g}.
\end{equation}
It is natural to assume that $M>0$ so that classical trajectories are stable to ripples, and we have already stated that $\gamma>0$ so that the counterclockwise direction is preferred. (These assumptions are relevant to quantization and thermodynamics rather than equations of motion.) Recall that in the original problem, the path length $L$ is the inverse temperature. Therefore, one may interpret $L^{-1}I_{\g}$ as free energy, $E_{\g}$ as energy, and $S_{\g}=-(I_{\g}-E_{\g}L)$ as entropy. Such interpretations are good semiclassically, but there are two caveats concerning their use in the quantum case. First, the action \eqref{Imod} has no minima and only saddle points, which are circles of a certain length $L$. Such a circle is minimal if $L$ is fixed, but represents a maximum with respect to $L$. For this reason, we will consider the fixed length variant of the path integral, then express the partition function and discuss energy and entropy. The second issue is that the path integral definition involves some renormalization of parameters, see Section~\ref{sec_Euc}.

To find the extremal paths, it is convenient to introduce an auxiliary time variable $\tau$ and write the action as $\int\calL_{\Euc}\,d\tau-2\pi\gamma$ with the Euclidean Lagrangian
\begin{equation}
\calL_{\Euc}= M|\dot{X}|-\gamma\omega_{\alpha}\dot{X}^{\alpha}.
\end{equation}
(Here we have used the notation $|v|=\sqrt{g_{\alpha\beta}v^{\alpha}v^{\beta}}$.) The Euclidean momentum is
\begin{equation}
(p_{\Euc})_{\alpha}=\frac{\partial\calL_{\Euc}}{\partial\dot{X}^{\alpha}}
=M\,\frac{\dot{X}_{\alpha}}{|\dot{X}|}-\gamma\omega_{\alpha},
\end{equation}
whereas the Hamiltonian is identically zero. Note the momentum satisfies the constraint
\begin{equation}
|p_{\Euc}+\gamma\omega|^2=M^2.
\end{equation}
The equation of motion,
\begin{equation}\label{eqmo}
MK=\gamma,
\end{equation}
is made intuitive using the balloon picture: $M$ is the tension of the balloon wall, and $\gamma$ is the air pressure inside.

The solutions of equation \eqref{eqmo} are curves with a constant curvature $K$. The thermodynamic interpretation requires that the curves be closed. Closed curves with constant curvature in the Poincare disk are circles; all circles with the same curvature are related to each other by symmetry transformations. Thus a representative solution is as follows:
\begin{equation}\label{std_circle}
z\bar{z}=r^2,\qquad\text{where}\quad
r=K-\sqrt{K^2-1},\quad\: K>1.
\end{equation}
Some of its characteristics are
\begin{equation}
L=\frac{2\pi}{\sqrt{K^2-1}},\qquad E_{\g}=\gamma-M=\gamma(1-K^{-1}),\qquad
S_{\g}=-(I_{\g}-E_{\g}L)=2\pi\gamma\sqrt{1-K^{-2}}.
\end{equation}
Using these relations, we can replace one of the conditions of the Schwarzian limit with equivalent ones,
\begin{equation} \label{Schlim2}
L\gg 1 \quad\:\Leftrightarrow\quad\:
M/\gamma\approx 1 \quad\:\Leftrightarrow\quad\:
E_{\g}\ll\gamma.
\end{equation}

To describe all circles with a given curvature, let us use the variables
\begin{equation}\label{z1z2}
z_1=z,\qquad z_2=\bar{z}^{-1}.
\end{equation}
In this notation, the standard circle \eqref{std_circle} is the set of points such that $z_2=r^{-2}z_1$. The transformation $z_1\mapsto z_2$ is a linear fractional map; symmetries of the Poincare disk take it to conjugate maps because they act on $z_1$ and $z_2$ in the same way. Thus the relation between $z_1$ and $z_2$ assumes the form
\begin{equation}\label{z2Vz1}
z_2=V(z_1),\qquad\: \text{where}\quad\:V(z)=\frac{az+b}{cz+d},\quad\:
a+d=2K,\quad ad-bc=1.
\end{equation}
A more careful analysis gives the additional conditions
\begin{equation}\label{abcd_cond1}
a>1,\qquad d\in\RR,\qquad b=-\bar{c}.
\end{equation}

To establish a correspondence between Euclidean and Lorentzian spacetimes, we embed both $\HH^2$ and $\tAdS_2$ into a suitable complex manifold $\calM$. The latter may be regarded as a complexification of the hyperbolic plane. It consists of all pairs of distinct points on the Riemann sphere $\CC\cup\{\infty\}$, whereas $\HH^2$ is the subset of pairs $(z_1,z_2)=(z,\bar{z}^{-1})$ with $|z|<1$. The embedding of anti-de Sitter space is chosen such that some time slice coincides with a diameter of the Poincare disk. This is the embedding $\mathring{J}$ from Ref.~\cite{SL2R}, which we will now describe.

The space $\AdS_2$ consists of pairs of distinct points on the unit circle. Its universal cover $\widetilde{\AdS}_2$ is parametrized by real variables $\vp_1$, $\vp_2$ such that $0<\vp_1-\vp_2<2\pi$. A more standard description uses global anti-de Sitter time $\phi$ and spatial coordinate $\theta$,
\begin{equation}
\phi=\frac{\vp_1+\vp_2}{2},\qquad
\theta=\frac{\pi-\vp_1+\vp_2}{2},
\end{equation}
in terms of which the metric is
\begin{equation} \label{gAdS_2}
ds^2=\frac{-d\phi^2+d\theta^2}{\cos^2\theta}.
\end{equation}
The embedding of $\tAdS_2$ in the complex manifold $\calM$ is given by the following equations, where we have also introduced an analogue of Schwarzschild coordinates $(r,t)$ covering the shaded region:
\begin{equation}\label{embed}
\figbox{1.0}{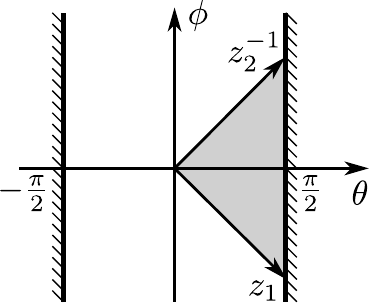}\hspace{2cm}
\begin{aligned}
z_1&=\tan\biggl(\frac{\pi}{4}-\frac{\vp_1}{2}\biggr)
=\tan\biggl(\frac{\theta-\phi}{2}\biggr)=re^{-t},\\[10pt]
z_2&=\tan\biggl(\frac{\pi}{4}-\frac{\vp_2}{2}\biggr)
=\cot\biggl(\frac{\theta+\phi}{2}\biggr)=\bigl(re^{t}\bigr)^{-1}.
\end{aligned}
\end{equation}
For certain purposes, functions on $\HH^2$ are not analytically continued to the whole of $\widetilde{\AdS}_2$, but only to the Schwarzschild patch; Euclidean coordinates $r$ and $\vp$ correspond to $r$ and $it$, respectively.

Now we describe classical Lorentzian trajectories. The symmetric ones are given by the equation $r=\const$. They consist of two disjoint pieces as shown in Figure~\ref{fig_geometry}b on page~\pageref{fig_geometry}, and may be viewed as lines on a topological cylinder, the complex trajectory embedded in $\calM$. The Euclidean section of the cylinder $\lt(z\bar{z}=r^2 \text{ in } \HH^2\rt)$ is a circle crossing both lines, see Figure~\ref{fig:Phides}a on page~\pageref{fig:Phides}. In the semiclassical picture, it describes tunneling between propagating states. A general Lorentzian trajectory is given by equation \eqref{z2Vz1} with real coefficients $a,b,c,d$.

To conclude the geometric formalism, let us discuss the choice of gauge in Lorentzian spacetime. A nice property of the disk gauge is that it admits an analytic continuation to $\calM$ (albeit with singularities), and is real on both $\HH^2$ and $\tAdS_2$ if the above embeddings are used~\cite{SL2R}. However, its anti-de Sitter version is regular only for $|\phi\pm\theta|<\pi$. A so-called \emph{tilde gauge} does not have this drawback. The corresponding local frame is proportional to the $(\phi,\theta)$ coordinate frame,
\begin{equation}
\label{tgauge}
\begin{pmatrix} \tilde{v}_0^0 & \tilde{v}_1^0\\[2pt]
\tilde{v}_0^1 & \tilde{v}_1^1 \end{pmatrix}
=(\cos\theta) \begin{pmatrix} 1 & 0\\ 0 & 1 \end{pmatrix},\qquad\qquad
\figbox{1.0}{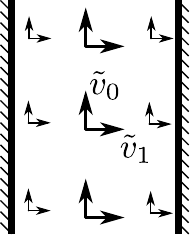}
\end{equation}
and the spin connection is $(\widetilde{\omega}_\phi,\widetilde{\omega}_\theta) =(\tan\theta,\,0)$. The full set of spin connection coefficients $\tensor{\omega}{_\mu^a_b}=\omega_{\mu}\tensor{\Xi}{^a_b}$ involves the Lorentz boost generator $\Xi=\left(\begin{smallmatrix}0&1\\ 1&0\end{smallmatrix}\right)$. Relative to the tilde gauge, the disk gauge is Lorentz boosted by $\ln\frac{\cos((\phi+\theta)/2)}{\cos((\phi-\theta)/2)}$. The tilde gauge is compatible with a different embedding of $\tAdS_2$ in $\calM$, namely $(\tz_1,\tz_2)=(e^{i\vp_1},e^{i\vp_2})$.

\section{Euclidean path integral}\label{sec_Euc}

Ideally, we would like to define a path integral version of problem \eqref{I0}. The most useful object is the propagator,
\begin{equation}\label{prop_I0}
G_{\g}(x_1,x_0;L)=
e^{\gamma(-L+2\pi)}\!\!\int\limits_{\text{paths}/\Diff[0,1]}\hspace{-0.5cm}
DX\,\,\delta\bigl(\len[X]-L\bigr)\,
\exp\biggl(\gamma\int\omega_{\alpha}dX^{\alpha}\biggr),
\end{equation}
where $X:\,[0,1]\to \HH^2$ is a path from $x_0$ to $x_1$ considered up to reparametrizations. However, path integrals of this type are sensitive to the UV cutoff. The simplest short-distance regularization procedure is to replace smooth paths with jagged ones, consisting of straight sections of length~$\epsilon$. When $\epsilon$ is small, path statistics are described by a quadratic action which generates the diffusion equation. The effective time $\beta$ in the diffusion problem is proportional to $L$ with an $\epsilon$-dependent coefficient. Thus the regularized action and corresponding propagator are
\begin{empheq}[box=\widebox]{gather}
\label{tI}
I[X]=\int_{0}^{\beta}d\tau\, 
\left(\frac{1}{2}\,g_{\alpha\beta}\dot{X}^{\alpha}\dot{X}^{\beta}
-\gamma\omega_{\alpha}\dot{X}^{\alpha}\right) ,
\\[5pt]
G(x_1,x_0;\beta)=\int_{\substack{X(0)=x_0\\ X(\beta)=x_1}} DX\, e^{-I[X]}.
\end{empheq}
The latter is well-defined, whereas the original propagator involves some non-universal parameters $b_1,b_2$:
\begin{equation}\label{G_vs_tG}
G_{\g}(x_1,x_0;L)=e^{(b_{2}-\gamma)L+2\pi\gamma}\,G(x_1,x_0;\beta),\qquad
\beta=b_1^{-1}L.
\end{equation}
We will consider three more specific regularization recipes:
\begin{enumerate}
\item \label{reg1} For general values of $\gamma$ and $L$, one has to take the $\epsilon\to 0$ limit, or at least to assume that $\epsilon\ll\min\{\gamma^{-1},1,L\}$. Then equation \eqref{G_vs_tG} holds for $b_1=2/\epsilon$ and $b_2=0$. A similar result is derived in Sections 9.1--9.2 of Polyakov's book~\cite{Polyakov}, but we will give a simpler argument. Unfortunately, the arbitrariness of $\epsilon$ complicates the comparison with the Schwarzian problem.
\item \label{reg2} In the Schwarzian limit where both $\gamma$ and $L$ are large, $\epsilon$ need not be very small. If we assume that $\gamma^{-1}\ll\epsilon\ll 1$, then $b_1$ and $b_2$ are $\epsilon$-independent, namely, $b_1\approx\gamma$ and $b_2\approx\gamma/2$. However, the accuracy of this approximation is not sufficient to match the Schwarzian partition function.
\item \label{reg3} The correct match is achieved if $b_1=\gamma$ and $b_2=\gamma/2+1/(8\gamma)$. This will be shown later by calculating the density of states.
\end{enumerate}

\begin{figure}
\centerline{\begin{tabular}{c@{\hspace{1.5cm}}c}
\includegraphics{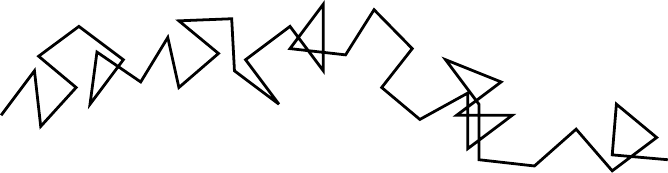} & \includegraphics{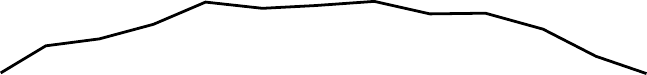}
\vspace{8pt}\\
a) & b)
\end{tabular}}
\caption{Typical path shapes, a) for $\epsilon M\ll 1$ and b) for $\epsilon M\gg 1$.}
\label{fig_jagged}
\end{figure}

The qualitative difference between cases~\ref{reg1} and~\ref{reg2} is in the shape of a typical path as we zoom in, see Figure~\ref{fig_jagged}. To justify both claims, we first separate path properties at small distances from those at intermediate and large distances. At distances $\Delta x\sim 1$ in the Schwarzian limit, one may use classical equations. We have already found from their analysis that $M\approx\gamma$. At short distances, the parameter $M$ is important, but $\gamma$ (as the coefficient in the area term) is not. Indeed, if $\Delta x\ll 1$, one may replace the hyperbolic plane with $\RR^2$. The contribution to the area from a ($\Delta x$)-size section of the path with fixed endpoints varies at most by $(\Delta x)^2$. Thus, if $\Delta x\ll\gamma^{-1/2}$, the area term is negligible. In the Schwarzian limit, the area term may actually be ignored if $\Delta x\ll 1$ because, as we will see shortly, typical paths are almost straight.

Let us discuss the short-distance behavior in more detail. For this purpose, we work in $\RR^2$ and neglect the area term. We also drop the trivial term $\gamma(L-2\pi)$ and simultaneously subtract $\gamma$ from $E_{\g}$ so that $I_{\g}[X]$ vanishes but the modified action remains the same (up to an additive constant). The number $E_{\g}=-M$ plays the role of a chemical potential for a small piece of a path. The simplified propagator $G_{\ch}$ (excluding the $e^{\gamma(-L+2\pi)}$ factor) is completely characterized by the integration measure
\begin{equation}
D\mu=\prod_{j=1}^{n}
\Bigl((2\pi\epsilon)^{-1}\delta(|X_{j}-X_{j-1}|-\epsilon)\Bigr)
\prod_{j=1}^{n-1}dx_j,\qquad\quad
X_0=x_0,\quad X_n=x_1,\quad n=\frac{L}{\epsilon}.
\end{equation}
A convenient analogy is a fluctuating polymer chain. Suppose that one end of a chain, $x_0$, is fixed at some location far away (compared to $\epsilon$). The probability density of the other end, $f_{\ch}(x_1,L)=G_{\ch}(x_1,x_0;L)$, satisfies the equation
\begin{equation}\label{polychain_eq}
f_{\ch}(x,L+\epsilon)=\frac{1}{2\pi\epsilon}
\int\delta\bigl(|x-y|-\epsilon\bigr)\,f_{\ch}(y)\,dy.
\end{equation}
Now imagine pulling on that end with force $p_{\Euc}$. Applying the force and passing to the grand canonical ensemble means multiplying $f_{\ch}(x,L)$ by $\exp((p_{\Euc})_{\mu}x^{\mu}-ML)$. If the chain is long enough to attain the thermodynamic limit, the modified $f_{\ch}$ should be constant. Thus, the original function is $f_{\ch}(x,L)\propto\exp(-(p_{\Euc})_{\alpha}x^{\alpha}+ML)$. Plugging this ansatz into \eqref{polychain_eq}, we find the dispersion relation
\begin{equation}\label{Euc_disp}
M=\frac{\ln I_{0}(\epsilon|p_{\Euc}|)}{\epsilon}
\approx\begin{dcases}
\frac{\epsilon p_{\Euc}^2}{4} & \text{if } \epsilon|p_{\Euc}|\ll 1\\[2pt]
|p_{\Euc}|& \text{if } \epsilon|p_{\Euc}|\gg 1
\end{dcases}
\end{equation}
where $I_0$ is the modified Bessel function. Typical path geometries in the two cases are shown in Figure~\ref{fig_jagged}. Thus in the $\epsilon\to 0$ limit, equation \eqref{polychain_eq} becomes $\partial_{L}f_{\ch}=(\epsilon/4)\nabla^2f_{\ch}$. It can be reduced to the standard diffusion equation that corresponds to the quadratic action $I[X] =\frac{1}{2}\int_{0}^{\beta}\dot{X}^2d\tau$:
\begin{equation}
\partial_{\beta}f_{\ch}=\frac{1}{2}\,\nabla^2f_{\ch},\qquad\qquad
\text{where}\quad
\beta=\frac{\epsilon}{2}\,L.
\end{equation}
This proves claim~\ref{reg1}. As for claim~\ref{reg2}, the conditions $\epsilon\gamma\gg 1$ and $M\approx\gamma$ are consistent with the second case of equation \eqref{Euc_disp}, namely, $M\approx|p_{\Euc}|$ for $|p_{\Euc}|\approx\gamma$. To first order in $|p_{\Euc}|-\gamma$, this dispersion relation can also be written as
\begin{equation}
M\approx \frac{p_{\Euc}^2}{2\gamma}+\frac{\gamma}{2}.
\end{equation}
Hence
\begin{equation}
f_{\ch}(x,L)=e^{(\gamma/2)L}f(x,\beta),\qquad
\text{where}\quad
\partial_{\beta}f=\frac{1}{2}\,\nabla^2f,\qquad
\beta=\frac{L}{\gamma}.
\end{equation}

From here on, we study the quadratic action \eqref{tI}. The propagator can be obtained by solving the diffusion equation with a suitable initial condition:
\begin{equation}\label{diffusion}
\wideboxed{
\partial_{\tau}G(x_1,x_0;\tau)=\frac{1}{2}\,\nabla_{x_1}^{2}G(x_1,x_0;\tau),
\qquad\quad
\lim_{\tau\to 0}G(x_1,x_0;\tau)=\frac{\delta(x_1-x_0)}{\sqrt{g(x_1)}}
}
\end{equation}
where the Laplacian involves the covariant derivative acting on $\nu$-spinors,
\begin{equation} \label{coder}
\nabla_{\alpha}\psi
=(\partial_{\alpha}-i\nu\omega_{\alpha})\psi,\qquad\quad
\nu=-i\gamma.
\end{equation}
The partition function is defined as the integral of $e^{-I[X]}$ over closed paths. To make the quantity finite, we divide it by the volume of the symmetry group $\PSL(2,\RR)$, which is $2\pi$ times the area of the hyperbolic plane:
\begin{equation} \label{Zdef}
\wideboxed{
Z(\beta)=\frac{1}{\operatorname{vol}(\PSL(2,\RR))}\,
\int_{\HH^2}d^2x\,\sqrt{g(x)}\,G(x,x;\beta)
=\frac{1}{2\pi}\,G(0,0;\beta).
}
\end{equation}

In the remainder of this section, we solve equation \eqref{diffusion} and analyze the resulting expression for the partition function. Without loss of generality, we may assume $x_0=0$; then the solution $\rG(x,0;\tau)$ is rotationally symmetric, \ie independent of the polar angle $\vp$. (The ring accent indicates the disc gauge; we generally put it only where it matters.) The Laplacian on the hyperbolic plane is related to the $\SL(2,\RR)$ Casimir operator $Q$,
\begin{equation} \label{nabla}
-\nabla^2=Q+\nu^2.
\end{equation}
The representation of $\tSL(2,\RR)$ by spinors on $\HH^2$ is described in~\cite{SL2R}. However, some results hold only for real $\nu$, so the corresponding arguments have to be redone. The disk gauge expression for the Casimir operator is
\begin{equation} \label{Q_d_gauge}
\rQ=-(1-u)^2\,\bigl(u\partial_{u}^2+\partial_{u}\bigr)
+\frac{1-u}{4u}\Bigl((m-\nu)^2-(m+\nu)^{2}u\Bigr),
\end{equation}
where
\begin{equation}
u=r^2,\qquad m=\nu-i\partial_{\vp}.
\end{equation}
While $\rQ$ is not Hermitian for imaginary values of $\nu$, it becomes Hermitian when restricted to the $m=\nu$ subspace, which consists of rotationally symmetric functions. In this special case,
\begin{equation}\label{Q0}
\rQ=-(1-u)^2\,\bigl(u\partial_{u}^2+\partial_{u}\bigr)+\gamma^2(1-u).
\end{equation}
The functions in question depend only on $u\in[0,1)$, but we should use the correct inner product and boundary condition at $u=0$. The inner product is given by the integral over the hyperbolic plane
\begin{equation}
\braket{f_1}{f_2}=2\pi\int_{0}^{1}f_1(u)^{*}f_2(u)\,\frac{2\,du}{(1-u)^2}.
\end{equation}
Therefore, normalizable functions vanish at $u\to 1$ faster than $(1-u)^{1/2}$. To determine the condition at the origin, we notice that eigenfunctions of $\rQ$ have the asymptotic form $f(u)\approx a+b\ln u$ for $u\to 0$. But in two dimensions, a singularity at the origin is not allowed; hence $b=0$. A more general condition is that $f(0)$ is finite and $\lim_{u\to 0} u\partial_{u}f(u)=0$. It guarantees the Hermicity of $\rQ$ because $\braket{f_1}{\rQ f_2}-\braket{\rQ f_1}{f_2}=4\pi u\bigl(f_1^*(\partial_{u}f_2)-(\partial_{u}f_1^*)f_2\bigr)\big|_{u=0}$.

Let us find an eigenbasis of the operator $\rQ$ acting in the Hilbert space we have just described. The $m=\nu$ eigenfunctions are as follows~\cite{SL2R}:
\begin{equation} \label{Eucpsi}
\rpsi^{\nu}_{\lambda}(u)=\rpsi^{\nu}_{\lambda}(0)\cdot
(1-u)^\lambda\,\hgfs(\lambda+\nu,\,\lambda-\nu,\,1;\,u),
\qquad\quad
\rQ\rpsi^{\nu}_{\lambda}=\lambda(1-\lambda)\rpsi^{\nu}_{\lambda},
\end{equation}
where $\hgfs(a,b,c;x)=\Gamma(c)^{-1}\hgf(a,b,c;x)$ is the scaled hypergeometric function and $\rpsi^{\nu}_{\lam}(0)$ is for now simply a normalization factor. The eigenvalue $\lambda(1-\lambda)$ must be real; hence $\lambda$ is real or has the form $\frac{1}{2}+is$ with a real $s$. Eliminating the $\lambda\leftrightarrow 1-\lambda$ redundancy, there are three mutually exclusive cases: $\lambda=\frac{1}{2}$,\, $\lambda>\frac{1}{2}$, and $\lambda=\frac{1}{2}+is$ with $s>0$. It follows from the asymptotic expression
\begin{equation}\label{Q0ef}
{\rpsi^{\nu}_{\lambda}(u) \ov \rpsi^{\nu}_{\lambda}(0)}\approx
\frac{\Gamma(1-2\lambda)}{\Gamma(1-\lambda+\nu)\,\Gamma(1-\lambda-\nu)}\,
(1-u)^{\lambda}
+\frac{\Gamma(2\lambda-1)}{\Gamma(\lambda+\nu)\,\Gamma(\lambda-\nu)}\,
(1-u)^{1-\lambda}\quad\:
\text{for }\, u\to 1
\end{equation}
that the first two sets of eigenfunctions are not normalizable or $\delta$-normalizable. Thus we restrict to the third case. Fixing $\rpsi^{\nu}_{\lambda}(0)=\bigl((2\pi)^{-1}\sinh(2\pi s)/(\cosh(2\pi\ga) + \cosh(2\pi s))\bigr)^{1/2}$, we have
\begin{equation}
\bbraket{\psi^{-i\gamma}_{1/2+is}}{\psi^{-i\gamma}_{1/2+is'}} 
=s^{-1}\delta(s-s').
\end{equation}
(Unlike equation \eqref{Q0ef}, the statements about normalization depend on the fact that $\nu$ is purely imaginary.) Thus the eigenfunctions $\bket{\psi^{-i\gamma}_{1/2+is}}$ form a basis in terms of which the identity decomposes as
\begin{equation}\label{decid0}
\unit=\int_{0}^{\infty}s\,ds\, \bket{\psi^{-i\gamma}_{1/2+is}}\bbra{\psi^{-i\gamma}_{1/2+is}}.
\end{equation}

We are now in a position to solve the diffusion equation. Let $E$ be the eigenvalue of the operator $-\frac{1}{2}\nabla^2=\frac{1}{2}(Q-\gamma^2)$, and
let $\rho(E)=(2\pi)^{-1}|\psi_E(0)|^2$:
\begin{equation} \label{E&rho}
\wideboxed{
E=\frac{1}{2}\,\biggl(s^2+\frac{1}{4}-\gamma^2\biggr),\qquad\quad
\rho(E)=(2\pi)^{-2}\frac{\sinh(2\pi s)}{\cosh(2\pi\gamma)+\cosh(2\pi s)}.
}
\end{equation}
Relabeling $\psi^{-i\gamma}_{1/2+is}$ as $\psi_{E}$, we can simplify some previous formulas,
\begin{equation}
\braket{\psi_{E}}{\psi_{E'}}=\delta(E-E'),\qquad\quad
\unit=\int dE\, \ket{\psi_{E}}\bra{\psi_{E}},
\end{equation}
and represent the solution to equation \eqref{diffusion} as
\begin{equation} \label{Eucprop}
\rG(x_1,x_0;\tau)=\int dE\,e^{-E\tau}\,\rG_{E}(x_1,x_0),\qquad
\rG_{E}(x,0)=\rpsi_{E}(u)\,\rpsi_{E}(0)^{*}.
\end{equation}
Working with rotationally symmetric functions, we are restricted to the $x_0=0$ case, but a general expression for $\rG_{E}(x_1,x_0)$ can be obtained using $\PSL(2,\RR)$ symmetry. Representing points of the Poincare disk as complex numbers $z=re^{i\vp}$ and following the argument at the end of section~5.3 in~\cite{SL2R}, we get:
\begin{equation}
\rG_{E}(z_1,z_0)=2\pi\rho(E)
\biggl(\frac{1-\bar{z}_1z_0}{1-z_1\bar{z}_0}\biggr)^{\nu}(1-w)^{\lambda}\,
\hgfs(\lambda+\nu,\,\lambda-\nu,\,1;\,w),
\end{equation}
where
\begin{equation}
w=\frac{(z_1-z_0)(\bar{z}_1-\bar{z}_0)}{(1-z_1\bar{z}_0)(1-\bar{z}_1z_0)}.
\end{equation}
In particular, the partition function regularized as in \eqref{Zdef} is given by
\begin{equation} \label{Zbeta}
Z(\beta)={1 \ov 2 \pi}\,\rG(0,0;\beta)=\int dE\, e^{-\beta E}\,\rho(E)
\end{equation}
so that $\rho(E)$ may be interpreted as the density of states. In the Schwarzian limit,
\begin{equation}\label{rho_rhoSch}
\rho(E)\approx e^{-2\pi\gamma}\rho_{\Sch}(E_{\Sch}),\qquad\quad
\rho_{\Sch}(E_{\Sch})=\lt(2\pi^2\rt)^{-1}\sinh\biggl(2\pi\sqrt{2E_{\Sch}}\biggr),
\end{equation}
where
\begin{equation} \label{Egsrel}
E_{\Sch} = \gamma E_{\g} = E+\frac{\gamma^2}{2}-\frac{1}{8} =\frac{s^2}{2}.
\end{equation}
This result justifies the regularization recipe~\ref{reg3}.

\section{Hilbert space and statistical mechanics}

Our Lorentzian problem is defined by Wick-rotating both the proper Euclidean time and spacetime in the regularized action \eqref{tI} on $\HH^2$. The new action is 
\be \label{LorS}
S=\int dT\, \lt( {1 \ov 2}\,g_{\al\bt}\dot{X}^{\al}\dot{X}^{\bt}+ \ga \om_{\al}\dot{X}^{\al}\rt)
\ee
where $T$ denotes proper time,  and we have replaced $\omega_{\al}\to i\omega_{\al}$ so as to preserve the spin connection $\tensor{\omega}{_\mu^a_b}=\omega_{\mu}\tensor{\Xi}{^a_b}$. Meanwhile, the spacetime is rotated as
\be \label{dgcont}
(\vp,r) \to (it,r),
\ee
where $(\vp,r)$ are polar coordinates on $\HH^2$ in which the metric is \eqref{Pdisk}, and $(t,r)$ are Schwarzschild coordinates on the patch \eqref{embed} of $\tAdS_2$ in which the metric is
\be \label{Smetric}
 ds^2={4 \ov (1-r^2)^2}(dr^2-r^2 dt^2).
\ee
The rotation may be understood as an analytic continuation from $\HH^2$ to $\tAdS_2$, where the former is embedded in the complex space $\calM$ as $(z_1, z_2)=(z, \bar{z}^{-1})$ and the latter as $(z_1,z_2)=(re^{-t},r^{-1}e^{-t})$. The second embedding is defined on the Schwarzschild patch, see \eqref{embed}, but we are also using the fact that the two-dimensional Schwarzschild spacetime can be extended to pure anti-de Sitter space.

In this section, we find---in the setting of our Lorentzian problem defined on global $\tAdS_2$---the Hilbert space of single particles, and the wavefunction at each energy of two particles corresponding to the boundaries of a two-sided black hole. We use the latter wavefunctions to construct the thermal density matrix and a variant of the thermofield double state for black holes in $\tAdS_2$. Throughout, the isometry group $\tSL(2,\RR)$ (the universal cover of $\SL(2,\RR)$) will play an important role.

It follows from standard rules of quantization applied to \eqref{LorS} that single-particle wavefunctions are spinors with spin $\nu=-i \ga$; we elevate the momentum to an operator as $p_{\al} =g_{\al\bt}\dot{X}^{\bt}+\ga \om_{\al} \to -i \p_{\al}$, from which it follows that 
\be \label{Ham}
H={1 \ov 2}\,g_{\mu\nu}\dot{X}^{\al}\dot{X}^{\bt} \to -{1 \ov 2}\nabla^2
\ee
where $\nabla_{\al}=\partial_{\al}+\nu\omega_{\al}$ is the covariant derivative acting on such spinors. Here, let us discuss our choice of gauge for the spinors. In our calculations on $\HH^2$ in the previous section, it was natural to use the disk gauge in which the local frame is non-singular at the origin, see \eqref{dgauge}. As noted previously, the disk gauge is compatible with \eqref{dgcont} in that frame vectors remain real after continuation. Thus we can consistently match the Euclidean propagator continued under \eqref{dgcont} to a two-point function for spinors in $\tAdS_2$ written in the disk gauge, and we do so to obtain the aforementioned wavefunctions for a two-sided black hole. We will also sometimes invoke the disk gauge in discussing $\tSL(2,\RR)$-invariant two-point functions of spinors, as it is naturally compatible with Schwarzschild coordinates covering different regions of $\tAdS_2$, on whose boundaries the two-point function---with one point fixed at the origin---diverges. For most other purposes in the current section, we work with global coordinates $(\phi,\theta)$ in which the metric is \eqref{gAdS_2}, and use the tilde gauge in which the local frame is smooth over the entirety of $\tAdS_2$, see \eqref{tgauge}. Sometimes $\tilde{u}=e^{i(\pi-2 \tht)}$ will be a convenient variable.\footnote{Wavefunctions written in terms of $\phi$ and $\tilde{u}$ satisfy the same equations as in the case of $\vp$,\, $u=r^2$, and the Euclidean version of the tilde gauge. This is due to an alternative analytic continuation, which will not be used in any serious way.} The action of $\sL_2$ generators on spinors is then given by \eqref{sl2gen}, and in particular, a spinor with $L_0=-m$ factorizes as
\be \label{tpsifact}
\tpsi(\phi, \tht)=f(\tht)\,e^{im\phi}.
\ee
In the following, spinors will be implicitly in the tilde gauge unless indicated otherwise.

\subsection{Single-particle wavefunctions} \label{sec:partwf}

\begin{figure}[t]
\centerline{
\includegraphics[scale=1.1]{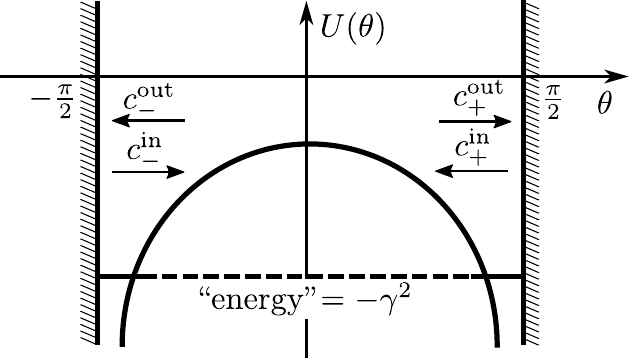} 
}
\caption{Propagation of physical states in asymptotic regions of $\tAdS_2$. We show coefficients defined in \eqref{asymc} corresponding to amplitudes of ingoing and outgoing waves.}
\label{fig:prop}
\end{figure}

The Schrodinger equation for a stationary single-particle wavefunction, $-\frac{1}{2}\nabla^2\psi=E\psi$, reduces via \eqref{nabla} to the Casimir eigenvalue equation $Q\psi=\lam(1-\lam)\psi$ with
\be \label{EQrel}
E={1 \ov 2}\lt(\lam(1-\lam)-\ga^2\rt).
\ee
Let us look for a basis of single-particle wavefunctions consisting of Casimir eigenfunctions $\psi^{\nu}_{\lam, m}$ organized into irreducible representations of $\tSL(2,\RR)$. The parameters $\lambda$ and $\mu$ (possible choices for $\mu$, which depend on $\lam$, are discussed below) specify a unique irreducible representation type, while $m \in \mu+\ZZ$ indexes states within that representation.\footnote{Notice from \eqref{tpsifact} that $\mu$ characterizes the periodic behavior of $\psi$ in $\phi$, namely, $\psi(\phi+2\pi n)=\psi(\phi)\,e^{2\pi i\mu n}$.} In the Euclidean problem, we saw that spinors which account for the density of states---the Green functions $G_E$ with one point fixed at the origin---were eigenfunctions with $\lambda={1 \ov 2}+is$ for $s>0$. Here, we identify the single-particle Hilbert space as consisting of Lorentzian wavefunctions $\psi^{\nu}_{\lam, m}$ organized into representations with the same values of $\lam$. Note that for each $\nu$, $s$, and $m$, there are two linearly independent Casimir eigenfunctions; thus the sequences $\bigl(\psi^{\nu}_{\lam,m}:\, m\in\mu+\ZZ\bigr)$ form a two-dimensional vector space. From a physical point of view, these wavefunctions are each subject to an inverted potential that falls off to $-\infty$ near the boundaries of $\tAdS_2$ (see Figure~\ref{fig:prop}); they describe particles which propagate freely near an asymptotic boundary, but must tunnel through a potential barrier to reach the opposite near-boundary region. We will see that the tunneling probability calculated from these wavefunctions reproduces the density of states found in the previous section.

To define the Hilbert space, we use the inner product 
\begin{equation} \label{innprod}
\bbraket{\psi_1} {\psi_2}=\int_{\tAdS_2} d^2 x \sqrt{-g}\,\psi_1^*(x) \psi_2(x)
\end{equation}
for spinor wavefunctions. It is invariant under the action of $\sL_2$ generators $L_{-1}, L_{0}, L_{1}$ (see \eqref{sl2gen}) on the wavefunctions. The physical interpretation of the wavefunctions and inner product is as follows: the probability for a particle with spin $\nu=-i \ga$---describing a boundary of nearly-$\tAdS_2$ spacetime---to be in the state corresponding to wavefunction $\psi$ is given by the integral of $\|\psi\|^2$ over $\tAdS_2$. From the point of view of quantum mechanics on the boundary, the $\tAdS_2$ coordinates are auxiliary variables, while $E$, the energy conjugate to proper time, is the dynamical variable. We can think of the boundary particle as an observer in $\tAdS_2$ with a clock that measures proper time. In the most general setting, the observer can emit and absorb excitations which change $E$, and which are described by fields second-quantized on $\tAdS_2$. In the absence of such interactions with bulk fields, $E$ is conserved.

Casimir eigenfunctions which are normalizable with respect to \eqref{innprod} fall into $\tSL(2,\RR)$ representations in either the principal series $\calC^{\mu}_{\lam(1-\lam)}$ with $\lambda={1 \ov 2}+i s$,\, $s>0$ and $\mu \in \RR/\ZZ$, or the discrete series $\calD_{\lambda}^{\pm}$ with $\lambda>1/2$,\, $\mu=\pm \lam$.\label{princdiscr} See Appendix \ref{app:repth} for a complete discussion. Let us consider the Casimir eigenvalue equation $Q \psi=\lambda(1-\lambda)\psi$ for such a normalizable wavefunction $\psi^{\nu}_{\lambda,m}$. The explicit form of $Q$ is as in \eqref{tgQ}. It follows that the spatial part of the wavefunction $f(\tht)$ in the decomposition in \eqref{tpsifact} satisfies a time-independent Schrodinger equation with a certain potential and energy,
\be \label{pot}
\left( -\p_\tht^2 + U(\tht)\right)f = \left(m^2-\ga^2\right)f, \qquad U(\tht)=-{\lambda(1-\lambda)\ov \cos^2 \tht} + 2 \ga m \tan \tht.
\ee
Note the first term in the potential $U$ dominates sufficiently close to the two boundaries of $\tAdS_2$ at $\tht=\pm{\pi \ov 2}$. For a wavefunction in a principal series representation with $\lam(1-\lam)={1 \ov 4}+s^2$, the potential falls off to $-\infty$ near the boundaries. Thus the corresponding particle is classically allowed in some asymptotic regions near each boundary, where it can move in or out, but must tunnel through a potential barrier to go from one asymptotic region to another, see Figure \ref{fig:prop}. On the other hand, for a wavefunction in the discrete series with a generic value of $\lam(1-\lam)<{1 \ov 4}$, the particle is bound in the interior of $\tAdS_2$. A precise characterization of wavefunctions in the principal series, as opposed to the discrete series, is that only the former have non-vanishing Klein-Gordon flux in the $\tht$ direction $F=\int d\phi\,  J_{\tht}$, $J_{\mu}={i \ov 2}\left(\nabla_\mu \psi^*  \cdot \psi -  \psi^* \nabla_\mu \psi\right)$. They correspond to propagating states whose energies are greater than some threshold, $E > {1 \ov 2}\lt({1 \ov 4}-\ga^2\rt)$.\footnote{Normalizable wavefunctions in the discrete series have complementary characteristics; their flux is non-vanishing in the $\phi$-direction of $\tAdS_2$ and their energies are below the threshold, $E <{1 \ov 2}\lt( {1 \ov 4}-\ga^2\rt)$. Because their frequencies with respect to $\phi$ are bounded, they are appropriate for describing matter fields quantized on global $\tAdS_2$, although in the case of matter fields $\nu$ must take integer and half-integer values rather than $\nu=-i \ga$. See Section \ref{sec:corr} for an application.\label{ftn:discf}} We take them to be the physical single-particle wavefunctions in our problem. 

Given a physical wavefunction $\psi_{\lam, m}^{\nu}$ with some $\lam$ and $\mu$ such that
\be
\lam={1 \ov 2}+i s,\quad s>0, \qquad \mu\in \RR/\ZZ,
\ee
we may define coefficients of ingoing and outgoing waves in each of the asymptotic regions as
\be \label{asymc}
\psi_{\lam,m}^{\nu}(\phi,\theta)\approx\begin{dcases}
\left(c_{+}^{\IN}(\pi-2\theta)^{\lambda}
+c_{+}^{\OUT}(\pi-2\theta)^{1-\lambda}\right)e^{im\phi}
&\text{for }\theta\to\tfrac{\pi}{2}\\[3pt]
\left(c_{-}^{\IN}(\pi+2\theta)^{\lambda}
+c_{-}^{\OUT}(\pi+2\theta)^{1-\lambda}\right)e^{im\phi}
&\text{for }\theta\to-\tfrac{\pi}{2}
\end{dcases}.
\ee
Furthermore, it is natural to define a scattering matrix using the in and out coefficients,  
\be
\begin{pmatrix} 
c^{\OUT}_+ \\
c^{\OUT}_- 
\end{pmatrix} = S \begin{pmatrix}c^{\IN}_+ \\
c^{\IN}_-
\end{pmatrix}, \qquad S=\begin{pmatrix} S_{++} & S_{+-} \\
S_{-+} & S_{--}
 \end{pmatrix},
 \ee
and to calculate the tunneling probability $p=|S_{+-}|^2$. To obtain the coefficients $c_{\pm}^{\text{in}}$, $c_{\pm}^{\text{out}}$, and thus $S$, we solve for the two linearly independent solutions to the Casimir eigenvalue equation in the complex $\tilde{u}=e^{i(\pi-2 \tht)}$ plane; see \eqref{asymco}. 

We find that in fact $S_{\pm\pm}$ and $|S_{+-}|^2$ are independent of $m$ for $m\in\mu+\ZZ$, or in other words, well-defined for a particle whose state belongs to a given representation type. In particular, the probability for the particle to tunnel is given by
\be \label{tunp}
p(s, \mu)={\sinh^2(2 \pi s) \ov 4 a b}, \qquad a={1 \ov 2}\left( \cosh(2\pi s)+\cosh(2\pi\ga)\right),\quad b={1 \ov 2}\left( \cosh(2\pi s)+\cos (2\pi\mu)\right).
\ee
Integrating over the non-observable parameter $\mu$ to obtain the total tunneling probability at a given energy, we find that it coincides with the density of states $\rho$ found in \eqref{E&rho} up to a constant,
\be
p(s)=\int d\mu\, p(s, \mu)=(2\pi)^2 \rho(E).
\ee
In this context, the factor $e^{-2\pi\ga}$ in the Schwarzian limit of $\rho$, isolated in \eqref{rho_rhoSch}, expresses the exponential suppression of tunneling probability in the height of the potential barrier (relative to the ``energy" in the Schrodinger equation), which grows like $\ga^2$. Using \eqref{Schlim2} and \eqref{Egsrel}, the Schwarzian limit can also be written as
\be \label{Schlim3}
\ga \gg 1, \qquad s^2 \ll \ga^2.
\ee
Then we also see, from the potential in \eqref{pot}, that in this limit particles are constrained to stay very close to the boundary, where $\pi/2-|\tht| \ll 1$.

\subsection{Two-sided wavefunctions and density matrices}

\begin{figure}[t]
\centerline{\begin{tabular}{@{}c@{\hspace{4cm}}c@{}}
{\includegraphics[scale=0.853]{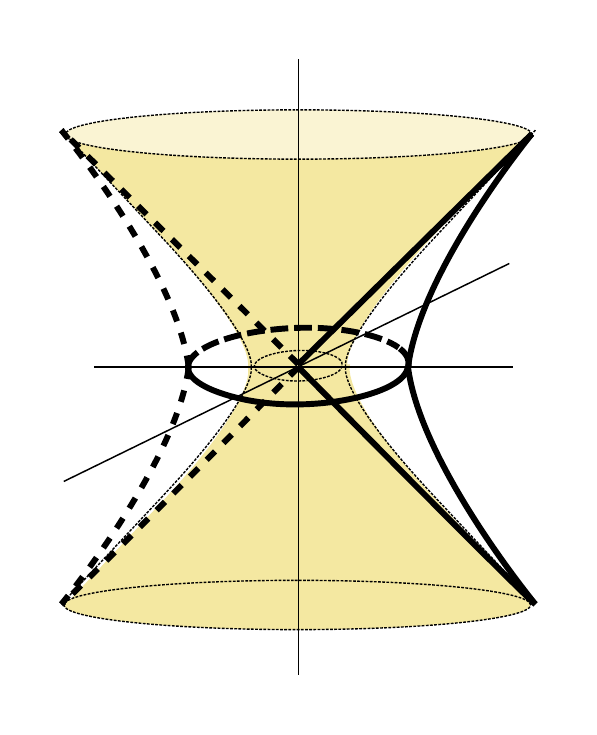}} &
\includegraphics[scale=0.92]{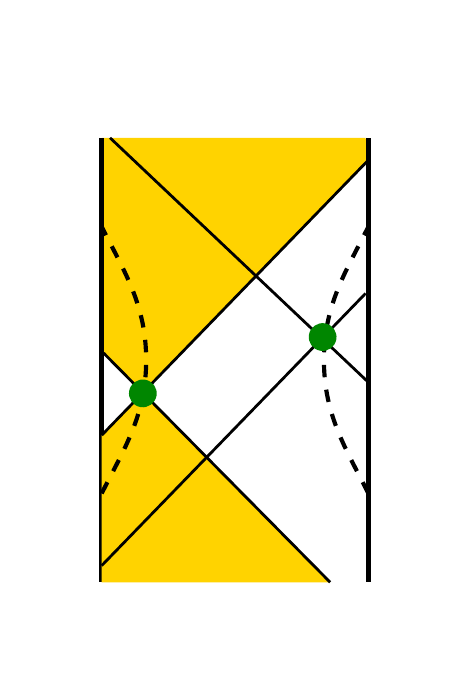}\\
a) & b)
\end{tabular}}
\caption{a) Relation between $\HH^2$ (horizontal disk) and $\tAdS_2$ (vertical cross section) embedded in the complex space $\calM$. Euclidean and Lorentzian classical trajectories consist of intersections of $\HH^2$ and $\tAdS_2$ with a complex classical trajectory, shown as a colored hyperboloid. To obtain $\mathring{\Phi}_E(x;0)$, the Euclidean Green function $\mathring{G}_E(x,0)$ is continued from $\HH^2$ to the right Schwarzschild patch of $\tAdS_2$. b) We can view $\Phi_E(x;x')$ as a wavefunction for the two boundaries of a two-sided black hole in $\tAdS_2$, which are space-like at any given instant of proper time.}
\label{fig:Phides}
\end{figure}

In the above, we saw that the density of states in our system appeared as a probability of tunneling computed from asymptotic coefficients of single-particle wavefunctions. It turns out that the density is also encoded in the square of an $\tSL(2,\RR)$-invariant two-point function $\Phi_E(x;x')$ determined by the characteristics that i) $\Phi_E(x;0)$ on an exterior Schwarzschild patch agrees with the analytic continuation of the Euclidean Green function $\mathring{G}_E(x, 0)=\mathring{\psi}_E(x)\mathring{\psi}_E(0)^*$ from $\HH^2$, and ii) $\Phi_E(x;x')$ is non-vanishing only at space-like separation. To satisfy the first condition, we analytically continue $\mathring{G}_{E}(x,0)$ from $\HH^2$ to the right exterior Schwarzschild patch of $\tAdS_2$, then continue the resulting $\Phi_E(x;0)$ to the rest of $\tAdS_2$ using the spinor wave equation, and finally extend it to $\Phi_E(x;x')$ using $\tSL(2,\RR)$ symmetry. The function $\Phi_E(x;x')$ may be interpreted as a tunneling amplitude. Alternately, we can identify it as the physical wavefunction of a two-sided black hole with definite energy. The space-like support of the wavefunction implies that the two sides of the black hole, viewed as particles in $\tAdS_2$, are causally disconnected; see Figure \ref{fig:Phides}.

After introducing a regularized notion of trace in which we quotient out by $\tSL(2,\RR)$, the thermal partition function we found by Euclidean methods \eqref{Zbeta} can be reconstructed in Lorentzian signature as $Z(\beta)=\frac{1}{2}\tr(e^{-\beta H} \Rho)$,\, $\Rho=\Phi \Phi^{\dagger}$, where $\Phi=\int dE\, \Phi_E$, and $\Phi_E$ is the operator acting on the single-particle Hilbert space for which $\Phi_E(x;x')=\bra{x}\Phi_E \ket{x'}$ is a matrix element in the position basis. More generally, any density matrix for a one-sided black hole in $\tAdS_2$ without matter fields will take the form $\varrho = \int dE\, f(E) \Rho_E$, where the weight function $f$ satisfies the trace condition $\int dE\, f(E) \rho(E)=1$.

Let us first describe the Hilbert space of single-particle states $\calH^{\nu}_{\p}$ (we use the subscript $\p$ which stands for boundary, as a particle describes a boundary of nearly-$\tAdS_2$ spacetime) more precisely, and also the space of $\tSL(2,\RR)$-invariant operators acting on it. We will then specify the two-sided black hole wavefunction $\Phi_E$ and proceed to construct the thermal partition function and general density matrices for  a one-sided black hole.

\subsubsection{Single-particle Hilbert space and $\tSL(2,\RR)$-invariant operators}

In the previous section, we found that single-particle wavefunctions in our problem consist of $\nu$-spinors on $\tAdS_2$ which fall into principal series representations of $\tSL(2,\RR)$ with $\lam={1 \ov 2}+i s$,\, $s>0$, and $\mu \in \RR/\ZZ$. In fact, the space of intertwiners $\psi$ which map states $\ket{\lam, m}$ in such a representation to wavefunctions $\psi^{\nu}_{\lam, m}$ in $\calH^{\nu}_{\p}$ is two-dimensional. In other words, there are two independent solutions to the equations \eqref{compeq} with $Q$ and $L_0$ given by \eqref{Qdef}, \eqref{sl2gen}, and both are normalizable under the inner product \eqref{innprod}. It follows that an $\tSL(2,\RR)$-invariant operator acting on the subspace $\calH^{\nu}_{\p; \lam, \mu}\subset \calH^{\nu}_{\p}$ with quantum numbers $\lam$ and $\mu$ takes the form
\be \label{subop}
\Psi^{\nu}_{\lam,\mu}[R]=\sum_{\alpha,\beta} R_{\alpha\beta} \sum_{m\in\mu+\ZZ}
\bket{\lt( \psi_{\alpha} \rt)^{\nu}_{\lambda,m}}
\bbra{\lt( \psi_{\beta} \rt)^{\nu}_{\lambda,m}}
\ee
where $R$ is best understood as an operator on the space of intertwiners with matrix elements $R_{\al\beta}$ with respect to some basis $\psi_{\al}$, $\al=1,2$. Given $R$ as a function of $s$ and $\mu$, we may integrate $\Psi^{\nu}_{\lam, \mu}[R(s,\mu)]$ as
\be \label{Mgenop}
\wideboxed{
\Psi^{\nu}[R]=\int dE \int d\mu\, \rPl(E, \mu)\, \Psi^{\nu}_{1/2+is,\mu}[R(s,\mu)], \qquad \rPl(E, \mu)=(2\pi)^{-2}\,{\sinh(2\pi s) \ov 2 b}}
\ee
to obtain an arbitrary $\tSL(2,\RR)$-invariant operator acting on $\calH^{\nu}_{\p}$. Here $E$ is related to $s$ by \eqref{EQrel} and $b$ was defined in \eqref{tunp}. The Plancherel measure $dE\,d\mu\,\rPl(E,\mu)$ is used because it is a natural measure on $\widetilde{\SL}(2,\RR)$ irreps; it plays the role of effective dimension and enters the definition \eqref{M_tr_def} of trace.\footnote{In Appendix \ref{app:opalg}, we represent the same measure as $ds\,d\mu\,\rcont(s,\mu)$ and also include a discrete series part $d\lambda\,\rdisc(\lambda)$. The latter is not needed for the present problem.} For consistency, wavefunctions in \eqref{subop} are normalized as in \eqref{op_ip} with an inverse Plancherel factor, so that the total operator $\Psi^{\nu}$ is independent of the normalization and the multiplication rule $\Psi^{\nu}[R]\cdot\Psi^{\nu}[R']=\Psi^{\nu}[RR']$ holds. It will be convenient to separately label the operator at fixed energy,
\begin{equation}\label{subop_E}
\Psi^{\nu}_E[R]=\int d\mu\, \rPl(E, \mu)\,\Psi^{\nu}_{1/2+is, \mu}[R(s,\mu)].
\end{equation}

\begin{figure}[t]
\centerline{\begin{tabular}{@{}c@{\hspace{4cm}}c@{}}
\includegraphics{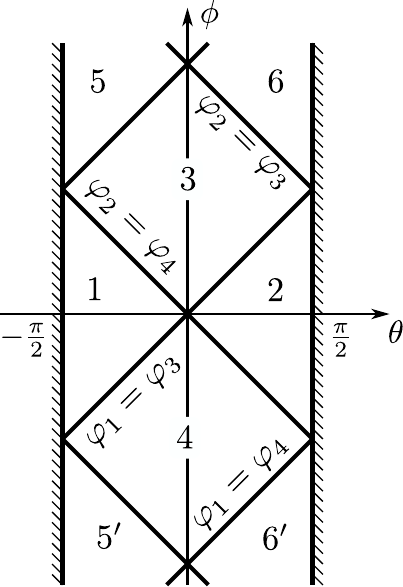} &
\includegraphics{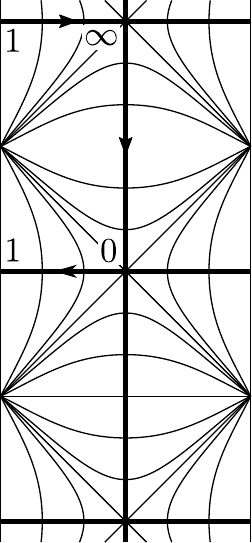}\vspace{5pt}\\
a) & b)
\end{tabular}}
\caption{a) Division of $\AdS_2$ into regions bounded by light rays from the point $x'=0$. There are infinitely many copies of regions $1,2,\dotsc,6$ that are translations by $\phi \to \phi+ 2\pi n$ for $n\in\ZZ$.\, b)~Orbits of points under the subgroup $H \subset \tSL(2,\RR)$ preserving the origin (thin lines), and points in a skeleton representation $S$ of the quotient space $H \backslash \tAdS_2$ (thick lines). The coordinate $u$ is shown in regions $1$, $3$, $5$.}
\label{fig:regionsm}
\end{figure}

An arbitrary $\tSL(2,\RR)$-invariant two-point function $\Psi^{\nu}(x; x')$ transforming as a $(\nu, -\nu)$ spinor is then a representation of some operator $\Psi^{\nu}$ with respect to the position basis, $\Psi^{\nu}(x; x')=\bra{x}{\Psi^{\nu}\ket{x'}}$. (Note it follows from our inner product for wavefunctions \eqref{innprod} that $\unit=\int d^2x \sqrt{-g}\,\ket{x}\bra{x}$ is representation of the identity operator.) On symmetry grounds, $\Psi^{\nu}$ has the general structure
\be \label{twopoint}
\Psi^{\nu}(x;x')=\left|\frac{\vp_{23}}{\vp_{14}}\right|^{\nu}f_{j}(w),\qquad
w=\frac{\vp_{13}\vp_{24}}{\vp_{14}\vp_{23}}
\ee
where $\vp_1=\phi-\theta+\frac{\pi}{2}$,\, $\vp_2=\phi+\theta-\frac{\pi}{2}$ and $\vp_3=\phi'-\theta'+\frac{\pi}{2}$,\, $\vp_4=\phi'+\theta'-\frac{\pi}{2}$ are coordinates of points $x$ and $x'$,\, $\vp_{kl}=2\sin\frac{\vp_k-\vp_l}{2}$,\, and $j$ in $f_j$ points to a region bounded by light rays from $x'$ to which $x$ belongs. Let us elaborate further. Given the pair of points $(x;x')$, we may use an element of $\tSL(2,\RR)$ to map $x'$ to the origin. At the same time, $x$ is mapped to some point with Schwarzschild-like coordinates $(t,u)$ (using the notation $u=r^2$) in some region $j$ bounded by light rays from the origin reflected at the boundaries of $\tAdS_2$, see Figure~\ref{fig:regionsm}a. Elements of the subgroup $H \subset \tSL(2,\RR)$ fixing the origin act as boosts within each region by shifting $t$ and preserving $u$, so we can further boost $(t,u)$ to $(0,u)$. The union $S$ of $t=0$ slices over all regions constitutes a representation of the quotient $H\backslash \tAdS_2$ and is shown in Figure~\ref{fig:regionsm}b by thick lines (vertical in regions $3$, $4$ and their copies, horizontal in other regions). We have
\be \label{urange}
0<u<1\, \text{ in regions }1,2,\qquad
u<0\, \text{ in regions }3,4,\qquad
u>1\, \text{ in regions }5,6
\ee
and the same for translations of each region under $\phi \to \phi+2\pi n$. Notice that $u$ is equal to $w$ in \eqref{twopoint}, which is invariant under $\tSL(2,\RR)$ transformations:
\begin{equation} \label{wurel}
\lt. w \rt|_{(x;x')}=\lt. w \rt |_{((0,u);0)}=u.
\end{equation}
The parameter $w$ measures the geodesic distance of $x$ from $x'$, and is related to the cross ratio $\chi=(\vp_{12}\vp_{34})/(\vp_{13}\vp_{24})$ as $\chi=1-w^{-1}$. The function $f_j(w)$ in \eqref{twopoint} is the two-point function $\Psi^{\nu}((0,u);0)$ between the final image of $x$ and the origin, and the phase factor in front of it represents the Lorentz transformation of spinors. In the case where $x'=0$, the phase factor also corresponds to the transition between the tilde gauge and the disk gauge; hence 
\begin{equation}
\rPsi^{\nu}(x;0)=f_j(u)\qquad \text{in region }j.
\end{equation}
Note that $f_j(u)$ is analytic inside region $j$ but in general singular on its non-asymptotic boundaries, and the function $\Psi^{\nu}(x;0)$ is continued across boundaries between regions by the condition that it satisfies the wave equation for a $\nu$-spinor.

\subsubsection{Main results}\label{sec_tswf}

Now, let us choose an $\tSL(2,\RR)$-invariant operator $\Phi=\Psi^{\nu}[R]$ based on physical requirements for the two-point function $\Phi_E=\Psi^{\nu}_E[R]$ at each energy $E$ (see \eqref{subop_E}) suitable for it being a wavefunction of the two boundaries of a two-sided black hole in $\tAdS_2$:

\begin{enumerate}

\item
In the right Schwarzschild patch, $\Phi_E(x;0)$ is the analytic continuation of the Euclidean propagator $\rG_E(x,0)=\rpsi_E(u)\,\rpsi_E(0)^*$ on $\HH^2$ (see \eqref{Eucprop}, \eqref{Eucpsi}) under the Wick rotation \eqref{dgcont}, \ie
\be \label{match}
\mathring{\Phi}_E(x;0)=\mathring{\psi}_E(u)\mathring{\psi}_E(0)^* \qquad \text{in region $2$.}
\ee

\item
The support of $\Phi_E(x;x')$ is at space-like separation, \ie $\mathring{\Phi}_E(x;0)$ is non-vanishing only in regions $1,2$.

\end{enumerate}
In Appendix \ref{app:twopt}, we compute the  two-point function $\Psi^{\nu}_{\lam, \mu}[R]$ associated with an arbitrary $R$ at fixed $s$ and $\mu$. The function vanishes in regions $3$, $4$, $5$, $6$ and their copies if $R$ is proportional to a certain operator $Z$, which is expressed in different ways by \eqref{XZ}, \eqref{XpZ}, and by \eqref{Zmatrix} as a $2\times 2$ matrix, using bases of single-particle wavefunctions defined in \eqref{normfs}. Note that $Z$ is Hermitian with eigenvalues $1$ and $-1$. To satisfy both conditions~1 and~2, we must set $R=-i \sqrt{b/a}\,Z$ so that 
\be \label{Phi}
\wideboxed{\begin{aligned}
&\Phi=\Psi^{\nu}\bigl[-i \sqrt{b/a}\,Z\bigr],\\[3pt]
&\mathring{\Phi}_E(x;0)=\begin{dcases}
\pm 2\pi\rho(E) A_{\lam, \nu, -\nu}(u) & \text{``$+$'' in region $2$,\, ``$-$'' in region $1$}\\
0 & \text{in all other regions}
\end{dcases}
\end{aligned}}
\ee
where 
\begin{equation} \label{MAf}
A_{\lambda,l,r}(u)=u^{(l+r)/2}(1-u)^{\lambda}\,
\hgfs\bigl(\lambda+l,\,\lambda+r,\,1+l+r;\,u\bigr).
\end{equation}

To derive \eqref{Phi} from the expression \eqref{Zwf} for the function $\Psi^{\nu}_{\lam,\mu}[Z]$, we integrate over $\mu$ with the Plancherel factor. It is nontrivial that the integral $\int d\mu\,\rPl\,\Psi^{\nu}_{\lambda,\mu}\bigl[-i \sqrt{b/a}\,Z\bigr]$ vanishes in copies of regions $1$ and $2$---this is due to the integrand not depending on $\mu$ in regions $1,2$ and to twisted periodicity, $\Psi^{\nu}_{\lambda,\mu}[R](\phi+2\pi,\theta;\phi',\theta') =e^{2\pi i\mu}\kern1pt \Psi^{\nu}_{\lambda,\mu}[R](\phi,\theta;\phi',\theta')$, which follows from the $\phi$-dependence of wavefunctions \eqref{tpsifact} and the condition $m \equiv \mu\,(\text{mod }1)$. Incidentally, the two-point function $\Psi^{\nu}\bigl[-i \sqrt{p}Z\bigr]$ is identical to $\Phi$ in regions $1$, $2$ and  agrees with the Euclidean propagator in the same sense as $\Phi$.\,\footnote{The operator $\Psi^{\nu}\bigl[-i \sqrt{p}Z\bigr]$ measures the Klein-Gordon flux of single-particle states in the spatial $\tht$-direction of $\tAdS_2$, see \eqref{fluxop}.} However, it does not vanish in copies of regions $1$, $2$ and thus does not satisfy our second criterion of vanishing at time-like separation. Nor does its square encode the density of states in the way that $\Phi \Phi^{\dagger}$ does and which we explain below.

As already mentioned, the space-like support of $\Phi_E(x; x')$ allows us to alternatively interpret it as the tunneling amplitude of a boundary particle. Its relation to the density of states of a black hole can first be seen by inserting an integral over an intermediate point in the path integral that is the Euclidean partition function, $2\pi Z(\beta)=\mathring{G}(0,0;\beta)=\int_{X(0)=X(\beta)=0} DX\, e^{-I[X]}$ (see \eqref{tI}):
\begin{equation}\label{Zlong}
\begin{aligned}
Z(\beta)&=\lt(2\pi\rt)^{-1}\int_{\HH^2} d^2x \sqrt{g(x)}\, \mathring{G}(0, x; \tau)\,\mathring{G}(x,0;\beta-\tau)\\
&=\int dE\, dE'\, e^{-E\tau}e^{-E'(\beta-\tau)}\underbrace{\int{2du \ov (1-u)^2}\, \lt(\mathring{\psi}_E(u)^* \mathring{\psi}_E(0)\rt) \lt(\mathring{\psi}_{E'}(u) \mathring{\psi}_{E'}(0)^*\rt)}_{\de(E-E')\rho(E)}.
\end{aligned}
\end{equation}
We can view the integral in the last line,
\be \label{trans}
\int{2du \ov (1-u)^2}\, \lt(\mathring{\psi}_E(u)^* \mathring{\psi}_E(0)\rt) \lt(\mathring{\psi}_{E'}(u) \mathring{\psi}_{E'}(0)^*\rt)=\int{2du \ov (1-u)^2}\, \left.\mathring{\Phi}_E(x;0)^* \mathring{\Phi}_{E'}(x;0) \right|_{\text{region $2$}},
\ee 
as one half the result of a trace performed in Lorentzian signature in which we quotient out the infinite volume of $\tSL(2,\RR)$:
\bea \label{trexp}
\tr\lt(\Phi_E^{\dagger}\Phi_{E'}\rt)&=&\int_{\tSL(2,\RR)\backslash \tAdS_2 \times \tAdS_2}\Phi^{\dagger}_E(x'; x)\Phi_{E'}(x; x')\nn
&=&\sum_{\text{regions $1,2$}}\int{2 du \ov (1-u)^2}\, \mathring{\Phi}_{E}(x;0)^*\, \mathring{\Phi}_{E'}(x;0).
\eea
In the first line, we have inserted two factors of the identity $\unit=\int d^2x \sqrt{-g}\,\ket{x}\bra{x}$ and quotiented the domain of integration for the two points by $\tSL(2,\RR)$. In the second line, we have represented the quotient space $\tSL(2,\RR)\backslash \tAdS_2 \times \tAdS_2=H\backslash \tAdS_2$ as $\{(x, x'):\, x\in S,\, x'=0\}$ where $S$ is the skeleton set depicted by thick lines in Figure \ref{fig:regionsm}b. (Recall that $H$ is the group of boosts fixing the origin, or simply translations in Schwarzschild time $t$.) The quotient space comes with the measure $2du/(1-u)^2$, equal to the ratio of $d^2x\sqrt{-g}$ and $dt$. To perform the integral, we use that $\Phi_E^{\dagger}(x'; x)=\Phi_E(x; x')^*$ and that the phase factor in \eqref{twopoint} cancels between $\Phi_E(x;x')^*$ and $\Phi_{E'}(x;x')$, so that $\Phi_{E}(x;0)^* \Phi_{E'}(x;0)=\mathring{\Phi}_E(x;0)^*\mathring{\Phi}_{E'}(x;0)$.
The integrand is nonzero only in regions $1$ and $2$, which contribute equally.

More formally, let us define a trace operation on an operator \eqref{Mgenop} acting on $\mathcal{H}^{\nu}_{\p}$ as
\be \label{M_tr_def}
\wideboxed{
\tr\bigl(\Psi^{\nu}[R]\bigr)
\equiv\int dE \int d\mu\,\rPl(E,\mu)\Tr(R(s,\mu)).}
\ee
Unlike in \eqref{Mgenop}, the Plancherel factor here is not canceled by the normalization of wavefunctions. It represents the dimension of the $\tSL(2,\RR)$ irrep with given $s$ and $\mu$ divided by the volume of the group. In Appendix \ref{app:opalg}, we show that the trace defined as such of the product of two operators can indeed be computed as in \eqref{trexp}---in other words, for operators $F$ and $G$ with
\begin{equation}
F(x;x')=\left|\frac{\vp_{23}}{\vp_{14}}\right|^{\nu}f_j(w),\qquad
G(x;x')=\left|\frac{\vp_{23}}{\vp_{14}}\right|^{\nu}g_j(w),
\end{equation}
\be \label{trFG}
\tr(F^{\dagger}G)=\sum_{j}\int \frac{2du}{(1-u)^2}\,f_j(u)^{*}\,g_j(u).
\ee
This has to do with the fact that matrix elements of $R$ appear in the asymptotic behavior of $\Psi^{\nu}_{\lam, \mu}[R]$ near the boundaries of each region $j$, which applied to $F$ and $G$ determine the integral in \eqref{trFG}. In fact, we also find that for a sufficiently regular operator $F$,\, $\tr(F)$ itself can be extracted from the coefficient of the logarithmic singularity in $f_{j}(u)$,\, $j=1,2,3,4$ as $u \to 0$, namely, $f_j(u) =-\tr (F) \ln |u|+\cdots$; see \eqref{F_asymp}.

We conclude the thermal partition function of a one-sided nearly-$\tAdS_2$ black hole \eqref{Zlong} can be constructed in Lorentzian signature as
\be \label{LorZ}
\wideboxed{
Z(\beta)={1 \ov 2}\tr\bigl(e^{-\beta H}\Rho\bigr), \qquad
\Rho=\Phi \Phi^{\dagger},\quad\: H=\Psi^{\nu}(EI).
}
\ee 
The factor of ${1 \ov 2}$ is due to the fact that our trace is over the entirety of $\tAdS_2$ with two boundaries, whereas the partition function is for a one-sided system (in particular, $\Rho$ is a one-sided operator that maps states of one boundary of $\tAdS_2$ to those of the same boundary). Then as compared to the expected formula from standard statistical mechanics, we have the additional insertion of $\Rho$. Note our starting point \eqref{Zdef}, \eqref{Zlong} was to apply the standard formula in an appropriately regularized Euclidean problem \eqref{tI}.\footnote{Note the trace we used in \eqref{Zdef} was regularized analogously to our Lorentzian trace, by quotienting out $\PSL(2,\RR)$.} In passing to Lorentzian signature, $\int_{\PSL(2,\RR)\backslash\HH^2 \times \HH^2} \mathring{G}_E(x';x)\mathring{G}_{E'}(x;x')={1 \ov 2}\int_{\tSL(2,\RR)\backslash \tAdS^2 \times \tAdS^2}\Phi^{\dagger}_{E}(x';x)\Phi_{E'}(x;x')$, it became necessary to insert the operator $\Rho$. 

The explicit form of $\Rho$ is given by
\be \label{Rho}
\wideboxed{\begin{aligned}
&\Rho=\Psi^{\nu}\bigl[(b/a)I\bigr],\\
&\mathring{\Rho}_E(x;0)= \rho(E)\begin{dcases}
 -2 C_{\lam, \nu}(u) & \text{in regions $1, 2$}\\
-2  \bC_{\lam, \nu}(u)&  \text{in regions $3, 4$}\\
\Gamma(\lam+\nu)\Gamma(1-\lam+\nu)A_{\lam, \nu, \nu}\bigl(u^{-1}\bigr) & \text{in regions $5, 6'$}\\
\Gamma(\lam-\nu)\Gamma(1-\lam-\nu)A_{\lam, -\nu, -\nu}\bigl(u^{-1}\bigr) & \text{in regions $5', 6$}\\
0 & \text{in all other regions}
\end{dcases},
\end{aligned}}
\ee 
where the functions $C_{\lam, \nu}$, $\bC_{\lam, \nu}$, which are defined in \eqref{Cf}, \eqref{bCf}, \eqref{bAf} diverge logarithmically as $u \to 0$, being $\approx \ln |u|$. The operator $\Rho_E$ encodes the density of states of a one-sided black hole at a given energy as ${1 \ov 2}\tr(\Rho_E)=\rho(E)$. (In comparison, $\tr(\Phi_E)=0$.) We can extrapolate that any density matrix for such a black hole will take the form 
\be \label{dmatrix}
\varrho=\Psi^{\nu}[fI]\cdot\Rho=\int dE\, f(E)\,\Rho_E, \qquad {1 \ov 2}\tr(\varrho)=\int dE\, f(E)\,\rho(E)=1,
\ee
where $f$ is some weight function over energies---for example, $f(E)=Z^{-1}e^{-\beta E}$ at thermal equilibrium. This is valid in the absence of particles in the bulk, so that the left and right boundaries form an $\tSL(2,\RR)$ singlet with the wavefunction $\int dE\,\sqrt{f(E)}\,\Phi_E(x;x')$. The quantum entropy of the density matrix \eqref{dmatrix} should be taken as
\be
S=-{1 \ov 2}\tr(\varrho(\ln \varrho-\ln \Rho))=-\int dE\, \rho(E)\,f(E)\ln f(E),
\ee
where $\ln \Psi^{\nu}[R]=\Psi^{\nu}[\ln R]$ is an invariant definition of the logarithm of an operator.

\section{Correlation functions of external operators}\label{sec:corr}

\subsection{Statement of the problem and some results}\label{sec:corrgen}

Let us discuss correlators of matter fields in black hole states, in two different settings. The first one is best understood in the Euclidean case. Let us consider some field theory in $\HH^2$ with local observables $\calX(x)$, $\calY(x)$, etc. In addition, there is a fluctuating curve $X$ with the regularized geometric action \eqref{tI}, and we are interested in correlation functions of the fields with respect to the (regularized) proper time $\tau$. For simplicity, we focus on a two-point function of fields with zero spin,
\begin{equation}
\label{corr_Euc}
\figbox{1.0}{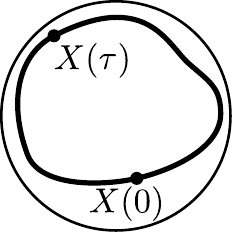}\hspace{1.5cm}
\calF_{\calX,\calY}(\tau,0)
=Z^{-1}\int DX\,e^{-I[X]}\,
\bcorr{\calX(X(\tau))\,\calY(X(0))}.
\end{equation}
The correlator is easily expressed in terms of the Euclidean propagator for the curve,
\begin{equation} \label{Euc_corr_1}
\calF_{\calX,\calY}(\tau,0)
=Z^{-1}\int_{\PSL(2,\RR) \backslash \HH^2 \times \HH^2 }  
G(x_0,x_1;\beta-\tau)\,G(x_1,x_0;\tau)\,\corr{\calX(x_1)\kern1pt\calY(x_0)}.
\end{equation}
Here the quotiented domain indicates the same regularization of \eqref{corr_Euc} as in \eqref{Zdef}.

Assuming that the field theory admits an analytic continuation to $\tAdS_2$ and a Hilbert space description, it should also be possible to define Lorentzian correlators. However, there is one ambiguity---whether matter fields in global $\tAdS_2$ should be quantized with respect to time $\phi$ or $-\phi$. In the former case, an excited state $\calO\ket{0}$ (where $\calO$ is some field operator) evolves in time with positive frequencies with respect to $\phi$, whereas in the latter, it evolves with negative frequencies with respect to $\phi$.\footnote{We use the convention that the phase $e^{-i \om t}$ has frequency $\om$ with respect to $t$.} Let us label the Hilbert space of excitations in the two cases $\calH_{\text{fields}}$ and $\calH_{\text{fields}}^*$, respectively. We will only resolve the choice between them in the Schwarzian limit; however, there will be correlators which do not depend on the choice, and thus are well-defined in general.

We now attempt to define the total Hilbert space of a black hole, consisting of matter fields and two boundaries represented by particles with spin $\nu$ and $-\nu$. The matter fields are decoupled from the boundaries, which imposes the structure
\be \label{HS_gen}
\calH \subseteq (\calH_{\text{fields}} \oplus \calH_{\text{fields}}^*)\otimes\calH^{\nu}_{\p}\otimes\calH^{-\nu}_{\p}.
\ee
Here, we have not resolved the ambiguity regarding quantization of fields; in addition, it should be understood that only $\tSL(2,\RR)$ singlet states are physical. Now, in the Schwarzian limit, the two-dimensional Hilbert space $\calH^{\nu}_{\p;\lambda,\mu}$ of a particle with definite quantum numbers splits into two one-dimensional subspaces $\calH^{\nu}_{\ra;\lambda,\mu}$ and $\calH^{\nu}_{\la;\lambda,\mu}$---localized near the right and left boundaries, respectively---because tunneling is suppressed. As explained in the next subsection, the correct choice of time for matter fields is such that we should choose $\calH_{\text{fields}}$ if the spin-$\nu$ particle is on the right, and $\calH_{\text{fields}}^*$ if it is on the left. (The $(-\nu)$-particle is always on the opposite side for black hole states.) This leads us to the total Hilbert space of a two-sided black hole
\begin{equation}\label{HS_fRL}
\calH = \bigl(\calH_{\text{fields}}
\otimes\calH^{\nu}_{\ra}\otimes\calH^{-\nu}_{\la}\bigr)
\oplus\bigl(\calH_{\text{fields}}^*
\otimes\calH^{\nu}_{\la}\otimes\calH^{-\nu}_{\ra}\bigr).
\end{equation}
Note the Hamiltonian, for say the $\nu$-particle, does not mix the spaces $\calH^{\nu}_{\ra}$ and $\calH^{\nu}_{\la}$. Thus in the above the two terms in the direct sum do not mix under dynamics, and the quantization of fields is well-defined for any given state.

We proceed to find the thermofield double state in the Hilbert space \eqref{HS_fRL}. Note $\Phi_E$ can be viewed as a vector in $\calH^{\nu}_{\p} \otimes \calH^{-\nu}_{\p}$---this justifies assigning the spins $\nu$ and $-\nu$ to the two particles of a black hole.\footnote{Mathematically, $\Phi_E$ is a vector in $\calH^{\nu}_{\p}\otimes\bigl(\calH^{\nu}_{\p}\bigr)^*$, but $\bigl(\calH^{\nu}_{\p}\bigr)^*$ (the Hilbert space dual to $\calH^{\nu}_{\p}$) is isomorphic to  $\calH^{-\nu}_{\p}$ because complex conjugation flips the imaginary spin $\nu$.} Then the state
\begin{equation}\label{Xi}
\ket{\Xi}=Z^{-1/2}\int dE \,e^{-\beta E/2} \ket{\Phi_E}.
\end{equation}
describes the thermofield double state of just the particle system. In general, we would like to take the tensor product of the field theory vacuum (which gives rise to a thermal state in the Schwarzschild patch) with the above. But to find the total state in \eqref{HS_fRL}, we recall that $\Phi_E(x;x')$ is supported in region $2$, where the $\nu$-particle is to the right of the $(-\nu)$-particle, and in region $1$, in which relative positions are flipped; this leads to the decomposition of $\ket{\Xi}$ into two orthonormal vectors $\ket{\Xi_{\ra\la}}$ and $\ket{\Xi_{\la\ra}}$, and to the total state 
\begin{equation}\label{TFD}
\ket{\TFD}=\ket{0}_{\text{fields}}\otimes\ket{\Xi_{\ra\la}}
+\ket{0^*}_{\text{fields}}\otimes\ket{\Xi_{\la\ra}}.
\end{equation}

Now, let us represent an operator $\calO$ acting at the position of the $\nu$-particle as
\begin{equation}\label{O_R}
\hat{\calO}^{\nu}
=\int d^2x \sqrt{-g(x)}\, \calO(x)\otimes \ket{x}\bra{x}\otimes \unit,\qquad
\hat{\calO}^{\nu}(T)=e^{iH^{\nu}T}\hat{\calO}^{\nu}e^{-iH^{\nu}T},
\end{equation}
where $H^{\nu}=\unit_{\text{fields}}\otimes H\otimes \unit$. The operators $\hat{\calO}^{-\nu}$, $\hat{\calO}^{-\nu}(T)$ acting at the location of the $(-\nu)$-particle are defined similarly, with the replacement $H^{\nu}\to H^{-\nu}=-\unit_{\text{fields}}\otimes \unit\otimes H^{\Tt}$.\,\footnote{If the field $\calO$ has nonzero spin, it should be transformed by the PT symmetry in the definition of $\hat{\calO}^{-\nu}$, and its Euclidean version in \eqref{corr_Euc} should be taken in the tilde gauge.} Then we consider the correlation functions
\begin{align}
\label{two-sided}
\tensor*{\calF}{*^{\nu,}_{\calX,}^{-\nu}_{\calY}}(T,0)
&=\frac{1}{2}\bbra{\TFD}\hat{\calX}^{\nu}(T)\kern1pt
\hat{\calY}^{-\nu}(0)\bket{\TFD},
\\[3pt]
\label{one-sided}
\tensor*{\calF}{*^{\nu,}_{\calX,}^{\nu}_{\calY}}(T,0)
&=\frac{1}{2}\bbra{\TFD}\hat{\calX}^{\nu}(T)\kern1pt
\hat{\calY}^{\nu}(0)\bket{\TFD}.
\end{align}
(We define the inner product of $\tSL(2,\RR)$-invariant functions as an integral over $\tSL(2,\RR)\backslash\tAdS_2\times\tAdS_2$ but multiply it by $\frac{1}{2}$ to obtain a physical quantity such as probability, see \eg \eqref{LorZ}.) We will see the two-sided correlator \eqref{two-sided} is not sensitive to the difference between fields quantized in $\calH_{\text{fields}}$ and $\calH_{\text{fields}}^*$;  thus we can replace $\ket{\text{TFD}}$ with $\ket{0}_{\text{fields}}\otimes \ket{\Xi}$ in its definition, and interpret it as a correlator in the thermofield double state quite generally, beyond the Schwarzian limit. It will be straightforward to show the correlator coincides with the analytic continuation of $\calF_{\calX,\calY}(\tau,0)$ at $\tau=\beta/2+iT$. In contrast, the one-sided correlator \eqref{one-sided} will turn out to be sensitive to the difference in quantization of fields. Making non-trivial use of the Schwarzian limit, we will show
\begin{equation}\label{ts-Euc}
\wideboxed{
\tensor*{\calF}{*^{\nu,}_{\calX,}^{\nu}_{\calY}}(T,0)
=\calF_{\calX,\calY}(iT,0).
}
\end{equation}

The second version of the problem is set in the context of the SYK or a similar quantum mechanics model. In this case, $\hat{\calX}$, $\hat{\calY}$ are understood as microscopic observables on a single copy of the system, but their Euclidean correlators can be expressed in a form similar to \eqref{corr_Euc} using the Schwarzian approximation. For example, if $\hat{\calX}=\hat{\calY}=\hat{\chi}_j$ is one of the Majorana modes in the SYK model, then
\begin{equation}
\begin{aligned}
\calF_{\calX,\calY}(\tau,0)
&=Z^{-1}\Tr\bigl(e^{-(\beta-\tau)H_{\text{mic}}}\hat{\calX}
e^{-\tau H_{\text{mic}}}\hat{\calY}\bigr)\\
&\approx\int D\vp\,e^{-I_{\Sch}[\vp]}\,
\bcorr{\calX(\vp(\tau))\kern1pt\calY(\vp(0))}\,
\vp'(\tau)^{\Delta}\vp'(0)^{\Delta},
\end{aligned}
\end{equation}
where $\corr{\calX(\vp_1)\calY(\vp_0)} \propto \vp_{10}^{-2\Delta}\sgn\vp_{10}$,\, $\vp_{10}=2\sin\frac{\vp_1-\vp_0}{2}$, and $I_{\Sch}$ is defined in \eqref{ISch}. The Lorentzian correlators are defined using the microscopic thermofield double
\begin{equation}
\ket{\TFD_{\text{mic}}}=Z^{-1/2}\sum_{n}e^{-\beta E_n/2}\ket{n,n}
\in\calH_{\text{mic}}\otimes\calH_{\text{mic}}^{*}.
\end{equation}
More specifically,
\begin{align}
\label{two-sided-mic}
\tensor*{\calF}{*^{\ra,}_{\calX,}^{\la}_{\calY}}(T,0)
&=\bbra{\TFD_\text{mic}}\hat{\calX}(T)\otimes\hat{\calY}^{\Tt}(0)
\bket{\TFD_\text{mic}},
\\
\label{one-sided-mic}
\tensor*{\calF}{*^{\ra,}_{\calX,}^{\ra}_{\calY}}(T,0)
&=\bbra{\TFD_\text{mic}}\hat{\calX}(T)\kern1pt\hat{\calY}(0)
\otimes \unit\bket{\TFD_\text{mic}},
\end{align}
so that the analogue of equation \eqref{ts-Euc}, $\tensor*{\calF}{*^{\ra,}_{\calX,}^{\ra}_{\calY}}(T,0)=\calF_{\calX, \calY}(iT,0)$, is trivial. Thus proving \eqref{ts-Euc} in the previous setting is a consistency check: it amounts to showing that the Schwarzian model has been correctly quantized such that $\ket{\TFD}$ is an adequate coarse-grained representation of $\ket{\TFD_{\text{mic}}}$.

The study of correlation functions in the Schwarzian limit can be framed in terms of asymptotic geometry. In the Euclidean case, we consider an infinitesimal neighborhood of the boundary of the Poincare disk with coordinates $\vp$ and
\begin{equation}
\zeta=2\gamma(1-r).
\end{equation}
(Note that $1-r\approx\vp'$, see the paragraph after equations \eqref{rpa1}, \eqref{rpa2}). Fields in this neighborhood are related to those at the boundary as
\begin{equation}
\calO(\vp,\zeta)\approx(\zeta/\gamma)^{\Delta}\calO(\vp).
\end{equation}
Although the Schwarzian approximation is valid only for $1-r\ll 1$, i.e.\ $\zeta\ll\gamma$, asymptotic expressions of relevant functions do not depend on $\gamma$ (except as an overall factor) and can be extrapolated to $\zeta\in(0,\infty)$. Thus the neighborhood of the boundary is $S^1\times(0,\infty)$, which is a topological cylinder. We will not define a metric on it, but rather use the functions $w$ and $\chi=1-w^{-1}$ of a pair of points as analogues of the geodesic distance. Two points on the Poincare disk may be specified as complex numbers $z_j=r_je^{i\vp_j}$,\, $j=0,1$. In this notation, $\chi$ is the cross-ratio of $(z_1,\bar{z}_1^{-1};z_0,\bar{z}_0^{-1})$:
\begin{equation}
\chi=-\frac{(1-z_1\bar{z}_1)(1-z_0\bar{z}_0)}{(z_1-z_0)(\bar{z}_1-\bar{z}_0)}
\approx -\frac{\zeta_1\zeta_0}{\gamma^2\vp_{10}^2}.
\end{equation}
The method of~\cite{BaAlKa16,BaAlKa17} corresponds to fixing one of the points, $(\vp_0,\zeta_0)=(-\pi,1)$, and using the variable $\phi=-\ln(-\chi)+\const$.

\begin{figure}[t]
\centerline{\begin{tabular}{@{}c@{\hspace{4cm}}c@{}}
\includegraphics{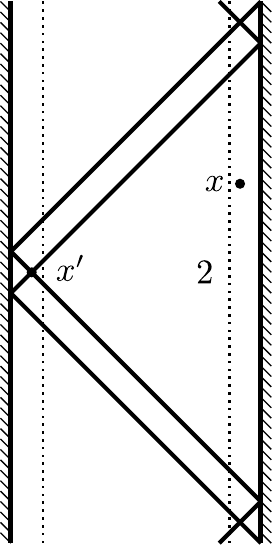} &
\includegraphics{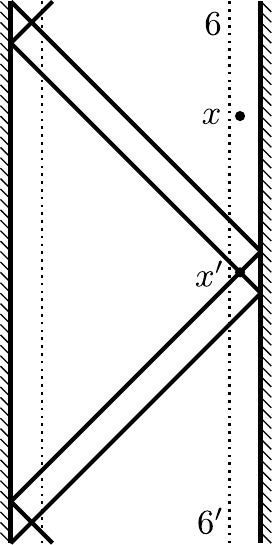}\vspace{5pt}\\
a) & b)
\end{tabular}}
\caption{Regions contributing to the asymptotic configuration space of a pair of points $(x,x')$: a) when $x$ is restricted to a neighborhood of the right boundary and $x'$ to a neighborhood of the left boundary of $\tAdS_2$ ; b) when both points are on the right.}
\label{fig_asympreg}
\end{figure}

In taking the Schwarzian limit in Lorentzian signature, we replace $\tAdS_2$ with the union of neighborhoods of the right and left boundaries, parametrized by $\phi$ and $\zeta=\gamma(\pi\mp 2\theta)$. Each component of the asymptotic space is a half-plane, $\RR\times(0,\infty)$. It can be represented as a quotient of $\tSL(2,\RR)$ by the subgroup generated by a parabolic element (\eg $\Lambda_2\pm\Lambda_0$ using the notation of~\cite{SL2R}). Functions on the asymptotic space provide a rigorous model of the $\tSL(2,\RR)$ representation $\calH^{\nu}_{\ra}\oplus\calH^{-\nu}_{\la}$ for $\nu=-i\gamma$,\, $\gamma\gg 1$, which is actually independent of $\gamma$.

Adapting the theory of $\tSL(2,\RR)$ invariant operators to the asymptotic setting involves the reduction of the relative configuration space $\tSL(2,\RR)\backslash \tAdS_2\times\tAdS_2$ to certain regions. The result would be more obvious if we studied functions on the asymptotic space from scratch, but let us give an informal argument based on what we already know. The space of operators on $\calH^{\nu}_{\ra}\oplus\calH^{\nu}_{\la}$ splits into four subspaces. To describe an operator in a particular subspace by a function $\Psi^{\nu}(x;x')$, we need to specify which side of $\tAdS_2$ each point is on. For example, if $x$ is on the right and $x'$ is on the left, then only region~2 and its copies are substantial, see Figure~\ref{fig_asympreg}a. Similarly, if both $x$ and $x'$ are on the right, then the asymptotic geometry includes only region~6 and its copies, as illustrated by Figure~\ref{fig_asympreg}b. These geometries are relevant to the two-sided correlator \eqref{two-sided} and one-sided correlator \eqref{one-sided}, respectively, as follows. We may consider only the first term in \eqref{TFD} in the expectation values, simultaneously eliminating the overall factor of $\frac{1}{2}$. Then in the two-sided correlator, the $\nu$- and $(-\nu)$-particles are restricted to be on the right and left, respectively. In the one-sided case, we integrate over the position of the left particle, obtaining a function of two points that are both on the right.

\subsection{Evaluation of Lorentzian correlators}

\begin{figure}[t]
\centerline{\begin{tabular}{@{}c@{\hspace{1.5cm}}c@{\hspace{2.5cm}}c@{}}
\includegraphics{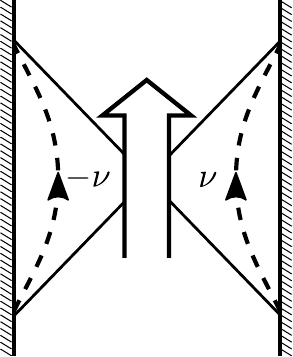} &
\includegraphics{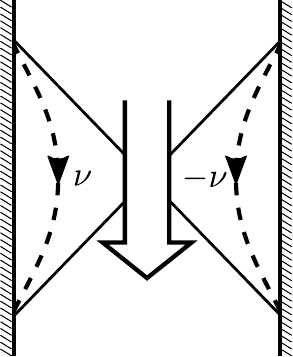} &
\includegraphics{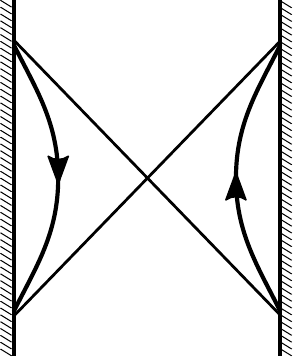} \\
a) & b) & c)
\end{tabular}}
\caption{a,b) The time with respect to which matter fields should be quantized, determined by the natural direction of propagation for $\nu$- and $(-\nu)$-particles on opposite sides of a classical trajectory; c) the direction corresponding to evolution by $H_{\text{TFD}}$.}
\label{fig:quant}
\end{figure}

In this section, we focus on correlators of matter fields in the coarse-grained thermofield double state $\ket{\text{TFD}}$. Before evaluating \eqref{two-sided} and \eqref{one-sided}, let us describe the physical motivation for the times chosen for quantizing matter fields in \eqref{TFD}. The reasoning is that it should agree with the direction of proper time of boundary particles on classical trajectories. Recall that a classical particle with spin $\nu=-i\gamma$,\, $\gamma>0$ moves counter-clockwise on circles in the Poincare disk. Hence the Euclidean proper time $\tau$ runs in the same direction as the polar coordinate $\vp$, and the Lorentzian proper time $T=-i\tau$, in the same direction as Schwarzschild time $t=-i\vp$; the last statement means that the particle traverses a pair of hyperbola-like trajectories counterclockwise. A particle with spin $-\nu$ moves in the opposite direction. Thus if the particles with spin $\nu$ and $-\nu$ stay on opposite sides of $\tAdS_2$, they move in the same direction, either up or down, as shown in Figure~\ref{fig:quant}\,a,b. By quantizing matter fields in that direction, we will obtain a complete agreement between different correlators in the Schwarzian limit. Note that there is another possible choice of time direction, based on the Hamiltonian
\be \label{H_TFD}
H_{\text{TFD}}=\unit_{\text{fields}} \otimes \lt(H \otimes \unit - \unit \otimes H^{\Tt}\rt),
\ee
which is a symmetry of the thermofield double. Since the second term, acting on the ($-\nu$)-particle, has a minus sign, $H_{\text{TFD}}$ pushes that particle in the opposite direction. As a result, the proper time for both $\nu$- and ($-\nu$)-particles is in the same direction as Schwarzschild time, see Figure~\ref{fig:quant}c.

We now proceed to use the state \eqref{TFD} in \eqref{two-sided} and \eqref{one-sided}. Each expectation value can be expressed as a trace of operators acting on $\calH^{\nu}_{\p}$:
\bea
\label{2sided-tr}\tensor*{\calF}{*^{\nu,\, -\nu}_{\calX,\calY}}(T,0)&=&\lt \langle\,{1 \ov 2}\tr \left( \sqrt{Z^{-1}e^{-\beta H}}\,\Phi^{\dagger}\, \hat{\calX}(T)\, \sqrt{Z^{-1}e^{-\beta H}}\, \Phi \,\hat{\calY}(0) \right) \rt \rangle_{\text{fields}}, \\
\label{1sided-tr}
\tensor*{\calF}{*^{\nu,}_{\calX,}^{\nu}_{\calY}}(T,0)&=&\lt \langle{1 \ov 2}\tr \lt(Z^{-1}e^{-\beta H}\Rho\, \hat{\calX}(T)\,\hat{\calY}(0) \rt) \rt \rangle_{\text{fields}}.
\eea
(The matter operators are defined as $\hat{\calO}(T)=e^{iH^{\nu}T}\hat{\calO}e^{-iH^{\nu}T}$,\, $\hat{\calO}=\int d^2x \sqrt{-g(x)}\, \calO(x)\otimes \ket{x}\bra{x}$.) Note $\Phi=-\Phi^{\dagger}$ commutes with $H= \Psi^{\nu}[E I]$. Compared to standard expressions, we have the substitutions $\varrho \to \varrho\, \Rho$, $\sqrt{\varrho} \to \pm \sqrt{\varrho}\, \Phi$. This is a natural extension of our prescription for density matrices given in \eqref{dmatrix}. Now, we may expand each trace as an integral over the quotient $\tSL(2,\RR)\backslash \tAdS_2 \times \tAdS_2$, or the space $S$ shown in Figure \ref{fig:regionsm}b, as in \eqref{trexp}; the expectation value $\corr{\calX(x)\calY(x')}$ of matter fields in their vacuum state will appear in the integrand, along with two-point functions that are position space representations of operators such as $\Phi$ and $\Rho$. The resulting integrals are a prescription for evaluating matter correlators in black hole states. We use them to establish the equivalence of the matter correlators to analytic continuations of the Euclidean correlator $\calF_{\calX,\calY}$ given by \eqref{Euc_corr_1}. 


Let us first consider the two-sided correlator \eqref{2sided-tr}. Its integral expansion is given by
\be \label{2sided_int}
\wideboxed{
\tensor*{\calF}{*^{\nu,\, -\nu}_{\calX,\calY}}(T,0)=Z^{-1}\int dE dE' \,e^{-(\bt/2-i T)E}e^{-(\bt/2+iT)E'} W_{\calX, \calY}(E,E'),
}
\ee
\be 
\wideboxed{\begin{aligned}\label{W1}
W_{\calX, \calY}(E,E')&= {1 \ov 2}\int_{\tSL(2,\RR)\backslash \tAdS^2 \times \tAdS^2}\Phi^{\dagger}_{E}(x'; x)\Phi_{E'}(x;x')\corr{\calX(x)\calY(x')}\\
&={1 \ov 2}\sum_{\text{regions $1,2$}}\int{2 du \ov (1-u)^2}\, \mathring{\Phi}_{E}(x;0)^*\, \mathring{\Phi}_{E'}(x;0)\corr{\calX(x)\calY(0)}.
\end{aligned}} 
\ee
Note that in \eqref{W1}, because of the space-like support of $\Phi$, the Wightman function $\corr{\calX(x)\calY(0)}$ is only used in space-like regions so does not depend on whether it is evaluated in $\calH_{\text{fields}}$ or $\calH_{\text{fields}}^*$. Thus as claimed in the previous subsection, we may replace $\ket{\text{TFD}}$ with $\ket{0}_{\text{fields}}\otimes \ket{\Xi}$ in the definition of $\tensor*{\calF}{*^{\nu,\, -\nu}_{\calX,\calY}}$ in \eqref{two-sided}, and interpret it as a correlator in the thermofield double state in general, not just in the Schwarzian limit. In fact, $W_{\calX, \calY}$ is just the kernel that appears in the Euclidean correlator
\be \label{WEuc}
\calF_{\calX, \calY}(\tau,0)=Z^{-1}\int dE dE'\, e^{-(\beta-\tau)E}e^{-\tau E'} \underbrace{\int {2du \ov (1-u)^2}\,\rG_E(0,x)\rG_{E'}(x,0)\corr{\calX(x)\calY(0)}}_{W_{\calX, \calY}(E,E')},
\ee
where to take the quotient with respect to $\PSL(2,\RR)$ in the domain of \eqref{Euc_corr_1}, we have restricted $x_0=0$ and further divided the integral over $x_1$ by $2\pi$. (To express $W_{\calX, \calY}$ in the form in \eqref{WEuc}, we use the fact that the integrand in \eqref{W1} is symmetric between regions $1,2$, as well as \eqref{match} and the analogous condition for $\corr{\calX(x)\calY(0)}$ that it is analytically continued from $\HH^2$ to region $2$ of $\tAdS_2$. We made a similar transition between Euclidean and Lorentzian integrals in \eqref{trans}.) It follows that
\be
\tensor*{\calF}{*^{\nu,\, -\nu}_{\calX,\calY}}(T,0)=\calF_{\calX, \calY}(\bt/2+i T,0).
\ee

As an aside, let us note that $W_{\calX, \calY}(E,E')$ is invariant under $E \leftrightarrow E'$ and $\calX \leftrightarrow \calY$, independently. The former follows from $\mathring{\Phi}_{E}(x;0)$ being real, see \eqref{Phi}. To see the latter, in the first line of \eqref{W1}, we replace $\corr{\calX(x)\calY(x')} \to \corr{\calY(x')\calX(x)}$ using that $\Phi$ has space-like support, then note the rest of the integrand $\Phi^{\dagger}_{E}(x';x)\Phi_{E'}(x;x')=-\Phi_E(x';x)\Phi_{E'}(x;x')$ is invariant under $x \leftrightarrow x'$. These symmetries imply an emergent time-reversal symmetry in our correlators $\tensor*{\calF}{*^{\nu,\, -\nu}_{\calX,\calY}}(T,0)$ and  $\tensor*{\calF}{*^{\nu,}_{\calX,}^{\nu}_{\calY}}(T,0)$ (the function $W_{\calX, \calY}$ also determines the latter via \eqref{ts-Euc} which we prove below), in the sense that in a generic quantum mechanical system, analogous correlators \eqref{two-sided-mic}, \eqref{one-sided-mic} will be invariant under $\calX \leftrightarrow \calY$ only if there the Hamiltonian $H$ and operators $\calX, \calY$ are invariant under time-reversal.

\begin{figure}[t]
\centerline{\begin{tabular}{@{}c@{\hspace{4cm}}c@{}}
\includegraphics{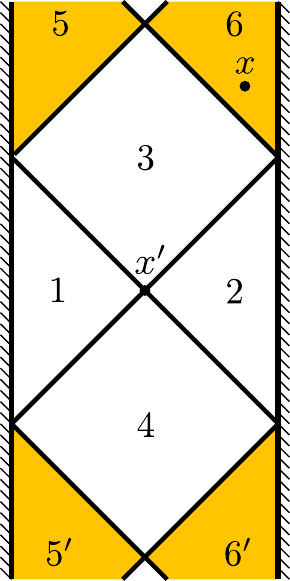} & \includegraphics{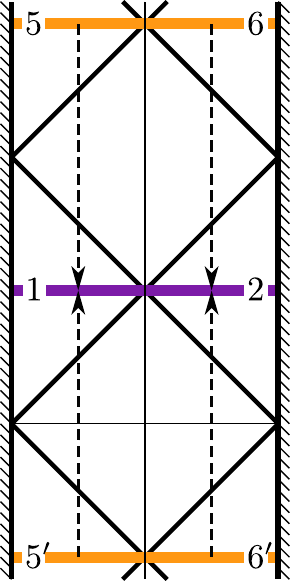} \\
a) & b)
\end{tabular}}
\caption{a): In the Schwarzian limit, one-sided propagators $\Rho(x;x')$ and $\Iota(x;x')$ are only supported in time-like, near-boundary regions $5,6$ and their copies; b) Analytic continuation between the time-like support of \eqref{W2} (orange) and space-like support of \eqref{W1} (purple).}
\label{fig:Sch}
\end{figure}

Next, we turn to expanding the one-sided correlator \eqref{1sided-tr} as an integral. To do so we need, besides matrix elements of $\Rho$, those of the identity operator on $\calH^{\nu}_{\p}$; the latter operator is physically just the propagator for a $\nu$-particle.\footnote{As for $\Rho$, its matrix elements give the amplitude for a $\nu$-particle to propagate via tunneling to and back from the other side of the black hole.} It is given by
\be \label{Id}
\wideboxed{\begin{aligned}
&\Iota=\Psi^{\nu}\bigl[I\bigr],\\[2pt]
&\mathring{\Iota}_E(x;0)= (2\pi)^{-2}\\
 &\hspace{55pt}\begin{dcases}
 -2 C_{\lam, \nu}(u)e^{2\pi i \lam |n|}& \text{in regions $(1,n)$ and $(2,n)$}\\
-2  \bC_{\lam, \nu}(u)e^{2\pi i \lam |n|} & \text{in regions $(3,n)$ and $(4,n)$}\\
\Gamma(\lam+\nu)\Gamma(1-\lam+\nu)A_{\lam, \nu, \nu}\bigl(u^{-1}\bigr)e^{2\pi i \lam |n|} \\
+\Gamma(\lam-\nu)\Gamma(1-\lam-\nu)A_{\lam, -\nu, -\nu}\bigl(u^{-1}\bigr)e^{2\pi i \lam |n+1|}& \text{in regions $(5,n)$ and $(6, -n-1)$}\\
\end{dcases}
\end{aligned}}
\ee 
where we have denoted the translation of region $j$, $j=1,\dotsc,6$ by $\phi \to \phi+2\pi n$ as $(j,n)$. Note $\Rho(x; x')$ and $\Iota(x; x')$ are non-vanishing at space-like separation. In the Schwarzian limit \eqref{Schlim3}, however, the two-point functions conform to our usual intuition as to how massive particles behave, in that they are suppressed in space-like regions $1,2$ (and in fact also their copies), exponentially in $\ga$. Furthermore, to leading order, they are also suppressed in interior regions---regions $3,4$ and their copies---which is a manifestation of the tendency of a particle to localize near a boundary, first seen in single-particle wavefunctions. See Figure \ref{fig:Sch}a. Thus we have
\be\label{1sided_int}
\wideboxed{
\tensor*{\calF}{*^{\nu,}_{\calX,}^{\nu}_{\calY}}(T,0)=Z^{-1}\int dE dE' \,e^{-\bt E}e^{i (E-E')T} \,W^{\text{1-sided}}_{\calX, \calY}(E,E'),
}
\ee
\be
\wideboxed{\begin{aligned} \label{W2}
W^{\text{1-sided}}_{\calX, \calY}(E,E')&={1 \ov 2}\int_{\tSL(2,\RR)\backslash \tAdS^2 \times \tAdS^2}\Rho_E(x'; x)\Iota_{E'}(x; x')\corr{\calX(x)\calY(x')}, \hspace{25pt}\\
&\underset{\ga \to \infty}{\approx}{1 \ov 2}\sum_{\text{regions }5, 5', 6, 6'}\int{2 du \ov (1-u)^2}\, \mathring{\Rho}_{E}(x;0)^*\, \mathring{\Iota}_{E'}(x;0)\corr{\calX(x)\calY(0)}.
\end{aligned} \quad  \figbox{1.1}{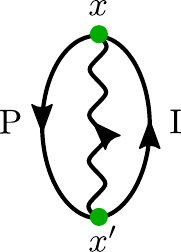}}
\ee
We may compare $W^{\text{1-sided}}_{\calX, \calY}$, which is a kind of spectral function, with its analogue in a microscopic theory---there, the one-sided correlator $\tensor*{\calF}{*^{\ra,}_{\calX,}^{\ra}_{\calY}}$ would be given by \eqref{1sided_int} but with $W^{\text{1-sided}}_{\calX, \calY}$ replaced by $\sum_{n,m}\braket{n | \hat{\calX}}{m}\de(E_m-E')\braket{m | \hat{\calY}}{n}\de(E_n-E)$. In both spectral functions, there is a propagation of intermediate states and a trace over initial/final states, but in our case the trace is performed  with a factor of the density of states, \ie we have the operator $\Rho$ completing the diagram in \eqref{W2} rather than another insertion of $\Iota$.

Now let us note that the integral in \eqref{W2}, which includes the evaluation of $\corr{\calX(x)\calY(x')}$ in the state \eqref{TFD}, is defined only in the Schwarzian limit: it is only after taking the Schwarzian limit of $\Rho$ and $\Iota$, which reduces the support of the integral to regions $5, 5', 6, 6'$, then interpreting the regions in the context of the asymptotic geometry described at the end of the last subsection (consisting of disconnected left and right components), that we can impose the quantization in \eqref{TFD} on $\corr{\calX(x)\calY(x')}$, which depends on whether the points $x$, $x'$ (which are positions of $\nu$-particles) are in the right or left component. As shown in Figure \ref{fig_asympreg}b, in the Schwarzian geometry, the relative configuration $(x;x')$ being in region $6$ or $6'$ implies that $x,x'$ are in the right component, and similarly, the relative configuration being in region $5$ or $5'$ implies that the points are in the left component. Then it follows from \eqref{TFD} that in regions $6,6'$, fields should be quantized with respect to time $\phi$, and in regions $5,5'$, time $-\phi$. In the remainder of this section, we will show that using the quantization prescribed as such in \eqref{W2}, $W^{\text{1-sided}}_{\calX, \calY}$ is equal to $W_{\calX, \calY}$, which implies \eqref{ts-Euc}. The equality between $W^{\text{1-sided}}_{\calX, \calY}$ and $W_{\calX, \calY}$ follows from analytic continuation between the time-like support of \eqref{W2} and space-like support of \eqref{W1}, see Figure \ref{fig:Sch}b.

As a first step in the proof, let us obtain the Schwarzian limit of the particle two-point functions $\Phi$, $\Rho$, and $\Iota$. In taking $\ga \gg 1$, it is convenient to decompose the functions $A_{\lam, \nu, -\nu}(u)$ and $A_{\lam, \pm \nu, \pm \nu}(u)$ appearing in \eqref{Phi}, and \eqref{Rho}, \eqref{Id}, in terms of the basis functions $B_{\lam, \nu, -\nu}(u), B_{1-\lam, \nu, -\nu}(u)$, and $B_{\lam, \nu, \nu}(u^{-1}), B_{1-\lam, \nu, \nu}\bigl(u^{-1}\bigr)$, respectively---where $B_{\lambda,l,r}(u)=u^{(l+r)/2}(1-u)^{\lambda}\,\hgfs\bigl(\lambda+l,\,\lambda+r,\,2\lambda;\,1-u\bigr)$.
This is so that $\ga$ appears in only the first and second arguments of hypergeometric functions; the decompositions are given in \eqref{ABtrans12} and \eqref{ABtrans56}. We then use the identity $\lim_{a, b\to \infty}\hgfs\bigl(a, \, b, \, c; \, {z^2 \ov 4 a b} \bigr)=\lt({z \ov 2}\rt)^{1-c}I_{c-1}(z)$ where $I_{\upsilon}(z)$ is the modified Bessel function of the first kind. After also taking $s \ll \ga$---recall \eqref{Schlim3}---and restricting to the near-boundary region $|1-u| \ll 1$, we find that using the rescaled coordinate
\be \label{ydef}
y=2\ga \sqrt{|1-u|},
\ee
\bea \label{SchPhi}
\mathring{\Phi}_{E}(x; 0)&\approx& \pm \ga^{-1}e^{-\pi \ga}{\sinh(2\pi s) \ov 2 \pi^2}y K_{2 is }(y) \qquad \text{$+$: in region $1$, $-$: region $2$},\\ \label{SchRho}
\mathring{\Rho}_{E}(x; 0)&\approx&   \ga^{-1}e^{-2\pi \ga}{\sinh(2\pi s) \ov 2 \pi^2} \begin{dcases}
y K_{2 is }(-iy) & \text{in regions $5,6'$}\\
y K_{2 is }(iy) & \text{in regions $5',6$}
\end{dcases},\\ \label{SchG}
\mathring{\Iota}_{E}(x; 0)&\approx&   {\ga^{-1} \ov 4 \pi} \begin{dcases}
i y I_{2 is }(iy) & \text{in regions $5,6'$}\\
-i y I_{-2 is }(-iy) & \text{in regions $5',6$},
\end{dcases}
\eea
where $K_{\upsilon}(z)={\pi \ov 2 \sin(\pi \upsilon)}\lt(I_{-\upsilon}(z)-I_{\upsilon}(z) \rt)$ is the modified Bessel function of the second kind. We have shown $\Iota_E$ in regions entering \eqref{W2}; more generally, it is supported in regions $(5,n)$ and $(6, -n-1)$, where it is given by
\be \label{genSchG}
\mathring{\Iota}_E(x;0)\approx {\ga^{-1} \ov 4 \pi^2} y\lt(\lt( -e^{-2\pi s}\rt)^{|n|}K_{2is}(-i y)+\lt(-e^{-2\pi s}\rt)^{|n+1|}K_{2i s}(i y)\rt).
\ee

\begin{figure}[t]

\centerline{\begin{tabular}{@{}c@{\hspace{1pt}}c@{\hspace{1pt}}c@{}}
\includegraphics[scale=0.95]{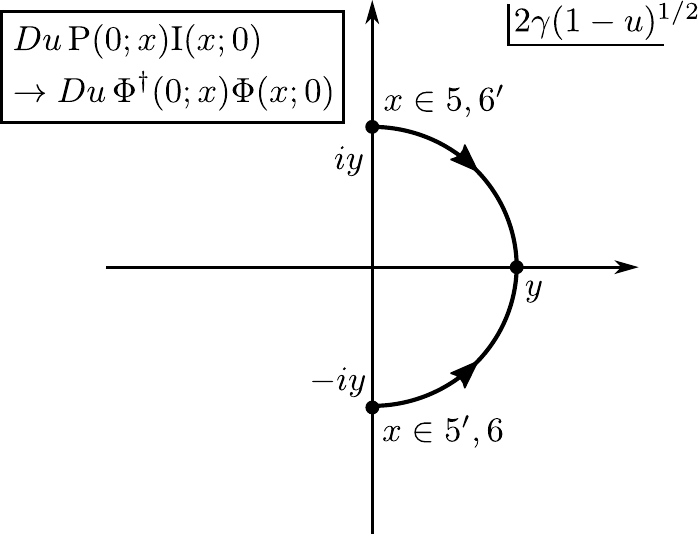} & \includegraphics[scale=0.95]{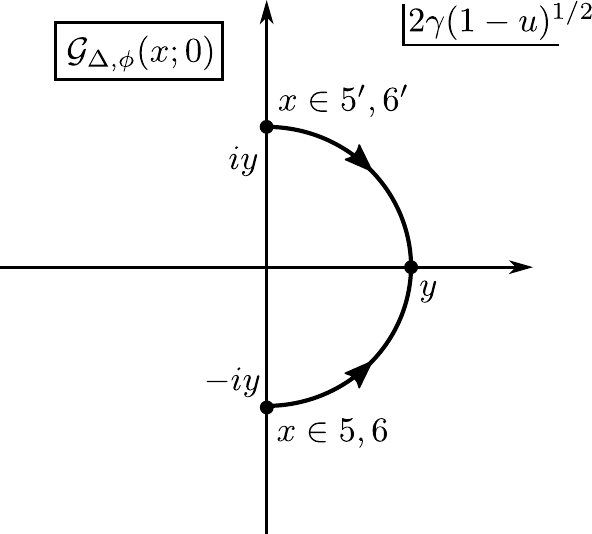} & \includegraphics[scale=0.95]{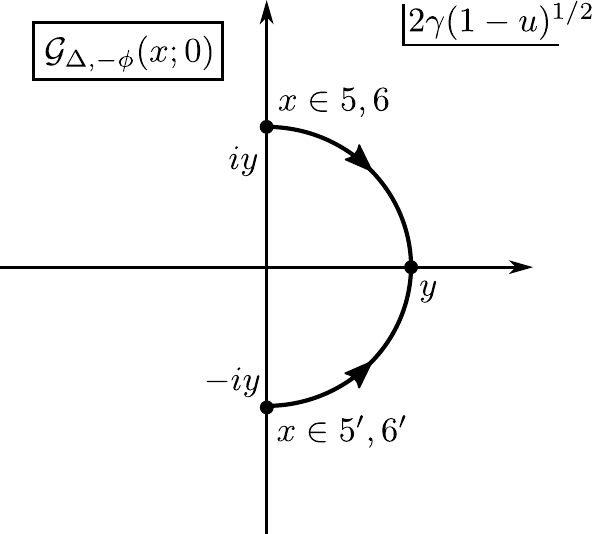}\\
a) & b) & c)
\end{tabular}}
\caption{a) We analytically continue $Du\,\Rho_{E}(0;x)\Iota_{E'}(x;0)$ in the coordinate $2\ga(1-u)^{1/2}$---which appears as the argument of Bessel functions in \eqref{SchPhi}, \eqref{SchRho}---to the real axis or $u<1$, \ie $x \in \text{regions }1,2$. The sum of continuations from regions $5$ and $5'$ (or $6$ and $6'$) gives $Du\,\Phi^{\dagger}_E(0;x)\Phi_{E'}(x;0)$. b, c) The analytic continuation between regions $5, 5', 6,6'$ and $1,2$ of $\calG_{\De, \phi}(x;0)$ and $\calG_{\De, -\phi}(x;0)$.}
\label{fig:condir}
\end{figure}

Now, the measure for integration on $S$ in each region is written using the $y$ coordinate \eqref{ydef} as $Du=16\ga^2 y^{-3}dy$.\footnote{We integrate near the boundary $u \approx 1$ in each region, so that given \eqref{urange}, the range and measure for integration is $\int_{1} 2du \,(1-u)^{-2}$ in regions $5,5', 6,6'$, and $\int^{1}2du\,(1-u)^{-2}$ in region $1,2$. Using the coordinate $y$, they can be uniformly expressed as $\int_0^{c} 16 \ga^2 dy\, y^{-3}$, where $c$ is a cutoff proportional to $\ga$.} Then also using \eqref{SchPhi}, \eqref{SchRho}, and \eqref{SchG}, we find that the sum of analytic continuations shown in Figure \ref{fig:condir}a, of $Du\,\Rho_{E}(0;x)\Iota_{E'}(x;0)$ in the pair of regions $(5,5')$ or $(6,6')$, equals $Du\,\Phi^{\dagger}_{E}(0;x)\Phi_{E'}(x;0)$ in region $1$ or $2$,\footnote{Recall that $\Rho_E(0;x)\Iota_{E'}(x;0)=\Rho_{E}^{\dagger}(x;0)^* \Iota_{E'}(x;0)=\Rho_{E}(x;0)^* \Iota_{E'}(x;0)$, and similarly $\Phi_{E}^{\dagger}(0;x)\Phi_{E'}(x;0)$, are gauge-invariant combinations of two-point functions.}
\bea
&&{16\,dy \ov y}\cdot{e^{-2\pi \ga}\sinh(2\pi s) \ov (2\pi^2)^2}\cdot{i \pi \ov 2}\lt(\underbrace{K_{2is}(iy)I_{2is'}(iy)}_{\text{$x \in$ region $5 \,(6')$}}-\underbrace{K_{2is}(-i y)I_{-2i s'}(-i y)}_{\text{$x \in$ region $5'\,(6)$}}\rt)\nn
&&\quad \underset{\text{cont. in Figure \ref{fig:condir}a}}{\longrightarrow}{16\, dy \ov y}\cdot {e^{-2\pi \ga} \sinh(2\pi s) \ov \lt(2\pi^2\rt)^2}\cdot{i \pi \ov 2}\cdot K_{2is}(y)\lt( \underbrace{I_{2is'}(y)-I_{-2i s'}(y)}_{{2\ov i \pi} \sinh(2\pi s')K_{2i s'}(y)} \rt).
\eea
Note expressions for $\Rho_E(x;0)$, $\Iota_E(x;0)$ in each region given in \eqref{SchRho}, \eqref{SchG} can be moved to the opposite imaginary axis in the argument of Bessel functions as
\be
\begin{pmatrix}
K_{\upsilon}(-iy) \\
I_{-\upsilon}(-i y)
\end{pmatrix}=\begin{pmatrix}
e^{-i \pi \upsilon} & i \pi \\
0 & e^{i \pi \upsilon
}\end{pmatrix}\begin{pmatrix}
K_{\upsilon}(i y) \\
I_{-\upsilon}(i y)
\end{pmatrix}.
\ee 
But then $\Rho_E(x;0)$ becomes a linear combination of $K_{2is}$ and $I_{\mp 2is}$ functions, and $Du\,\Rho_{E}(0;x)\Iota_{E'}(x;0)$ acquires a term quadratic in $I_{2is}$ (regions $5,6'$) or $I_{-2is}$ (regions $5', 6$). These functions grow exponentially at infinity, $I_{\upsilon}(z) \approx (2\pi z)^{-1/2}e^z$ for $\Re z>0$ (in comparison, $K_{\upsilon}(z) \approx (\pi/2z)^{1/2}e^{-z}$), so the quadratic term diverges at infinity and prohibits $Du\,\Rho_{E}(0;x)\Iota_{E'}(x;0)$ from being continued to the real axis. It follows that the direction of analytic continuation we show in Figure \ref{fig:condir}a is the only viable one from each of the regions $5, 5', 6, 6'$ to regions $1,2$.

It remains to consider analytic properties of the two-point function $\corr{\calX(x)\calY(0)}$ appearing in \eqref{W1} and \eqref{W2}. We have already determined that in \eqref{W2}, matter fields should be quantized with respect to $\phi$ in regions $6,6'$, and $-\phi$ in regions $5,5'$. By definition, quantization with respect to a time variable $\tilde{t}$ means that the corresponding Hamiltonian $\tilde{H}$ is positive. Ignoring the spatial dependence of $\calX(x)$ for simplicity, we may write $\calX(x)=e^{i\tilde{H}\tilde{t}}\calX e^{-i\tilde{H}\tilde{t}}$, and hence, $\corr{\calX(x)\calY(0)} =\sum_{m}\bra{0}\calX\ket{m} e^{-i\tilde{E}_m\tilde{t}} \bra{m}\calY\ket{0}$ with $\tilde{E}_m\ge0$. To be concrete, let us assume that the matter fields $\calX$, $\calY$ are free and consider single-particle excitations. By symmetry, these are the basis vectors $\ket{m}$,\, $\pm m=\Delta,\Delta+1,\ldots$ of some discrete series representation $\calD^{\pm}_{\Delta}$. By setting $\tilde{t}$ to $\phi$ or $-\phi$, we identify the frequency $\tilde{E}_{m}$ with $-m$ or $m$, respectively;\footnote{On page \pageref{princdiscr}, we described how $\tSL(2,\RR)$ representations, and spinors within a representation, are indexed by certain parameters. Here, it is relevant that a spinor depends on $\phi$ as $e^{im\phi}$.} this number is positive if we use the representation $\calD^{-}_{\Delta}$ in the first case and $\calD^{+}_{\Delta}$ in the second case.

Matter fields have integer or half-integer spins. For fields with zero spin, we can reuse the results of Appendix \ref{app:disc} (which are generally applicable to boundary particles with $\nu=-i\gamma$). Thus the Wightman function $\calG(x;x')=\corr{\calX(x)\calY(x')}$ evaluated for fields with dimension $\Delta$ and quantized in $\calH_{\text{fields}}$ or $\calH_{\text{fields}}^*$---which we denote $\calG_{\Delta, \phi}$ and $\calG_{\Delta, -\phi}$, respectively---are obtained by setting $\nu=0$ and $\lambda=\Delta$ in \eqref{tpf-disc}. In the Schwarzian limit, we have, up to a constant,
\be\label{SchG}
\calG_{\De,\pm \phi}(x;0) \approx \begin{dcases}
\lt( {y \ov 2 \ga}\rt)^{2\De} & \text{in regions $1,2$} \\
\lt( {\mp i y \ov 2 \ga}\rt)^{2\De} & \text{in regions $5,6$}\\
\lt( {\pm i y \ov 2 \ga}\rt)^{2\De} & \text{in regions $5',6'$}
\end{dcases}.
\ee
But then the direction of analytic continuation of the matter two-point function $\corr{\calX(x)\calY(0)}$, from each of the regions $5,5', 6,6'$ to regions $1,2$, is aligned with that of the rest of the integrand $Du\,\Rho_{E}(0;x)\Iota_{E'}(x;0)$ in \eqref{W2}, see Figure \ref{fig:condir}. This completes our proof that $W^{\text{1-sided}}_{\calX, \calY}=W_{\calX, \calY}$.

Before concluding, we obtain for completeness an explicit expression for the spectral function $W_{\calX, \calY}(E,E')$ in the Schwarzian limit. Using \eqref{SchPhi} and \eqref{SchG} in \eqref{W1}, we get
\begin{equation}\label{W_explicit}
W_{\calX, \calY}(E,E')\approx \gamma^{-2\Delta}\frac{e^{-2\pi\gamma}}{2\pi^4}
\sinh(2\pi s)\sinh(2\pi s')
\frac{\Gamma(\Delta\pm is\pm is')}{\Gamma(2\Delta)}
\end{equation}
where $\Gamma(\Delta\pm is\pm is') =\Gamma(\Delta+is+is')\,\Gamma(\Delta+is-is')\, \Gamma(\Delta-is+is')\,\Gamma(\Delta-is-is')$. We have used the identity
\begin{equation}
\int_{0}^{\infty}K_a(x)K_b(x)x^{1-c}\,dx =2^{c-3}\,
\frac{\Gamma\bigl(\frac{c+a+b}{2}\bigr)\Gamma\bigl(\frac{c+a-b}{2}\bigr)
\Gamma\bigl(\frac{c-a+b}{2}\bigr)\Gamma\bigl(\frac{c-a-b}{2}\bigr)}
{\Gamma(c)},
\end{equation}
which follows from the integral representation of the modified Bessel function of the second kind, $K_a(x)=\frac{1}{2}\int_{-\infty}^{+\infty}e^{-x\cosh\xi-a\xi}\,d\xi$. Plugging \eqref{W_explicit} into \eqref{WEuc} yields an expression for the Euclidean correlator that coincides with (4.10) in~\cite{MeTuVe17} up to a constant factor and is also consistent with equations (22), (23) in~\cite{BaAlKa16}.

\section{Summary and discussion}

Our main result is the construction of the two-point wavefunction $\Phi_E(x;x')$ with a fixed energy $E$ for a two-sided black hole. It may be viewed as a coarse-grained analogue of the microcanonical thermofield double state for the SYK model, which is proportional to $\sum_{n}\ket{n,n}\bra{n,n}$ with the sum taken over all Hamiltonian eigenstates in a narrow energy window. Both $\Phi_E$ and the eigenstate sum are highly entangled. In the SYK case, we assume that the energy window is much smaller than the temperature, but still contains exponentially many eigenstates. Naively, $\Phi_E$ has an infinite amount of entanglement because it includes all states in some infinite-dimensional representations of $\tSL(2,\RR)$. Much of the work was related to factoring out this infinity.

It is important that our geometric model and its limiting case, the Schwarzian model, have a complete Hilbert space description in Lorentzian spacetime. However, we used some input from the Euclidean version of the problem. To avoid this, we can reformulate condition~1 in Section~\ref{sec_tswf} as follows:
\begin{equation}\label{normPhi}
\frac{1}{2}\int_{\tSL(2,\RR)\backslash \tAdS_2 \times \tAdS_2}
\Phi^{\dagger}_E(x'; x)\,\Phi_{E'}(x; x')
=\rho(E)\,\delta(E-E'),\qquad
\rho(E)=(2\pi)^{-1}\Phi_E(x;x),
\end{equation}
where the integral is taken with the standard measure on the quotient space, $2du/(1-u)^2$. In the Schwarzian limit, we assume that $x$ and $x'$ are close to opposite boundaries of $\tAdS_2$; therefore $\Phi_E(x;x)$ is undefined. Instead of $\Phi_E(x;x)$, one may use the asymptotics of $\Phi_E(x;x')$ in the classically forbidden region to express $\rho(E)$ up to a constant factor. This seems to be the simplest and most robust interpretation of the density of states, which is implicit in~\cite{BaAlKa16,BaAlKa17}.

One of our motivations was to elucidate the meaning of wavefunctions on spaces with indefinite signatures, which appear in connection with the Wheeler-DeWitt equation. For single-particle wavefunctions on $\tAdS_2$, the inner product as the integral over the entire spacetime (rather than a time slice) is well justified. Indeed, the parameter $E$---which is conjugate to proper time, and thus may also be regarded as a particle's mass---is in general a dynamical variable; therefore, our particle has more degrees of freedom than the usual one. However, the integral in \eqref{normPhi} is essentially over a time slice. So we cannot draw a definite conclusion right now, but hope that our results will be useful in this context. 

Another open question is concerned with correlation functions. Our theory of (two-point) correlators is valid only in the Schwarzian limit; we do not know how to extend it to the general case. Perhaps one should abandon the idea that the Hilbert space factors into the spaces of fields and two individual boundaries. A more general principle is that ``particles'' with coordinates $x$, $x'$ representing the boundaries are always space-like separated, with $x$ on the right of $x'$. This allows for connecting $x$ and $x'$ by a space-like curve that may be regarded as a time slice of the physical spacetime. 

Finally, the construction of higher-order correlators in the Schwarzian limit seems straightforward, but it is still a nontrivial exercise to check the consistency between Euclidean and Lorentzian cases.

\section*{Acknowledgments}
We thank Daniel Jafferis, Juan Maldacena, Douglas Stanford, Herman Verlinde, and Zhenbin Yang for discussions. We gratefully acknowledge the support by the Simons Foundation through the ``It from Qubit'' program. A.K.\ is supported by the Simons Foundation under grant~376205 and by the Institute of Quantum Information and Matter, a NSF Frontier center funded in part by the Gordon and Betty Moore Foundation. This work was performed in part at Aspen Center for Physics, which is supported by National Science Foundation grant PHY-1607611.

\appendix

\section{Representation of $\widetilde{\SL}(2,\RR)$ by $\widetilde{\AdS}_2$ spinors} \label{app:repth}

We use the notation and definitions from Ref.~\cite{SL2R}. Let us give a quick summary and set up some further conventions. The standard $\widetilde{\AdS}_2$ coordinates are $(\phi,\theta)$, whereas the appropriate complex embedding is given by
\begin{equation}\label{vp1vp2}
(\tz_1,\tz_2)=(e^{i\vp_1},e^{i\vp_2}),\qquad \text{where}\quad \vp_1=\phi-\theta+\frac{\pi}{2},\quad\: \vp_2=\phi+\theta-\frac{\pi}{2}.
\end{equation}
We will also use the variable
\begin{equation}
\tu=\tz_1/\tz_2=e^{i(\pi-2\theta)}.
\end{equation}
On $\widetilde{\AdS}_2$ itself (rather than the bigger complex space), $\tu$ takes values in the unit circle without point $1$.

The Lie algebra of the symmetry group $\tGG\cong\widetilde{\SL}(2,\RR)$ is generated by three elements: $\Lambda_0$ (an infinitesimal shift in the $\phi$ direction), $\Lambda_1$ (a certain vector field preserving the $\phi=0$ slice), and $\Lambda_2$ (the Lorentz boost at the origin). It is often convenient to use the complex generators $L_{0}=-i\Lambda_0$ and $L_{\pm1}=\mp\Lambda_1-i\Lambda_2$, which satisfy the commutation relations $[L_n,L_m]=(n-m)L_{n+m}$. Note that symmetries act on each of the variables $\vp_1$, $\vp_2$ (and hence, $\tz_1$, $\tz_2$) separately, but in the same way.

In order to define spinors, consider the principal fiber bundle $\tGG\to\widetilde{\AdS}_2$ with the structure group $H$ generated by $\Lambda_2$. The fiber over point $x$ consists of the elements $g\in\tGG$ such that $g(0)=x$. Each point of the fiber may be identified with the local frame at point $x$ that is obtained from the standard frame at the origin by the symmetry transformation $g$. A $\nu$-spinor on $\widetilde{\AdS}_2$ is a function $\psi$ on $\tGG$ that has a special form on each fiber: $\psi(ge^{-\vt\Lambda_2})=e^{\nu\vt}\psi(g)$ for all $\vt$. For calculational purposes, spinors are represented as functions on $\widetilde{\AdS}_2$ by restricting $\psi$ to a particular cross section, called a ``gauge''. The standard nonsingular gauge is the tilde gauge defined in equation \eqref{tgauge} and surrounding text. Its relation to the disk gauge is described by this formula:
\begin{equation}\label{tilde_to_disk}
\tpsi(\phi,\theta)
=\left|\frac{\cos((\phi+\theta)/2)}{\cos((\phi-\theta)/2)}\right|^{\nu}
\rpsi(\phi,\theta).
\end{equation}
One may also view a spinor as a $\bigl(\frac{\nu}{2}, -\frac{\nu}{2}\bigr)$-form, that is, the formal expression
\begin{equation}\label{nunu-form}
\tpsi(\phi,\theta)\,(d\vp_1)^{\nu/2}(d\vp_2)^{-\nu/2}
\end{equation}
which behaves like an ordinary function if $\tpsi$ is transformed appropriately under $\widetilde{\SL}(2,\RR)$ and the differentials $d\vp_1$, $d\vp_2$ obey the standard transformation rules.

From now on, all spinors are implicitly given in the tilde gauge, unless indicated otherwise.
The action of the $\sL_2$ generators $L_{-1}$, $L_{0}$, $L_{1}$ on $\nu$-spinors in the $(\phi,\tu)$ and $(\phi,\theta)$ coordinates is given by these equations, where $\tu^{1/2}$ is understood as $e^{i(\pi/2-\theta)}$:
\begin{equation} \label{sl2gen}
\begin{aligned}
L_0&=i\partial_{\phi},\\[6pt]
L_{\pm1}&=e^{\pm i\phi}
\biggl(\pm(1-\tu)\tu^{1/2}\partial_{\tu}
+\frac{\tu^{-1/2}+\tu^{1/2}}{2}\,(i\partial_{\phi})
+\frac{\tu^{-1/2}-\tu^{1/2}}{2}\,\nu\biggr)\\[3pt]
&=e^{\pm i\phi}
\bigl(\pm\cos\theta\cdot\partial_{\theta}
+\sin\theta\cdot(i\partial_{\phi})-\cos\theta\cdot(i\nu)\bigr).
\end{aligned}
\end{equation}
If the spin value $\nu=-i\gamma$ is purely imaginary, the $\widetilde{\SL}(2,\RR)$ action is unitary, meaning that $L_{-n}$ is adjoint to $L_{n}$ with respect to the inner product
\begin{equation}\label{ip_AdS}
\braket{\psi_1}{\psi_2}=\int d^2 x \sqrt{-g}\,\psi_1(x)^*\kern1pt \psi_2(x).
\end{equation}
Our goal is to split this representation into isotypic components and study them individually. Recall that the $\widetilde{\SL}(2,\RR)$ irreps are characterized by parameters $\lambda$ and $\mu$ such that the Casimir operator
\begin{equation} \label{Qdef}
Q=-L_0^2+\frac{1}{2}(L_{-1}L_1+L_1L_{-1})
\end{equation}
is equal to $\lambda(1-\lambda)$ and the central element $e^{2\pi iL_0}$ to  $e^{-2\pi i\mu}$. (Note that $\mu$ is defined up to an integer and $\lambda$ up to the transformation $\lambda\leftrightarrow 1-\lambda$.) Thus the $(\lambda,\mu)$ isotypic component consists of solutions to these equations:
\begin{equation} \label{compeq}
Q\ket{\psi}=\lambda(1-\lambda)\ket{\psi},\qquad
L_0\ket{\psi}=-m\ket{\psi},\qquad m\in\mu+\ZZ.
\end{equation}
We will first find all solutions, and then select those that are normalizable or $\delta$-normalizable. 

The first part amounts to searching for functions of the form $\psi(\phi,\tu)=f(\tu)\,e^{im\phi}$ satisfying the equation $Qf=\lambda(1-\lambda)f$ with
\begin{equation} \label{tgQ}
Q=-(1-\tu)^2\,\bigl(\tu\partial_{\tu}^2+\partial_{\tu}\bigr)
+\frac{1-\tu}{4\tu}\Bigl((m-\nu)^2-(m+\nu)^{2}\tu\Bigr).
\end{equation}
This equation is closely related to the hypergeometric equation. Its solution space is two-dimensional, and one can define fundamental solutions by their asymptotic form at the regular singular points:
\begin{equation}
\begin{alignedat}{3}
\ww_1&\sim (-\tu)^{\frac{m-\nu}{2}},\qquad\quad&
\ww_2&\sim (-\tu)^{-\frac{m-\nu}{2}}\qquad\quad&
&\text{for }\, \tu\to-0, \\[3pt]
\ww_3&\sim (-\tu)^{\frac{m+\nu}{2}},\qquad\quad&
\ww_4&\sim (-\tu)^{-\frac{m+\nu}{2}}\qquad\quad&
&\text{for }\, \tu\to-\infty, \\[3pt]
\ww_5^{\pm}&\sim (1-\tu)^{\lambda},\qquad\quad&
\ww_6^{\pm}&\sim (1-\tu)^{1-\lambda}\qquad\quad&
&\text{for }\, \tu\to 1\pm i0.
\end{alignedat}
\end{equation}
These functions are defined on the complex plane with a branch cut from $0$ to $+\infty$. The first four solutions are more conveniently written in terms the variable
\begin{equation}
y=\frac{\tu}{\tu-1}=\frac{1}{2}-\frac{i}{2}\tan\theta,\qquad
y\notin(-\infty,0]\cup[1,\infty)
\end{equation}
so that the conditions $\tu\to-0$ and $\tu\to-\infty$ become $y\to+0$ and $y\to1-0$, respectively. The concrete expressions are as follows:
\begin{equation}\label{ww1234}
\begin{aligned}
\ww_1(\phi,\tu)&=y^{\frac{m-\nu}{2}}\,(1-y)^{-\frac{m+\nu}{2}}\,
\hgfs\bigl(\lambda-\nu,\,1-\lambda-\nu,\,1+m-\nu;\,y\bigl)\, e^{im\phi},\\[3pt]
\ww_2(\phi,\tu)&=y^{-\frac{m-\nu}{2}}\,(1-y)^{\frac{m+\nu}{2}}\,
\hgfs\bigl(\lambda+\nu,\,1-\lambda+\nu,\,1-m+\nu;\,y\bigl)\, e^{im\phi},\\[3pt]
\ww_3(\phi,\tu)&=y^{\frac{m-\nu}{2}}\,(1-y)^{-\frac{m+\nu}{2}}\,
\hgfs\bigl(\lambda-\nu,\,1-\lambda-\nu,\,1-m-\nu;\,1-y\bigl)\, e^{im\phi},\\[3pt]
\ww_4(\phi,\tu)&=y^{-\frac{m-\nu}{2}}\,(1-y)^{\frac{m+\nu}{2}}\,
\hgfs\bigl(\lambda+\nu,\,1-\lambda+\nu,\,1+m+\nu;\,1-y\bigl)\, e^{im\phi}.
\end{aligned}
\end{equation}
The other fundamental solutions and their more accurate $\tu\to 1\pm i0$ asymptotics are
\begin{equation}\label{ww56}
\begin{array}{l@{}c@{}l}
\ww_5^{\pm}(\phi,\tu)={}&
i^{\pm\lambda}\,B_{\lambda,m,-\nu}^{\pm}(\tu)\,e^{im\phi}
&\displaystyle
{}\approx\frac{1}{\Gamma(2\lambda)}\,\bigl(\pm i(1-\tu)\bigr)^{\lambda}
e^{im\phi},
\vspace{5pt}\\
\displaystyle\ww_6^{\pm}(\phi,\tu)={}&
i^{\pm(1-\lambda)}\,B_{1-\lambda,m,-\nu}^{\pm}(\tu)\,e^{im\phi}
&\displaystyle{}
{}\approx\frac{1}{\Gamma(2-2\lambda)}\,
{\ubrace{\bigl(\pm i(1-\tu)\bigr)}
_{\approx\,\pi\mp 2\theta\text{ for }\theta\to\pm\pi/2}}^{1-\lambda}\,
e^{im\phi}.
\end{array}
\end{equation}
Here we have used the notation
\begin{equation}\label{B}
B_{\lambda,l,r}(u)=u^{(l+r)/2}(1-u)^{\lambda}\,
\hgfs\bigl(\lambda+l,\,\lambda+r,\,2\lambda;\,1-u\bigr),
\end{equation}
whereas $B_{\lambda,l,r}^{\pm}(u)$ is the analytic continuations of $B_{\lambda,l,r}(u)$ from $u\in(0,1)$ to the domain $\CC-[0,+\infty)$ through the upper ($+$) or lower ($-$) half-plane.

The $\widetilde{\SL}(2,\RR)$ action is completely characterized by the operators $L_{\pm 1}$, which raise or lower $m$:
\begin{equation}\label{wwtrans}
\begin{aligned}
L_{-1}\ww_{1,m}&=i\,\ww_{1,m-1},&
L_{1}\ww_{1,m}&=-i(m+\lambda)(m+1-\lambda)\,\ww_{1,m+1},\\
L_{-1}\ww_{2,m}&=i(m-\lambda)(m-1+\lambda)\,\ww_{2,m-1},&
L_{1}\ww_{2,m}&=-i\,\ww_{2,m+1},\\
L_{-1}\ww_{3,m}&=-i(m-\lambda)(m-1+\lambda)\,\ww_{3,m-1},&
L_{1}\ww_{3,m}&=i\,\ww_{3,m+1},\\
L_{-1}\ww_{4,m}&=-i\,\ww_{4,m-1},&
L_{1}\ww_{4,m}&=i(m+\lambda)(m+1-\lambda)\,\ww_{4,m+1},\\
L_{-1}\ww_{5,m}^{\pm}&=\mp(m-\lambda)\,\ww_{5,m-1}^{\pm},&
L_{1}\ww_{5,m}^{\pm}&=\mp(m+\lambda)\,\ww_{5,m+1}^{\pm},\\
L_{-1}\ww_{6,m}^{\pm}&=\mp(m-1+\lambda)\,\ww_{6,m-1}^{\pm},&
L_{1}\ww_{6,m}^{\pm}&=\mp(m+1-\lambda)\,\ww_{6,m+1}^{\pm}.
\end{aligned}
\end{equation}

For given $\lambda$ and $m$, the $8$ fundamental solutions are related by these connection formulas:
\begin{equation}\label{conrel}
\begin{aligned}
\frac{\sin(2\pi\lambda)}{\pi}\,\ww_{1}
&=\frac{i^{\pm(-\lambda-m+\nu)}}{\Gamma(1-\lambda+m)\,\Gamma(1-\lambda-\nu)}\,
\ww_{5}^{\pm}
-\frac{i^{\pm(\lambda-1-m+\nu)}}{\Gamma(\lambda+m)\,\Gamma(\lambda-\nu)}\,
\ww_{6}^{\pm},\\[3pt]
\frac{\sin(2\pi\lambda)}{\pi}\,\ww_{2}
&=\frac{i^{\pm(-\lambda+m-\nu)}}{\Gamma(1-\lambda-m)\,\Gamma(1-\lambda+\nu)}\,
\ww_{5}^{\pm}
-\frac{i^{\pm(\lambda-1+m-\nu)}}{\Gamma(\lambda-m)\,\Gamma(\lambda+\nu)}\,
\ww_{6}^{\pm},\\[3pt]
\frac{\sin(2\pi\lambda)}{\pi}\,\ww_{3}
&=\frac{i^{\pm(\lambda-m-\nu)}}{\Gamma(1-\lambda-m)\,\Gamma(1-\lambda-\nu)}\,
\ww_{5}^{\pm}
-\frac{i^{\pm(1-\lambda-m-\nu)}}{\Gamma(\lambda-m)\,\Gamma(\lambda-\nu)}\,
\ww_{6}^{\pm},\\[3pt]
\frac{\sin(2\pi\lambda)}{\pi}\,\ww_{4}
&=\frac{i^{\pm(\lambda+m+\nu)}}{\Gamma(1-\lambda+m)\,\Gamma(1-\lambda+\nu)}\,
\ww_{5}^{\pm}
-\frac{i^{\pm(1-\lambda+m+\nu)}}{\Gamma(\lambda+m)\,\Gamma(\lambda+\nu)}\,
\ww_{6}^{\pm}.
\end{aligned}
\end{equation}
In fact, any element $\psi$ of the two-dimensional solution space can be expressed as a linear combination of $\ww_{5}^{+}$, $\ww_{6}^{+}$ and as a linear combination of $\ww_{5}^{-}$, $\ww_{6}^{-}$ with the coefficients proportional to the numbers $c_{+}^{\IN}$, $c_{+}^{\OUT}$ and $c_{-}^{\IN}$, $c_{-}^{\OUT}$ in this equation:
\begin{equation}\label{asform}
\psi(\phi,\theta)\approx \begin{dcases}
\left(c_{+}^{\IN}(\pi-2\theta)^{\lambda}
+c_{+}^{\OUT}(\pi-2\theta)^{1-\lambda}\right)e^{im\phi}
&\text{for }\theta\to\tfrac{\pi}{2},\\[3pt]
\left(c_{-}^{\IN}(\pi+2\theta)^{\lambda}
+c_{-}^{\OUT}(\pi+2\theta)^{1-\lambda}\right)e^{im\phi}
&\text{for }\theta\to-\tfrac{\pi}{2}.
\end{dcases}
\end{equation}
If $\lambda=\frac{1}{2}+is$ with $s>0$, then the terms $c_{\pm}^{\IN}(\pi\mp 2\theta)^{\lambda}$ and $c_{\pm}^{\OUT}(\pi\mp 2\theta\bigr)^{1-\lambda}$ may be interpreted as incoming and outgoing waves, respectively. The coefficients $c_{+}^{\IN}$, $c_{+}^{\OUT}$ are related to $c_{-}^{\IN}$, $c_{-}^{\OUT}$ by some transfer matrix $T$:
\begin{equation}
\begin{pmatrix} c_{+}^{\IN}\\ c_{+}^{\OUT} \end{pmatrix}
=T\begin{pmatrix} c_{-}^{\IN}\\ c_{-}^{\OUT} \end{pmatrix},\qquad\quad
T=\begin{pmatrix}
T^{\IN,\IN} & T^{\IN,\OUT}\\ T^{\OUT,\IN} & T^{\OUT,\OUT}
\end{pmatrix}.
\end{equation}
The explicit expression for $T$ is obtained from the connection formulas \eqref{conrel}:
\begin{equation}
\begin{aligned}
T^{\IN,\IN} &=
\frac{e^{i\pi\nu}\sin(\pi(\lambda-m))+e^{-i\pi\nu}\sin(\pi(\lambda+m))}
{\sin(2\pi\lambda)}\,,\\[4pt]
T^{\IN,\OUT} &=
\frac{2\pi\,\Gamma(1-2\lambda)\,\Gamma(2-2\lambda)}
{\Gamma(1-\lambda+\nu)\,\Gamma(1-\lambda-\nu)\,
\Gamma(1-\lambda+m)\,\Gamma(1-\lambda-m)}\,,\\[4pt]
T^{\OUT,\IN} &=
-\frac{2\pi\,\Gamma(2\lambda-1)\,\Gamma(2\lambda)}
{\Gamma(\lambda+\nu)\,\Gamma(\lambda-\nu)\,
\Gamma(\lambda+m)\,\Gamma(\lambda-m)}\,,\\[4pt]
T^{\OUT,\OUT} &=
-\frac{e^{i\pi\nu}\sin(\pi(\lambda+m))+e^{-i\pi\nu}\sin(\pi(\lambda-m))}
{\sin(2\pi\lambda)}\,.
\end{aligned}
\end{equation}

A wavefunction with the asymptotic form \eqref{asform} is normalizable or $\delta$-normalizable in the following two cases (up to the $\lambda\leftrightarrow 1-\lambda$ ambiguity):
\begin{enumerate}
\item $\lambda=\frac{1}{2}+is$ with $s>0$.
\item $\lambda>\frac{1}{2}$ and $c_{+}^{\OUT}=c_{-}^{\OUT}=0$. The last condition is satisfied (for a one-dimensional subspace of functions) if $T^{\OUT,\IN}=0$, that is, if $m=\lambda,\lambda+1,\ldots$ or $m=-\lambda,-\lambda-1,\ldots$.
\end{enumerate}
The first case corresponds to continuous series representations $\calC_{q}^{\mu}$ with $q=\lambda(1-\lambda)>\frac{1}{4}$ and the second to the discrete series representations $\calD_{\lambda}^{+}$, $\calD_{\lambda}^{-}$ (using the notation from~\cite{SL2R}). Thus, the Hilbert space $\calH^\nu$ of square-integrable $\nu$-spinors with purely imaginary $\nu$ splits into the isotypic components $\calH^{\nu}_{\lambda,\mu}\cong \CC^{2}\otimes\calC_{\lambda(1-\lambda)}^{\mu}$ for $\lambda=\frac{1}{2}+is$,\, $s>0$ and $\calH^{\nu}_{\lambda,\pm}\cong \calD_{\lambda}^{\pm}$ for $\lambda>\frac{1}{2}$.

The rest of the analysis will be done separately for the continuous and discrete series. One goal is to find all intertwiners from each $\widetilde{\SL}(2,\RR)$ irrep to the space of spinors. An intertwiner $\psi$ takes each basis vector $\ket{m}$ of the irreducible representation space to some function $\psi_m$. These functions should transform as the vectors $\ket{m}$, namely
\begin{equation}\label{bvtrans}
\begin{aligned}
L_{-1}\ket{m}&=-\sqrt{(m-\lambda)(m-1+\lambda)}\,\ket{m-1},\\[3pt]
L_0\ket{m}&=-m\,\ket{m},\\[3pt]
L_1\ket{m}&=-\sqrt{(m+\lambda)(m+1-\lambda)}\,\ket{m+1}.
\end{aligned}
\end{equation}
The space of intertwiners (of dimension $2$ or $1$) is denoted by $\calL^{\nu}_{\lambda,\mu}$ or $\calL^{\nu}_{\lambda,\pm}$ so that one may write $\calH^{\nu}_{\lambda,\mu}= \calL^{\nu}_{\lambda,\mu}\otimes\calC_{\lambda(1-\lambda)}^{\mu}$ and $\calH^{\nu}_{\lambda,\pm}= \calL^{\nu}_{\lambda,\pm}\otimes\calD_{\lambda}^{\pm}$ in the continuous and discrete case, respectively.

We will also construct the decomposition of the identity operator into projectors $\Pi^{\nu}_{\lambda,\mu}$, $\Pi^{\nu}_{\lambda,\pm}$ onto the isotypic components:
\begin{equation}
\!\wideboxed{
\unit=\int_{0}^{\infty}\!ds\,{s \ov (2\pi)^2}
\int_{-1/2}^{1/2}d\mu\, \frac{\sinh(2\pi s)}{\cosh(2\pi s)+\cos(2\pi\mu)}\,
\Pi^{\nu}_{1/2+is,\mu}
+\int_{1/2}^{\infty}d\lam\,\frac{\lambda-1/2}{(2\pi)^2}\kern1pt
\Bigl(\Pi_{\lambda,+}^{\nu}+\Pi_{\lambda,-}^{\nu}\Bigr)\!
}\!
\end{equation}
The scalar factors in the integration measure are a matter of convention. Here, the Plancherel measure is used as it corresponds to standard short-distance asymptotics of the two-point functions representing the projectors; namely, the coefficient in front of a logarithm is minus the dimension of the intertwiner space. In the continuous series case, there are four linearly independent operators (including the projector) that act within the corresponding isotypic component and commute with the group action. Of particular interest is a certain operator $Z$ that represents the particle flux in the $\theta$ direction. Its discrete series analogue is $\Pi_{\lambda,+}^{\nu}-\Pi_{\lambda,-}^{\nu}$; this operator measures the flux through a time slice.

\subsection{Continuous series components } \label{app:cont}

Let
\begin{equation}
s>0,\qquad \lambda=\frac{1}{2}+is,\qquad -\frac{1}{2}<\mu\leq\frac{1}{2},\qquad
\nu=-i\gamma.
\end{equation}
We will also use these abbreviations:
\begin{gather}
\begin{aligned}
a&=\sin(\pi(\lambda+\nu))\sin(\pi(\lambda-\nu))
=\frac{1}{2}\,\bigl(\cosh(2\pi s)+\cosh(2\pi\gamma)\bigr),\\[3pt]
b&=\sin(\pi(\lambda+\mu))\sin(\pi(\lambda-\mu))
=\frac{1}{2}\,\bigl(\cosh(2\pi s)+\cos(2\pi\mu)\bigr).
\end{aligned}
\end{gather}

\subsubsection{Basis functions and asymptotic coefficients }

The isotypic component $\calH^{\nu}_{\lambda,\mu}$ is spanned by functions of the form $\psi(\phi,\theta)=f(\theta)\,e^{im\phi}$ such that $m\in\mu+\ZZ$ and $Qf=\lambda(1-\lambda)f$. All such functions have already been found; we just need to organize them into sequences that transform as the vectors $\ket{m}$ in \eqref{bvtrans}. To this end, we multiply each sequence of fundamental solutions, which transform according to \eqref{wwtrans}, by suitable coefficients that depend on $m$:
\begin{equation}\label{normfs}
\begin{aligned}
\lt( \psi^{\lar}_{+} \rt)^{\nu}_{\lambda,m}&=
\frac{\Gamma(\lambda+\nu)\,\Gamma(1-\lambda+\nu)}
{\sqrt{\Gamma(\lambda+m)\,\Gamma(1-\lambda+m)}}\,i^{m}\,\lt(\ww_{2}\rt)_{\lam,m}^{\nu},
\\[3pt]
\lt( \psi^{\lar}_{-} \rt)^{\nu}_{\lambda,m}&=
\frac{\Gamma(\lambda+\nu)\,\Gamma(1-\lambda+\nu)}
{\sqrt{\Gamma(\lambda-m)\,\Gamma(1-\lambda-m)}}\,i^{-m}\,\lt(\ww_{4}\rt)_{\lam,m}^{\nu},
\\[3pt]
\lt( \psi^{\rar}_{+} \rt)^{\nu}_{\lambda,m}&=
\frac{\Gamma(\lambda-\nu)\,\Gamma(1-\lambda-\nu)}
{\sqrt{\Gamma(\lambda+m)\,\Gamma(1-\lambda+m)}}\,i^{-m}\,\lt(\ww_{3}\rt)_{\lam,m}^{\nu},
\\[3pt]
\lt( \psi^{\rar}_{-} \rt)^{\nu}_{\lambda,m}&=
\frac{\Gamma(\lambda-\nu)\,\Gamma(1-\lambda-\nu)}
{\sqrt{\Gamma(\lambda-m)\,\Gamma(1-\lambda-m)}}\,i^{m}\,\lt(\ww_{1}\rt)_{\lam,m}^{\nu},
\\[3pt]
\lt( \psi^{\IN}_{\pm} \rt)^{\nu}_{\lambda,m}&=
\frac{\Gamma(1-\lambda+\nu)\,\Gamma(1-\lambda-\nu)}{\sqrt{2\pi}}\,
\sqrt{\frac{\Gamma(1-\lambda\mp m)}{\Gamma(\lambda\mp m)}}\,\lt(\ww^{\mp}_{6}\rt)_{\lam, m}^{\nu},
\\[3pt]
\lt( \psi^{\OUT}_{\pm} \rt)^{\nu}_{\lambda,m}&=
\frac{\Gamma(\lambda+\nu)\,\Gamma(\lambda-\nu)}{\sqrt{2\pi}}\,
\sqrt{\frac{\Gamma(\lambda\mp m)}{\Gamma(1-\lambda\mp m)}}\,\lt(\ww^{\mp}_{5}\rt)_{\lam,m}^{\nu}.
\end{aligned}
\end{equation}
The choice of normalization factors and the meaning of indices will be clear from the subsequent discussion.

As already mentioned, a sequence of functions $\psi_m$ transforming as the basis vectors $\ket{m}$ represents an intertwiner from the $\widetilde{\SL}(2,\RR)$ irrep with parameters $(\lambda,\mu)$ to the space of $\nu$-spinors. An arbitrary intertwiner $\psi$ can be expressed in any of the four standard bases $\lt( \psi^{\lar}_{+},\psi^{\lar}_{-} \rt)$,\, $\lt(\psi^{\rar}_{+},\psi^{\rar}_{-} \rt)$,\, $\lt( \psi^{\IN}_{+},\psi^{\IN}_{-} \rt)$,\, $\lt( \psi^{\OUT}_{+},\psi^{\OUT}_{-} \rt)$:
\begin{equation}
\psi=r^{\sigma}_{+}\psi^{\sigma}_{+}+r^{\sigma}_{-}\psi^{\sigma}_{-},\qquad
\sigma=\lar,\,\rar,\,\IN,\,\OUT.
\end{equation}
The corresponding numbers $r^{\sigma}_{\pm}$, termed ``asymptotic coefficients'', are related by transformation matrices:
\begin{equation}
\begin{pmatrix}r^{\sigma}_{+}\\[2pt] r^{\sigma}_{-}\end{pmatrix}
=I^{\sigma,\tau}
\begin{pmatrix}r^{\tau}_{+}\\[2pt] r^{\tau}_{-}\end{pmatrix},\qquad\quad
I^{\sigma,\tau}=\begin{pmatrix}
I^{\sigma,\tau}_{++} & I^{\sigma,\tau}_{+-}\\[2pt]
I^{\sigma,\tau}_{-+} & I^{\sigma,\tau}_{--}
\end{pmatrix}.
\end{equation}
It follows from the connection formulas \eqref{conrel} that
\begin{equation}
\begin{aligned}
I^{\IN,\lar}&= \frac{\Gamma(\lambda+\nu)}{\sqrt{2\pi}}
\begin{pmatrix}
\eta_{+}^{1/2}i^{-\lambda-\nu} & i^{\lambda+\nu}\\[2pt]
i^{\lambda+\nu} & \eta_{+}^{-1/2}i^{-\lambda-\nu} 
\end{pmatrix},
\\[8pt]
I^{\OUT,\lar}&= \frac{\Gamma(1-\lambda+\nu)}{\sqrt{2\pi}}
\begin{pmatrix}
\eta_{-}^{1/2}i^{\lambda-1-\nu} & i^{1-\lambda+\nu}\\[2pt]
i^{1-\lambda+\nu} & \eta_{-}^{-1/2}i^{\lambda-1-\nu}
\end{pmatrix},
\\[8pt]
I^{\IN,\rar}&= \frac{\Gamma(\lambda-\nu)}{\sqrt{2\pi}}
\begin{pmatrix}
\eta_{-}^{-1/2}i^{\lambda-\nu} & i^{-\lambda+\nu}\\[2pt]
i^{-\lambda+\nu} & \eta_{-}^{1/2}i^{\lambda-\nu} 
\end{pmatrix},
\\[8pt]
I^{\OUT,\rar}&= \frac{\Gamma(1-\lambda-\nu)}{\sqrt{2\pi}}
\begin{pmatrix}
\eta_{+}^{-1/2}i^{1-\lambda-\nu} & i^{\lambda-1+\nu}\\[2pt]
i^{\lambda-1+\nu} & \eta_{+}^{1/2}i^{1-\lambda-\nu}
\end{pmatrix},
\end{aligned}\qquad\quad
\eta_{\pm}=e^{2\pi i\mu}\,
\frac{\sin(\pi(\lambda\pm\mu)}{\sin(\pi(\lambda\mp\mu)},
\end{equation}
and also
\begin{align}
I^{\rar\lar}&=\frac{\Gamma(\lambda+\nu)\,\Gamma(1-\lambda+\nu)}{\pi}
\begin{pmatrix}
\dfrac{i}{2}(e^{i\pi\nu}-e^{-i\pi\nu}e^{2\pi i\mu}) & \sqrt{b}
\vspace{3pt}\\
\sqrt{b} & \dfrac{i}{2}\bigl(e^{i\pi\nu}-e^{-i\pi\nu}e^{-2\pi i\mu}\bigr)
\end{pmatrix},
\displaybreak[0]\\[12pt]
\label{I_in-out}
I^{\OUT,\IN}&=\frac{\Gamma(1-\lambda-\nu)\,\Gamma(1-\lambda+\nu)}{2\pi}
\begin{pmatrix}
e^{i\pi\nu} + \eta^{-1} e^{-i\pi\nu} &
\dfrac{i\sinh(2\pi s)}{\sqrt{b}}
\vspace{4pt}\\
\dfrac{i\sinh(2\pi s)}{\sqrt{b}} &
e^{i\pi\nu} + \eta e^{-i\pi\nu}
\end{pmatrix},
\\[3pt]
\nonumber
&\text{where}\quad \eta=\frac{\sin(\pi(\lambda+\mu))}{\sin(\pi(\lambda-\mu))}.
\end{align}
For each pair of bases, the transformation in one direction is shown. All transformation matrices are unitary, and therefore their inverses are obtained easily.

Let us now explain the meaning of the asymptotic coefficients. The numbers $r^{\IN}_{\pm}$ and $r^{\OUT}_{\pm}$ are related to the amplitudes $c^{\IN}_{\pm}$, $c^{\OUT}_{\pm}$ of incoming and outgoing waves for the functions $\psi_m$ (see \eqref{asform}):
\begin{equation} \label{asymco}
\wideboxed{
\begin{aligned}
c_{\pm}^{\IN}&=\sqrt{\frac{2b}{\pi}}\,\Gamma(1-2\lambda)\,
\sqrt{\frac{\Gamma(\lambda\pm m)}{\Gamma(1-\lambda\pm m)}}\,
r_{\pm}^{\IN},\\[3pt]
c_{\pm}^{\OUT}&=\sqrt{\frac{2b}{\pi}}\,\Gamma(2\lambda-1)\,
\sqrt{\frac{\Gamma(1-\lambda\pm m)}{\Gamma(\lambda\pm m)}}\,
r_{\pm}^{\OUT},
\end{aligned}\qquad\quad
\frac{|c_{\pm}^{\IN}|^2}{|r_{\pm}^{\IN}|^2}
=\frac{|c_{\pm}^{\OUT}|^2}{|r_{\pm}^{\OUT}|^2}
=\frac{b}{s\sinh(2\pi s)}
}
\end{equation}
In physical applications, an important object is the $S$-matrix connecting the in- and out-amplitudes. Its elements are equal to the coefficients $I^{\OUT,\IN}_{\alpha\beta}$ up to some phase factors. From the explicit formula \eqref{I_in-out} for $I^{\OUT,\IN}$, we extract the tunneling probability:
\begin{equation}
p=|I^{\OUT,\IN}_{+-}|^2=|I^{\OUT,\IN}_{-+}|^2=\frac{\sinh^2(2\pi s)}{4ab}.
\end{equation}
Meanwhile, the coefficients $r^{\lar}_{\pm}$ and $r^{\rar}_{\pm}$ appear in an $m\to\pm\infty$ asymptotic formula for $\psi_m$, which can be derived as follows. First, we express $\psi_m$ as a linear combination of $\ww_{1,m}$ and $\ww_{3,m}$. By equation \eqref{ww1234}, the question is reduced to asymptotic properties of the hypergeometric function. The basic one is this:\footnote{This formula holds for all $y$ in the domain $D=\CC-[1,+\infty)$. It also extends to the part of the Riemann surface that is obtained by gluing infinitely many copies of the half-plane $\Re y>\frac{1}{2}$ to $D$ and to each other along the branch cut $[1,+\infty)$. For our purposes, $y$ takes values on the line $\Re y=\frac{1}{2}$, which is contained in $D$.}
\begin{equation}
\lim_{m\to+\infty}\hgf(a,b,c+m;y)=1.
\end{equation}
The transition to the scaled hypergeometric function $\hgfs$ is straightforward, whereas the $m\to-\infty$ case is analyzed using the identity
\begin{equation}
\begin{aligned}
\frac{\hgfs(a,b,c;y)}{\Gamma(b-c+1)\,\Gamma(a-c+1)}
={}&\frac{y^{1-c}(1-y)^{c-a-b}\,
\hgfs(1-b,\,1-a,\,2-c;\,y)}{\Gamma(a)\,\Gamma(b)}\\
&+\frac{\sin(\pi c)}{\pi}\,\hgfs(a,\,b,\,a+b-c+1;\,1-y).
\end{aligned}
\end{equation}
This calculation yields the following result:
\begin{equation}\label{minfty}
\wideboxed{
\begin{gathered}
\psi_m(\phi,\theta)\approx
|m|^{-1/2}\Bigl(
r^{\lar}_{\pm}\,i^{|m|}\bigl(2|m|\cos\theta\bigr)^{\nu}e^{im(\phi-\theta)}
+r^{\rar}_{\pm}\,i^{-|m|}\bigl(2|m|\cos\theta\bigr)^{-\nu}e^{im(\phi+\theta)}
\Bigr)\\[2pt]
\text{for }\, m\to\pm\infty\,
\text{ in any finite region of $\widetilde{\AdS}_2$}.
\end{gathered}
}
\end{equation}

Since the standard bases are related by unitary matrices, one can define an inner product on the intertwiner space $\calL^{\nu}_{\lambda,\mu}$ such that all four bases are orthonormal. Specifically, if intertwiners $\psi$ and $\psi'$ are characterized by the coefficients $r^{\sigma}_{\pm}$ and ${r'\kern1pt}^{\sigma}_{\pm}$, then
\begin{equation}\label{itw_ip}
\braket{\psi}{\psi'}^{\nu}
=(r^{\sigma}_{+})^{*}\,{r'\kern1pt}^{\sigma}_{+}
+(r^{\sigma}_{-})^{*}\,{r'\kern1pt}^{\sigma}_{-},\qquad
\sigma=\lar,\,\rar,\,\IN,\,\OUT.
\end{equation}
The inner product \eqref{itw_ip} is related to the usual inner product \eqref{ip_AdS} between the corresponding functions $\psi_{m}$ and $\psi'_{m'}$. On general grounds, the latter is proportional to $\delta(s-s')\,\delta(m-m')$ with some coefficient that can be calculated using the $\theta\to\pm\frac{\pi}{2}$ asymptotics. Thus,
\begin{align}
\nonumber
\braket{\psi_m}{\psi'_{m'}}&=
\Bigl((c_{+}^{\IN})^{*}\,{c'\kern1pt}_{+}^{\IN}
+(c_{-}^{\IN})^{*}\,{c'\kern1pt}_{-}^{\IN}
+(c_{+}^{\OUT})^{*}\,{c'\kern1pt}_{+}^{\OUT}
+(c_{-}^{\OUT})^{*}\,{c'\kern1pt}_{-}^{\OUT}\Bigr)\,
4\pi^2\delta(s-s')\,\delta(m-m')
\\[3pt]
\label{op_ip}
&=\braket{\psi}{\psi'}\cdot
\frac{2b}{\sinh(2\pi s)}\,(2\pi)^2
s^{-1}\delta(s-s')\,\delta(m-m').
\end{align}
Note that the last expression contains the inverse of the Plancherel factor. This was arranged by a suitable normalization of the basis vectors.

Finally, we comment on the geometric arrangement of the vectors $\ket{\psi^{\sigma}_{\pm}}$. Recall that these vectors are associated with the fundamental solutions of the hypergeometric equation. We have normalized them in a particular way, but the phase factors are arbitrary. An up-to-phase unit vector $\ket{\psi}\in\CC^2$ is characterized by the Pauli-like operator $2\ket{\psi}\bra{\psi}-I$, or equivalently, by the associated Bloch vector $\V{n}\in\RR^3$. An orthonormal basis corresponds to a pair of antipodal points on the Bloch sphere. A pair of bases such as $\ket{\psi^{\lar}_{\pm}}$, $\ket{\psi^{\rar}_{\pm}}$ makes a configuration with two $180^{\circ}$ rotation symmetries. In this example, they are described by Pauli-like operators $X$ and $Z$:
\begin{equation}\label{XZ}
\figbox{1.0}{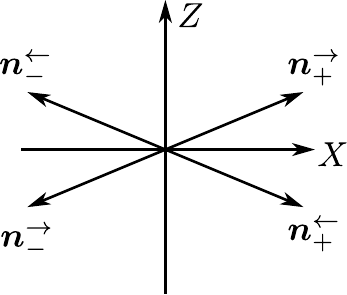}\qquad\quad
\begin{aligned}
\ket{\psi^{\rar}_{+}}\bra{\psi^{\rar}_{+}}
-\ket{\psi^{\lar}_{-}}\bra{\psi^{\lar}_{-}}
=\ket{\psi^{\lar}_{+}}\bra{\psi^{\lar}_{+}}
-\ket{\psi^{\rar}_{-}}\bra{\psi^{\rar}_{-}}
&=\sqrt{1-\frac{b}{a}}\,X,\\[3pt]
\ket{\psi^{\rar}_{+}}\bra{\psi^{\rar}_{+}}
-\ket{\psi^{\lar}_{+}}\bra{\psi^{\lar}_{+}}
=\ket{\psi^{\lar}_{-}}\bra{\psi^{\lar}_{-}}
-\ket{\psi^{\rar}_{-}}\bra{\psi^{\rar}_{-}}
&=\sqrt{\frac{b}{a}}\,Z.
\end{aligned}
\end{equation}
Similarly, for the $\ket{\psi^{\IN}_{\pm}}$, $\ket{\psi^{\OUT}_{\pm}}$ bases,
\begin{equation}\label{XpZ}
\begin{aligned}
\ket{\psi^{\OUT}_{+}}\bra{\psi^{\OUT}_{+}}
-\ket{\psi^{\IN}_{-}}\bra{\psi^{\IN}_{-}}
=\ket{\psi^{\IN}_{+}}\bra{\psi^{\IN}_{+}}
-\ket{\psi^{\OUT}_{-}}\bra{\psi^{\OUT}_{-}}
&=\sqrt{1-p}\,X',\\[3pt]
\ket{\psi^{\OUT}_{+}}\bra{\psi^{\OUT}_{+}}
-\ket{\psi^{\IN}_{+}}\bra{\psi^{\IN}_{+}}
=\ket{\psi^{\IN}_{-}}\bra{\psi^{\IN}_{-}}
-\ket{\psi^{\OUT}_{-}}\bra{\psi^{\OUT}_{-}}
&=\sqrt{p}\,Z,
\end{aligned}
\end{equation}
where $p=\frac{\sinh^2(2\pi s)}{4ab}$ is the tunneling probability. Importantly, $Z$ is the same in both cases.

The operator $\sqrt{p}\kern1pt Z$ measures the particle flux in the $\theta$ direction. Indeed, let us consider the Klein-Gordon current, whose matrix element between two wavefunctions is defined as follows:
\begin{equation}\label{J_def}
\bbra{\psi}J_{\alpha}(x)\bket{\psi'}
=\frac{i}{2}\Bigl(\nabla_{\alpha}\psi^{*}(x)\cdot\psi'(x)
-\psi^{*}(x)\cdot\nabla_{\alpha}\psi'(x)\Bigr).
\end{equation}
The current has zero divergence if $\psi$ and $\psi'$ are Casimir eigenfunctions with the same eigenvalue. To find the flux, we integrate the current over a vertical cross section, which can be placed at the right asymptotic boundary of $\widetilde{\AdS}_2$:
\begin{equation} \label{fluxop}
\begin{aligned}
\Bigl\langle\psi_m\Big|
\int_{\theta=\frac{\pi}{2}-0}\! dx^{\mu}\,
\epsilon_{\mu\nu}g^{\nu\al}J_{\al}(x)
\Big|\psi'_{m'}\,\Bigr\rangle
&=2s\lt( \lt( c_{+}^{\OUT} \rt)^{*}\,{c'\kern1pt}_{+}^{\OUT}
-\lt( c_{+}^{\IN} \rt)^{*}\,{c'\kern1pt}_{+}^{\IN} \rt)\,2\pi\,\delta(m-m')\\
&=2\pi\bbra{\psi}\sqrt{p}\kern1pt Z\bket{\psi'}\cdot
\frac{2b}{\sinh(2\pi s)}\,\delta(m-m').
\end{aligned}
\end{equation}
(Once again, the inverse Plancherel factor multiplying the delta-function is a consequence of the normalization convention.)

\subsubsection{$\widetilde{\SL}(2,\RR)$-invariant two-point functions} \label{app:twopt}

We now describe the functions that correspond to various operators acting in the two-dimensional space $\calL^{\nu}_{\lambda,\mu}$. Associated with an operator $R$ is the function $\Psi^{\nu}_{\lambda,\mu}[R]$ defined as follows:
\begin{equation}\label{tp_def}
\wideboxed{
\Psi^{\nu}_{\lambda,\mu}[R](\phi,\theta;\phi',\theta')
=\sum_{\alpha,\beta} R_{\alpha\beta} \sum_{m\in\mu+\ZZ}
\lt( \psi_{\alpha} \rt)^{\nu}_{\lambda,m}(\phi,\theta)\cdot
\lt( \psi_{\beta} \rt)^{\nu}_{\lambda,m}(\phi',\theta')^*
}
\end{equation}
Here $R_{\alpha\beta}$ are the matrix elements of $R$ in an arbitrary basis. A more specific notation (involving two bases) is $R^{\sigma\tau}_{\alpha\beta} =\bra{\psi^{\sigma}_{\alpha}}R \ket{\psi^{\tau}_{\beta}}$; the whole matrix is denoted by $R^{\sigma\tau}$.

Evaluating the sum \eqref{tp_def} presents some difficulty, so we take an indirect approach. Let us discuss some general properties of the two-point function $\Psi^{\nu}_{\lambda,\mu}[R]$. First, it is a $\nu$-spinor with respect to one point, $x=(\phi,\theta)$ and a $-\nu$-spinor with respect to the other point, $x'=(\phi',\theta')$. Furthermore, it is invariant under $\widetilde{\SL}(2,\RR)$ transformations. One may also regard $\Psi^{\nu}_{\lambda,\mu}[R]$ (or more exactly, the expression similar to \eqref{nunu-form}) as a $\bigl(\frac{\nu}{2},-\frac{\nu}{2}; -\frac{\nu}{2},\frac{\nu}{2}\bigr)$ form in the following variables:
\begin{equation}
\vp_1=\phi-\theta+\frac{\pi}{2},\qquad
\vp_2=\phi+\theta-\frac{\pi}{2};\qquad
\vp_3=\phi'-\theta'+\frac{\pi}{2},\qquad
\vp_4=\phi'+\theta'-\frac{\pi}{2}.
\end{equation}
Dividing it by another $\widetilde{\SL}(2,\RR)$-invariant form of the same type will produce an invariant scalar. Let
\begin{equation}
\vp_{jk}=2\sin\frac{\vp_j-\vp_k}{2},
\end{equation}
and let us use $|\vp_{14}|^{-\nu}|\vp_{23}|^{\nu}$ as a standard invariant form of the indicated type. Thus,
\begin{equation}\label{tPsi_f}
\Psi^{\nu}_{\lambda,\mu}[R](x;x')
=\left|\frac{\vp_{23}}{\vp_{14}}\right|^{\nu} f[R](x;x'),
\end{equation}
where $f[R]$ is an $\widetilde{\SL}(2,\RR)$-invariant scalar function.

\begin{figure}
\centerline{\begin{tabular}{@{}c@{\hspace{4cm}}c@{}}
\includegraphics{regions_a} &
\includegraphics{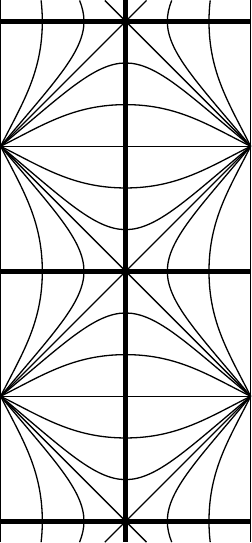}\vspace{5pt}\\
a) & b)
\end{tabular}}
\caption{a) Subdivision of $\widetilde{\AdS}_2$ into regions relative to $x'=0$;\, b) Orbits of points under the subgroup preserving $x'$ (thin lines) and a skeleton representation of the quotient set (thick lines).}
\label{fig_regions}
\end{figure}

To describe the position of $x$ relative to $x'$, let us place $x'$ at the origin, \ie set $\vp_3=\frac{\pi}{2}$ and $\vp_4=-\frac{\pi}{2}$. Then we consider $x$ up to residual symmetries preserving $x'=0$. Under such symmetries, the space splits into one-dimensional orbits filling two-dimensional regions. As shown in Figure~\ref{fig_regions}, there are non-equivalent regions $1$, $2$, $3$, $4$, $5$, $6$, and also their images under vertical translations, e.g.\ $5'$, $6'$. Since $\Psi^{\nu}_{\lambda,\mu}[R](\phi+2\pi,\theta;\phi',\theta') =e^{2\pi i\mu}\kern1pt \Psi^{\nu}_{\lambda,\mu}[R](\phi,\theta;\phi',\theta')$, it is sufficient to consider one copy of each region. We conclude that up to $\widetilde{\SL}(2,\RR)$ transformations, the pair $(x;x')$ is characterized by a discrete variable $j$ pointing to a particular region, as well as a continuous variable $w$. Hence, equation \eqref{tPsi_f} may be written as follows:
\begin{gather}
\label{tPsi_fj}
\wideboxed{
\Psi^{\nu}_{\lambda,\mu}[R](x;x')
=\left|\frac{\vp_{23}}{\vp_{14}}\right|^{\nu}f_{j}[R](w),\qquad
w=\frac{\vp_{13}\vp_{24}}{\vp_{14}\vp_{23}}
}\\[8pt]
0<w<1\, \text{ in regions }1,2,\qquad
w<0\, \text{ in regions }3,4,\qquad
w>1\, \text{ in regions }5,6.
\end{gather}
In regions $1$ and $2$, the points $x$ and $x'$ are space-like separated and $w=\tanh^2(\xi/2)$, where $\xi$ is the geodesic distance. These are some related quantities, including the familiar cross-ratio $\chi$:
\begin{equation}
1-w=\frac{\vp_{12}\vp_{34}}{\vp_{14}\vp_{32}},\qquad\quad
1-w^{-1}=\frac{\vp_{12}\vp_{34}}{\vp_{13}\vp_{24}}=\chi.
\end{equation}

Next, we use the fact that when $x'$ is fixed, $\Psi^{\nu}_{\lambda,\mu}[R](x,x')$ is a Casimir eigenfunction. To express this condition in terms of $f[R]$, we notice that $\Psi^{\nu}_{\lambda,\mu}[R](x,0)$ and $f[R](x,0)$ in \eqref{tPsi_f} are related by the same factor as $\tpsi$ and $\rpsi$ in \eqref{tilde_to_disk}. (Indeed, $\cos\bigl(\frac{\phi+\theta}{2}\bigr)=\vp_{32}$ and $\cos\bigl(\frac{\phi-\theta}{2}\bigr)=\vp_{14}$.) Therefore, $f[R](x,0)$ is the disk gauge variant of $\Psi^{\nu}_{\lambda,\mu}[R](x,0)$. The disk gauge allows for straightforward analytic continuation between the Schwarzschild patch, \ie region $2$, and the hyperbolic plane. In fact, the Casimir operator in the Schwarzschild patch  is obtained from its hyperbolic plane version \eqref{Q0} by simply replacing $u$ with $w$:
\begin{equation}
\rQ=-(1-w)^2\,\bigl(w\partial_{w}^2+\partial_{w}\bigr)-\nu^2(1-w).
\end{equation}
This expression for $\rQ$ is valid in all regions, although $f_{1}[R],\ldots,f_{6}[R]$ are not each other's analytic continuations. In each region $j$, the function $f_{j}[R]$ is a linear combination of two fundamental solutions with some coefficients. We will find them from equation \eqref{tp_def} by matching asymptotics.

The function $\Psi^{\nu}_{\lambda,\mu}[R](x;x')$ has singularities at all locations where $\vp_{13}$, $\vp_{14}$, $\vp_{23}$, or $\vp_{24}$ vanishes. These are exactly the region boundaries, which include the lines $\vp_1=\vp_3$,\, $\vp_2=\vp_4$ (representing the light cone), the lines $\vp_1=\vp_4$,\, $\vp_2=\vp_3$, and their translational copies. Since each term in \eqref{tp_def} is a smooth function, the singularities come from $m\to\pm\infty$. In this limit, each individual term is a product of a function of the form \eqref{minfty} and the complex conjugate of such a function. We may write
\begin{equation}\label{psip1}
\begin{aligned}
\psi_m(x)
&\approx |m|^{-1/2}\Bigl(
r^{\lar}_{\pm}\,i^{|m|}
\bigl(|m|\vp_{12}\bigr)^{\nu}e^{im(\vp_1-\pi/2)}
+r^{\rar}_{\pm}\,i^{-|m|}
\bigl(|m|\vp_{12}\bigr)^{-\nu}e^{im(\vp_2+\pi/2)}
\Bigr),\\[2pt]
\psi'_m(x')
&\approx |m|^{-1/2}\Bigl(
{r'\kern1pt}^{\lar}_{\pm}\,i^{|m|}
\bigl(|m|\vp_{34}\bigr)^{\nu}e^{im(\vp_3-\pi/2)}
+{r'\kern1pt}^{\rar}_{\pm}\,i^{-|m|}
\bigl(|m|\vp_{34}\bigr)^{-\nu}e^{im(\vp_4+\pi/2)}
\Bigr).
\end{aligned}
\end{equation}
The product $\psi_m(x)\,\psi'_m(x')^*$ involves the coefficients $r^{\sigma}_{\pm}({r'\kern1pt}^{\tau}_{\pm})^*$, but in the full function $\Psi^{\nu}_{\lambda,\mu}[R]$, they become $R^{\sigma\tau}_{\pm\pm} =\bra{\psi^{\sigma}_{\pm}}R \ket{\psi^{\tau}_{\pm}}$. Now the summation in $m$ is easy to perform, and we obtain the following expressions for the singular parts of $\Psi^{\nu}_{\lambda,\mu}[R](x;x')$ near the critical lines:
\begin{equation}\label{asymp_sing}
\Psi^{\nu}_{\lambda,\mu}[R](x;x')\approx \left|\frac{\vp_{23}}{\vp_{14}}\right|^{\nu}
\begin{dcases}
-(R^{\lar\lar}_{++}+R^{\lar\lar}_{--})\ln|\vp_{13}|
+i\frac{\pi}{2}(R^{\lar\lar}_{++}-R^{\lar\lar}_{--})\sgn\vp_{13}
&\text{if } \vp_1\approx\vp_3,
\\[2pt]
-(R^{\rar\rar}_{++}+R^{\rar\rar}_{--})\ln|\vp_{24}|
+i\frac{\pi}{2}(R^{\rar\rar}_{++}-R^{\rar\rar}_{--})\sgn\vp_{24}
&\text{if } \vp_2\approx\vp_4,
\\[2pt]
\Gamma(2\nu)\Bigl(e^{i\pi\nu\sgn\vp_{14}}R^{\lar\rar}_{++}
+e^{-2\pi i\mu}e^{-i\pi\nu\sgn\vp_{14}}R^{\lar\rar}_{--}\Bigr)
\left|\frac{\vp_{14}\vp_{23}}{\vp_{12}\vp_{34}}\right|^{-\nu}
&\text{if } \vp_1\approx\vp_4,
\\[2pt]
\Gamma(-2\nu)\Bigl(e^{-i\pi\nu\sgn\vp_{23}}R^{\rar\lar}_{++}
+e^{2\pi i\mu}e^{i\pi\nu\sgn\vp_{23}}R^{\rar\lar}_{--}\Bigr)
\left|\frac{\vp_{14}\vp_{23}}{\vp_{12}\vp_{34}}\right|^{\nu}
&\text{if } \vp_2\approx\vp_3.
\end{dcases}
\end{equation}

Equation \eqref{asymp_sing} and the Casimir eigenvalue equation are sufficient to reconstruct $\Psi^{\nu}_{\lambda,\mu}[R]$. However, let us also calculate the $\theta,\theta'\to\pm\frac{\pi}{2}$ asymptotics in order to allow for some cross-checks. By analogy with \eqref{psip1}, we write
\begin{equation}
\begin{aligned}
\psi_m(\phi,\theta) &\approx e^{im\phi}
\lt( c^{\IN}_{\alpha}\vp_{12}^{\lambda}
+c^{\OUT}_{\alpha}\vp_{12}^{1-\lambda} \rt)
&& \text{for }\, \theta\to\alpha\frac{\pi}{2},\quad \alpha=\pm 1,
\\[2pt]
\psi'_m(\phi',\theta') &\approx e^{im\phi'}
\bigl({c'\kern1pt}^{\IN}_{\beta}\vp_{34}^{\lambda}
+{c'\kern1pt}^{\OUT}_{\beta}\vp_{34}^{1-\lambda}\bigr)
&& \text{for }\, \theta'\to\beta\frac{\pi}{2},\quad \beta=\pm 1.
\end{aligned}
\end{equation}
In the expression for $\psi_m(\phi,\theta)\,\psi'_m(\phi',\theta')^*$, it is sufficient to keep the terms $c^{\IN}_{\alpha}\bigl({c'\kern1pt}^{\OUT}_{\beta}\bigr)^{*} e^{im(\phi-\phi')} \vp_{12}^{\lambda}\vp_{34}^{\lambda}$ and $c^{\OUT}_{\alpha}\bigl({c'\kern1pt}^{\IN}_{\beta}\bigr)^{*} e^{im(\phi-\phi' )} \vp_{12}^{1-\lambda}\vp_{34}^{1-\lambda}$. The other two terms may be neglected because they oscillate in $m$. When passing to the full function $\Psi^{\nu}_{\lambda,\mu}[R]$, the coefficients $c^{\IN}_{\alpha} \bigl( {c'\kern1pt}^{\OUT}_{\beta} \bigr)^{*}$ and $c^{\OUT}_{\alpha} \bigl( {c'\kern1pt}^{\IN}_{\beta} \bigr)^{*}$ should be replaced with
\begin{equation}
\begin{aligned}
c^{\IN,\OUT}_{\alpha\beta} &=\frac{2b}{\pi}\,\Gamma(1-2\lambda)^2
\sqrt{\frac{\Gamma(\lambda+\alpha m)\,\Gamma(\lambda+\beta m)}
{\Gamma(1-\lambda+\alpha m)\,\Gamma(1-\lambda+\beta m)}}\,
R^{\IN,\OUT}_{\alpha\beta},\\[3pt]
c^{\OUT,\IN}_{\alpha\beta} &=\frac{2b}{\pi}\,\Gamma(2\lambda-1)^2
\sqrt{\frac{\Gamma(1-\lambda+\alpha m)\,\Gamma(1-\lambda+\beta m)}
{\Gamma(\lambda+\alpha m)\,\Gamma(\lambda+\beta m)}}\,
R^{\OUT,\IN}_{\alpha\beta}.
\end{aligned}
\end{equation}
We consider four cases:
\begin{equation}
\begin{array}{@{}l@{\qquad}l@{\quad\,}l@{\quad\,}l@{}}
\text{region $1$:} & \alpha=-1, & \beta=1, & -\pi<\phi-\phi'<\pi,\\[2pt]
\text{region $2$:} & \alpha=1, & \beta=-1, & -\pi<\phi-\phi'<\pi,\\[2pt]
\text{region $5$:} & \alpha=-1, & \beta=-1, & 0<\phi-\phi'<2\pi,\\[2pt]
\text{region $6$:} & \alpha=1, & \beta=1, & 0<\phi-\phi'<2\pi.
\end{array}
\end{equation}
In each case, the summation in $m$ is reduced to the Fourier series
\begin{equation}
\biggl(2\sin\frac{\vp}{2}\biggr)^{-2\Delta} =\sum_{m\in\mu+\ZZ}
\frac{\Gamma(1-2\Delta)}{\Gamma(1-\Delta+m)\,\Gamma(1-\Delta-m)}\,
e^{im(\vp-\pi)},\qquad 0<\vp<2\pi.
\end{equation}
The result is as follows, where $w\to 1$:
\begin{equation}\label{asymp_infty}
\Psi^{\nu}_{\lambda,\mu}[R](x;x')\approx
\begin{dcases}
2\sqrt{b}\,\Bigl(\Gamma(1-2\lambda)\,R^{\IN,\OUT}_{-+}(1-w)^{\lambda}
+\Gamma(2\lambda-1)\,R^{\OUT,\IN}_{-+}(1-w)^{1-\lambda}\Bigr)
& \text{in region }1,
\\[2pt]
2\sqrt{b}\,\Bigl(\Gamma(1-2\lambda)\,R^{\IN,\OUT}_{+-}(1-w)^{\lambda}
+\Gamma(2\lambda-1)\,R^{\OUT,\IN}_{+-}(1-w)^{1-\lambda}\Bigr)
& \text{in region }2,
\\[2pt]
\begin{aligned}
2e^{i\pi\mu}\Bigl(
&\sin(\pi(\lambda+\mu))\,\Gamma(1-2\lambda)\,
R^{\IN,\OUT}_{--}\bigl(1-w^{-1}\bigr)^{\lambda}\\
&+\sin(\pi(\lambda-\mu))\,\Gamma(2\lambda-1)\,
R^{\OUT,\IN}_{--}\bigl(1-w^{-1}\bigr)^{1-\lambda}\,\Bigr)
\end{aligned}
& \text{in region }5,
\\[2pt]
\begin{aligned}
2e^{i\pi\mu}\Bigl(&\sin(\pi(\lambda-\mu))\,\Gamma(1-2\lambda)\,
R^{\IN,\OUT}_{++}\bigl(1-w^{-1}\bigr)^{\lambda}\\
&+\sin(\pi(\lambda+\mu))\,\Gamma(2\lambda-1)\,
R^{\OUT,\IN}_{++}\bigl(1-w^{-1}\bigr)^{1-\lambda}\,\Bigr)
\end{aligned}
& \text{in region }6.
\end{dcases}
\end{equation}

We now find the exact function $\Psi^{\nu}_{\lambda,\mu}[R]$ that matches the asymptotics \eqref{asymp_sing} and \eqref{asymp_infty}. It has the form \eqref{tPsi_fj} with $f_{1}[R],\ldots,f_{6}[R]$ satisfying the equation $\rQ f=\lambda(1-\lambda)f$. The concrete expressions involve fundamental solutions, which are chosen differently in three major cases.

\paragraph{Regions 1 and 2:} For $0<w<1$, the solutions with power-law behavior at $w\to 1$ make one suitable basis:
\begin{equation}
\begin{aligned}
B_{\lambda,\nu,-\nu}(w) &=(1-w)^{\lambda}\,
\hgfs\bigl(\lambda+\nu,\,\lambda-\nu,\,2\lambda;\,1-w\bigr),
\\[2pt]
B_{1-\lambda,\nu,-\nu}(w) &=(1-w)^{1-\lambda}\,
\hgfs\bigl(\lambda+\nu,\,\lambda-\nu,\,2-2\lambda;\,1-w\bigr),
\end{aligned}
\end{equation}
where the function $B_{\lambda,l,r}$ is defined by \eqref{B}. The $w\to 0$ solutions are constructed from
\begin{equation} \label{Af}
A_{\lambda,l,r}(w)=w^{(l+r)/2}(1-w)^{\lambda}\,
\hgfs\bigl(\lambda+l,\,\lambda+r,\,1+l+r;\,w\bigr).
\end{equation}
Specifically, we will use $A_{\lambda,\nu,-\nu}(w) =(1-w)^{\lambda}\, \hgfs(\lambda+\nu,\,\lambda-\nu,\,1;\,w)$ and
\begin{equation}
\begin{aligned} \label{Cf}
C_{\lambda,\nu}(w)={}&
\underbrace{\lim_{m\to\nu}\frac{A_{\lambda,m,-\nu}(w)-A_{\lambda,-m,\nu}(w)}
{m-\nu}}_{\approx\,\ln w-2\psi(1)\, \text{ for } w\to 0}\\[2pt]
&+\frac{\psi(\lambda+\nu)+\psi(1-\lambda+\nu)
+\psi(\lambda-\nu)+\psi(1-\lambda-\nu)}{2}\,A_{\lambda,\nu,-\nu}(w),
\end{aligned}
\end{equation}
where $\psi(x)=\frac{d}{dx}\ln(\Gamma(x))$. The last term is included so that the digamma function does not appear in the connection formulas:
\begin{equation}
\begin{aligned} \label{ABtrans12}
\frac{\sin(2\pi\lambda)}{\pi}\,A_{\lambda,\nu,-\nu}(w)
&=\frac{B_{\lambda,\nu,-\nu}(w)}{\Gamma(1-\lambda+\nu)\,\Gamma(1-\lambda-\nu)}
-\frac{B_{1-\lambda,\nu,-\nu}(w)}{\Gamma(\lambda+\nu)\,\Gamma(\lambda-\nu)},
\\[3pt]
-\frac{2a}{\pi^2}\,C_{\lambda,\nu}(w)
&=\frac{B_{\lambda,\nu,-\nu}(w)}{\Gamma(1-\lambda+\nu)\,\Gamma(1-\lambda-\nu)}
+\frac{B_{1-\lambda,\nu,-\nu}(w)}{\Gamma(\lambda+\nu)\,\Gamma(\lambda-\nu)}.
\end{aligned}
\end{equation}
In this notation,
\begin{equation}\label{tpf-12}
\wideboxed{
\begin{aligned}
f_{1}[R](w)&=-\Tr (R)\,C_{\lambda,\nu}(w)
+\pi\Tr\bigl( \bigl( D-i\sqrt{b/a}\,Z \bigr) R \bigr)\, A_{\lambda,\nu,-\nu}(w)
\\[3pt]
f_{2}[R](w)&=-\Tr (R)\,C_{\lambda,\nu}(w)
+\pi\Tr\bigl( \bigl(D+i\sqrt{b/a}\,Z\bigr)R\bigr)\, A_{\lambda,\nu,-\nu}(w)
\end{aligned}
}
\end{equation}
where $D=-i\sqrt{(b/a)(1-p)}\,X'Z$, see \eqref{XpZ}. Note that $D$ is traceless and anticommutes with $Z$ (which is also traceless). The comparison with the asymptotic formulas is best done by doing calculations in one particular basis, for example, $\ket{\psi^{\rar}_{\pm}}$. These are the expressions for $D$ and $Z$ in that basis:
\begin{align}
D^{\rar\rar}&=\frac{\sqrt{b}}{a}
\begin{pmatrix}
\dfrac{\sin(2\pi\mu)}{2\sqrt{b}} & 
\dfrac{e^{i\pi\nu}+e^{-i\pi\nu}e^{2\pi i\mu}}{2}\vspace{4pt}\\
\dfrac{e^{i\pi\nu}+e^{-i\pi\nu}e^{-2\pi i\mu}}{2} &
-\dfrac{\sin(2\pi\mu)}{2\sqrt{b}}
\end{pmatrix}, \label{Dmatrix}
\displaybreak[0]\\[12pt]
Z^{\rar\rar}&=\frac{1}{\sqrt{a}}
\begin{pmatrix}
\sqrt{b} & 
\dfrac{e^{i\pi\nu}-e^{-i\pi\nu}e^{2\pi i\mu}}{2i}\vspace{2pt}\\
-\dfrac{e^{i\pi\nu}-e^{-i\pi\nu}e^{-2\pi i\mu}}{2i} &
-\sqrt{b}
\end{pmatrix}.\label{Zmatrix}
\end{align}

\paragraph{Regions 3 and 4:} For $w<0$, we reuse $A_{\lambda,\nu,-\nu}(w)$ but modify $C_{\lambda,\nu}(w)$ so as to make it real. Let
\begin{equation} \label{bAf}
\bA_{\lambda,l,r}(w)=i^{-(l+r)}A^{+}_{\lambda,l,r}(w)
=y^{\frac{l+r}{2}}\,(1-y)^{\frac{-l+r}{2}}\,
\hgfs\bigl(\lambda+r,\,1-\lambda+r,\,1+l+r;\, y\bigr),\qquad y=\frac{w}{w-1},
\end{equation}
where $A^{+}_{\lambda,l,r}(w)$ is the analytic continuation of $A_{\lambda,l,r}(w)$ through the upper half-plane. The basis function complementary to $\bA_{\lambda,\nu,-\nu}(w)=A_{\lambda,\nu,-\nu}(w)$ is
\begin{equation}
\begin{aligned} \label{bCf}
\bC_{\lambda,\nu}(w)={}&
\underbrace{\lim_{m\to\nu}
\frac{\bA_{\lambda,m,-\nu}(w)-\bA_{\lambda,-m,\nu}(w)}{m-\nu}}
_{\approx\,\ln(-w)-2\psi(1)\, \text{ for } w\to 0}\\[2pt]
&+\frac{\psi(\lambda+\nu)+\psi(1-\lambda+\nu)
+\psi(\lambda-\nu)+\psi(1-\lambda-\nu)}{2}\,\bA_{\lambda,\nu,-\nu}(w).
\end{aligned}
\end{equation}
The solutions with power-law behavior at $w\to-\infty$ are:
\begin{equation}
\begin{aligned}
\bA_{\lambda,-\nu,-\nu} \big( w^{-1} \big)
&=(1-w)^{\nu}\,\hgfs\big(\lambda-\nu,\,1-\lambda-\nu,\,1-2\nu;\,
\tfrac{1}{1-w}\big),\\[3pt]
\bA_{\lambda,\nu,\nu} \bigl( w^{-1} \bigr)
&=(1-w)^{-\nu}\,\hgfs\bigl(\lambda+\nu,\,1-\lambda+\nu,\,1+2\nu;\,
\tfrac{1}{1-w}\bigr).
\end{aligned}
\end{equation}
The two bases are related by the connection formulas:
\begin{equation}
\begin{aligned}
\frac{\sin(2\pi\nu)}{\pi}\,\bA_{\lambda,\nu,-\nu}(w)
&=\frac{\bA_{\lambda,-\nu,-\nu}\big(w^{-1}\big)}
{\Gamma(\lambda+\nu)\,\Gamma(1-\lambda+\nu)}
-\frac{\bA_{\lambda,\nu,\nu}\big(w^{-1}\big)}
{\Gamma(\lambda-\nu)\,\Gamma(1-\lambda-\nu)},
\\[3pt]
-\frac{2a}{\pi^2}\,\bC_{\lambda,\nu}(w)
&=\frac{\bA_{\lambda,-\nu,-\nu}\big(w^{-1}\big)}
{\Gamma(\lambda+\nu)\,\Gamma(1-\lambda+\nu)}
+\frac{\bA_{\lambda,\nu,\nu}\big(w^{-1}\big)}
{\Gamma(\lambda-\nu)\,\Gamma(1-\lambda-\nu)}.
\end{aligned}
\end{equation}
Let us write the result in the first basis:
\begin{equation}\label{tpf-34}
\wideboxed{
\begin{aligned}
f_{3}[R](w)&=-\Tr (R)\,\bC_{\lambda,\nu}(w)
+\pi\Tr\bigl(\bigl(D+i\sqrt{1-b/a}\,X\bigr)R\bigr)\,\bA_{\lambda,\nu,-\nu}(w)
\\[3pt]
f_{4}[R](w)&=-\Tr (R)\,\bC_{\lambda,\nu}(w)
+\pi\Tr\bigl(\bigl(D-i\sqrt{1-b/a}\,X\bigr)R\bigr)\,\bA_{\lambda,\nu,-\nu}(w)
\end{aligned}
}
\end{equation}
where the operator $X$ was defined in \eqref{XZ}. It is traceless, anticommutes with $Z$, and has this matrix form:
\begin{equation}
X^{\rar\rar}=\sqrt{\frac{b}{a(a-b)}}\,
\begin{pmatrix}
\dfrac{a-b}{\sqrt{b}} & 
-\dfrac{e^{i\pi\nu}-e^{-i\pi\nu}e^{2\pi i\mu}}{2i}\vspace{4pt}\\
\dfrac{e^{i\pi\nu}-e^{-i\pi\nu}e^{-2\pi i\mu}}{2i} &
-\dfrac{a-b}{\sqrt{b}}
\end{pmatrix}.
\end{equation}

\paragraph{Regions 5 and 6:} For $w>1$, the two standard bases of the solution space are related as follows:
\begin{equation}
\begin{aligned} \label{ABtrans56}
\frac{\sin(2\pi\lambda)}{\pi}\,A_{\lambda,-\nu,-\nu}\big(w^{-1}\big)
&=\frac{B_{\lambda,\nu,\nu}\big(w^{-1}\big)}{\Gamma(1-\lambda-\nu)^2}
-\frac{B_{1-\lambda,\nu,\nu}\big(w^{-1}\big)}{\Gamma(\lambda-\nu)^2},
\\[3pt]
\frac{\sin(2\pi\lambda)}{\pi}\,A_{\lambda,\nu,\nu}\big(w^{-1}\big)
&=\frac{B_{\lambda,\nu,\nu}\big(w^{-1}\big)}{\Gamma(1-\lambda+\nu)^2}
-\frac{B_{1-\lambda,\nu,\nu}\big(w^{-1}\big)}{\Gamma(\lambda+\nu)^2}.
\end{aligned}
\end{equation}
Using the $A$ basis, the answer is:
\begin{equation}\label{tpf-56}
\wideboxed{
\begin{aligned}
f_{5}[R](w)&=\begin{aligned}[t]
&e^{2\pi i\mu} \Tr \bigl(\bigl( \tfrac{1}{2}I+G_{-} \bigr)R \bigr)\,
\Gamma(\lambda-\nu)\Gamma(1-\lambda-\nu)\,A_{\lambda,-\nu,-\nu}\big(w^{-1}\big)
\\[2pt]
&+\Tr\bigl(\bigl(\tfrac{1}{2}I-G_{+}\bigr)R \bigr)\,
\Gamma(\lambda+\nu)\Gamma(1-\lambda+\nu)\,A_{\lambda,\nu,\nu}\big(w^{-1}\big)
\end{aligned}
\\[8pt]
f_{6}[R](w)&=\begin{aligned}[t]
&\Tr\bigl(\bigl(\tfrac{1}{2}I+G_{+}\bigr)R \bigr)\,
\Gamma(\lambda-\nu)\Gamma(1-\lambda-\nu)\,A_{\lambda,-\nu,-\nu}\big(w^{-1}\big)
\\[2pt]
&+e^{2\pi i\mu} \Tr\bigl(\bigl(\tfrac{1}{2}I-G_{-}\bigr)R \bigr)\,
\Gamma(\lambda+\nu)\Gamma(1-\lambda+\nu)\,A_{\lambda,\nu,\nu}\big(w^{-1}\big)
\end{aligned}
\end{aligned}
}
\end{equation}
where
\begin{equation}
G_{\pm}=\frac{a}{\sin(2\pi\nu)}\,\lt(D\pm i\sqrt{1-b/a}\,X\rt).
\end{equation}

\subsubsection{Some special cases}

The most important cases are $R=I$ and $R=Z$. The function $\Psi^{\nu}_{\lambda,\mu}[I]=\Pi^{\nu}_{\lambda,\mu}$ represents the projector onto the $(\lambda,\mu)$ irrep. Its expression in the form \eqref{tPsi_fj} involves these functions:
\begin{equation}
\!\wideboxed{
\begin{aligned}
f_{1}[I](w)&=f_{2}[I](w)=-2C_{\lambda,\nu}(w)
\\[5pt]
f_{3}[I](w)&=f_{4}[I](w)=-2\bC_{\lambda,\nu}(w)
\\[5pt]
f_{5}[I](w)&= e^{2\pi i\mu}\,
\Gamma(\lambda-\nu)\Gamma(1-\lambda-\nu)\,
A_{\lambda,-\nu,-\nu}\bigl(w^{-1}\bigr)
+\Gamma(\lambda+\nu)\Gamma(1-\lambda+\nu)\,
A_{\lambda,\nu,\nu}\bigl(w^{-1}\bigr)
\\[5pt]
f_{6}[I](w)&=\Gamma(\lambda-\nu)\Gamma(1-\lambda-\nu)\,
A_{\lambda,-\nu,-\nu}\big(w^{-1}\big)
+e^{2\pi i\mu}\,\Gamma(\lambda+\nu)\Gamma(1-\lambda+\nu)\,
A_{\lambda,\nu,\nu}\big(w^{-1}\big)
\end{aligned}\!
}\!
\end{equation}
Remarkably, the function $\Psi^{\nu}_{\lambda,\mu}[Z]$ has support only in regions $1$ and $2$ and their copies:
\begin{equation} \label{Zwf}
\wideboxed{
\begin{aligned}
-f_{1}[Z](w)&=f_{2}[Z](w)=2\pi i\sqrt{b/a}\,A_{\lambda,\nu,-\nu}(w)
\\[5pt]
f_{3}[Z](w)&=f_{4}[Z](w)=f_{5}[Z](w)=f_{6}[Z](w)=0
\end{aligned}
}
\end{equation}

\subsection{Discrete series components} \label{app:disc}

The logic here is quite similar to that for the continuous series. Given $\nu=-i \ga$ and $\lambda >1/2$, there are two sequences of normalizable $\nu$-spinors $\psi^{\nu}_{m}$ that transform as the basis vectors $\ket{m}\in\calD^{\pm}_{\lambda}$: one for $m=\lambda+n$ and the other for $m=-(\lambda+n)$ with $n=0,1,\ldots$. They can be expressed in terms of the fundamental solutions \eqref{ww1234}, \eqref{ww56} in several ways:
\begin{align}
&\hspace{13pt}\begin{aligned}
\psi^{\nu}_{\lambda,\lambda+n}
&=c\,i^{-\nu}\sqrt{\frac{\Gamma(2\lambda+n)}{n!}}\,
\lt(\ww_{5}^{+}\rt)^{\nu}_{\lambda,\lambda+n}
=c\,i^{\nu}(-1)^{n}\sqrt{\frac{\Gamma(2\lambda+n)}{n!}}\,
\lt(\ww_{5}^{-}\rt)^{\nu}_{\lambda,\lambda+n}\\[2pt]
&=c\,\frac{\Gamma(1-\lambda+\nu)}{\sqrt{\Gamma(2\lambda+n)\,n!}}\,i^{n}
\lt(\ww_{2}\rt)^{\nu}_{\lambda,\lambda+n}
=c\,\frac{\Gamma(1-\lambda-\nu)}{\sqrt{\Gamma(2\lambda+n)\,n!}}\,i^{-n}
\lt(\ww_{3}\rt)^{\nu}_{\lambda,\lambda+n},
\end{aligned}
\displaybreak[0]\\[6pt]
&\begin{aligned}
\psi^{\nu}_{\lambda,-(\lambda+n)}
&=c\,i^{-\nu}\sqrt{\frac{\Gamma(2\lambda+n)}{n!}}\,
\lt(\ww_{5}^{-}\rt)^{\nu}_{\lambda,-(\lambda+n)}
=c\,i^{\nu}(-1)^{n}\sqrt{\frac{\Gamma(2\lambda+n)}{n!}}\,
\lt(\ww_{5}^{+}\rt)^{\nu}_{\lambda,-(\lambda+n)}\\[2pt]
&=c\,\frac{\Gamma(1-\lambda+\nu)}{\sqrt{\Gamma(2\lambda+n)\,n!}}\,i^{n}
\lt(\ww_{4}\rt)^{\nu}_{\lambda,-(\lambda+n)}
=c\,\frac{\Gamma(1-\lambda-\nu)}{\sqrt{\Gamma(2\lambda+n)\,n!}}\,i^{-n}
\lt(\ww_{1}\rt)^{\nu}_{\lambda,-(\lambda+n)}.
\end{aligned}
\end{align}
Let us also give an explicit formula and the expression for the normalization factor $c$ that corresponds to a nice inner product:
\begin{equation}\label{psi_disc}
\!\begin{gathered}
\psi^{\nu}_{\lambda,\pm(\lambda+n)}(\phi,\theta)
=c\,\sqrt{\frac{\Gamma(2\lambda+n)}{n!}}\,i^{\mp n}
\bigl(-\tu\bigr)^{\frac{-\lambda-n\pm\nu}{2}}\bigl(1-\tu\bigr)^{\lambda}\,
\hgfs\bigl(-n,\,\lambda\pm\nu,\,2\lambda;\,1-\tu\bigr)\,
e^{\pm i(\lambda+n)\phi},\\[3pt]
c=\sqrt{\Gamma(\lambda+\nu)\,\Gamma(\lambda-\nu)}.
\end{gathered}\!
\end{equation}
The inner product \eqref{ip_AdS} between such functions is proportional to $\de(\lam-\lam')\,\de_{n,n'}$ with some coefficient that depends on $\lambda$ but not on $n$. So it is sufficient to consider the case $n=n'=0$, where the scaled hypergeometric function is equal to $\Gamma(2\lambda)^{-1}$. The result is as follows:
\begin{equation}
\bbraket{\psi^{\nu}_{\lam,\pm(\lam+n)}}{\psi^{\nu}_{\lam',\pm(\lam'+n')}}
=(2\pi)^{2}\,\frac{\de(\lam-\lam')}{\lam-1/2}\,\de_{n,n'}.
\end{equation}

Unlike in the continuous series case, there is no flux in the $\theta$ direction because the functions $\psi^{\nu}_{\lam,\pm(\lam+n)}$ vanish at the boundaries of $\AdS_2$. Thus we may interpret them as bound states, as opposed to scattering states for the continuous series. However, they do have non-trivial flux in the $\phi$ direction, $\calF=\int d\tht \, J_{\phi}$ with $J_{\alpha}$ the Klein-Gordon current, see \eqref{J_def}. To calculate the flux, we first consider its matrix element between different Casimir eigenfunctions and then take the limit $\lambda'\to\lambda$. When $\lambda\not=\lambda'$, the current is not conserved, but rather, we have these equations:
\begin{gather}
\calF=\int_{\phi>0} d^{2}x\,\sqrt{-g}\,\nabla_{\mu}J^{\mu}(x),
\\[3pt]
\bbra{\psi}\nabla_{\mu}J^{\mu}(x)\bket{\psi'}
=\frac{i}{2}\,\Bigl(\nabla^{2}\psi(x)^{*}\cdot\psi'(x)
-\psi(x)^{*}\cdot\nabla^{2}\psi'(x)\Bigr).
\end{gather}
Let us plug $\psi=\psi^{\nu}_{\lam,\pm(\lam+n)}$,\, $\psi'=\psi^{\nu}_{\lam',\pm(\lam'+n')}$, and use the fact that $\nabla^2=-Q-\nu^2$:
\begin{equation}
\bbra{\psi^{\nu}_{\lam,\pm(\lam+n)}}\nabla_{\mu}J^{\mu}(x)
\bket{\psi^{\nu}_{\lam',\pm(\lam'+n')}}
=\frac{i}{2}\,\bigl(-\lambda(1-\lambda)+\lambda'(1-\lambda')\bigr)\,
\psi^{\nu}_{\lam,\pm(\lam+n)}(x)^{*}\cdot\psi^{\nu}_{\lam',\pm(\lam'+n')}(x).
\end{equation}
Integrating over the $\phi>0$ region and taking the $\lambda'\to\lambda$ limit, we get:
\begin{equation}
\bbra{\psi^{\nu}_{\lam,\pm(\lam+n)}}\calF
\bket{\psi^{\nu}_{\lam,\pm(\lam+n')}}
=\pm 2\pi\,\delta_{n,n'}.
\end{equation}

Let us now calculate the projector onto the isotypic component $\calH^{\nu}_{\lambda,\pm}$,
\begin{equation}\label{proj_disc}
\Pi^{\nu}_{\lambda,\pm}(\phi,\theta;\phi',\theta')
=\sum_{n=0}^{\infty}\psi^{\nu}_{\lam,\pm(\lam+n)}(\phi,\theta)
\cdot\psi^{\nu}_{\lam,\pm(\lam+n)}(\phi',\theta')^{*}.
\end{equation}
By the $\widetilde{\SL}(2,\RR)$ symmetry, $\Pi^{\nu}_{\lambda,\pm}(x;x')
=\bigl|\vp_{23}/\vp_{14}\bigr|^{\nu}f_{j}(w)$. As in the continuous series case, the functions $f_{j}$ can be found by matching asymptotics at the region boundaries. However, we will instead use analyticity in a complex domain. Let us view the coordinates $\tz_j=e^{i\vp_j}$\, ($j=1,\dots,4$) as specifying a point $\lt((\tz_1,\tz_2),(\tz_3,\tz_4)\rt) \in\calM\times\calM$, where $\calM$ is a complexified hyperbolic plane in which $\AdS_2$ is embedded. Plugging $e^{i\phi}=\sqrt{z_1z_2}$,\, $\tu=z_1/z_2$ into \eqref{psi_disc} and taking out the $n$-independent factor $(z_1/z_2)^{\nu}(z_1^{\pm1}-z_2^{\pm1})^{\lambda}$, we find that $\psi^{\nu}_{\lam,\pm(\lam+n)}(\phi,\theta)$ is a homogeneous degree $n$ polynomial in $z_1^{\pm1}$ and $z_2^{\pm1}$. We can also use the fact that $\bigl(\psi^{\nu}_{\lam,\pm(\lam+n)}\bigr)^* =(-1)^{n} \psi^{-\nu}_{\lam,\mp(\lam+n)}$. Hence, up to the indicated factor, $\Pi^{\nu}_{\lambda,\pm}$ is analytic in the domain
\begin{equation}
D_{\pm}=\bigl\{ (z_1, z_2):\, |z_1|^{\pm1}, |z_2|^{\pm1} < 1 \bigr\}
\times \bigl\{ (z_3, z_4):\, |z_3|^{\mp1}, |z_4|^{\mp1} <1 \bigr\}.
\end{equation}
The sum in \eqref{proj_disc} is easy to calculate in the limit $z_1^{\pm1} \to 0$,\, $z_3^{\mp 1} \to 0$; the result is extended to a function of the form $(\vp_{32}/\vp_{14})^{\nu}f(w)$ with $w=(z_1-z_3)(z_2-z_4)/((z_1-z_4)(z_2-z_3))$. Thus,
\begin{equation}\label{tPsip}
\Pi^{\nu}_{\lambda,+}(z_1, z_2; z_3, z_4)
=|c|^2\,\underbrace{i^{-2\nu}\biggl({z_1 z_3 \ov z_2 z_4}\biggr)^{\nu/2}
\biggl({1-z_2 / z_3 \ov 1-z_1/ z_4}\biggr)^{\nu}}_{(\vp_{32}/\vp_{14})^{\nu}}
\underbrace{(1-w)^\lam\, \hgfs\bigl(\lam-\nu,\,\lam + \nu,\,2\lam;\,1-w\bigr)}
_{B_{\lambda,\nu,-\nu}(w)}
\end{equation}
in the domain $D_{+}$, and $\Pi^{\nu}_{\lambda,+}(z_1,z_2;z_3,z_4)$ in $D_{-}$ is related by the symmetry $z_1 \leftrightarrow z_3$,\, $z_2 \leftrightarrow z_4$,\, $\nu \leftrightarrow -\nu$.

\begin{figure}
\centerline{
\includegraphics{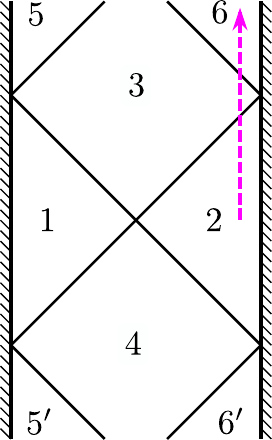}\hspace{2cm}\includegraphics{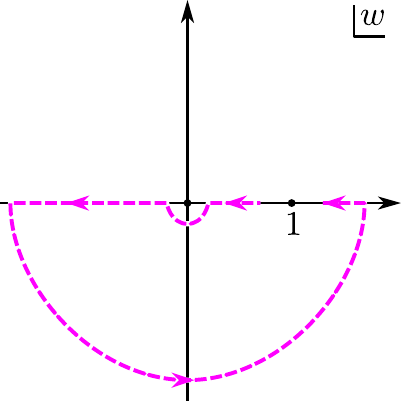}
}
\caption{Analytic continuation of the projector $\Pi^{\nu}_{\lambda,+}$.}
\label{fig_anacon}
\end{figure}

Finally, we analytically continue the functions $\Pi^{\nu}_{\lambda,\pm}$. For simplicity, let us focus on the ``$+$'' case. The expression on the right-hand side of \eqref{tPsip} is uniquely defined if $z_1\to+0$,\, $z_2\in(0,1)$,\, $z_3\to+\infty$,\, $z_4\in(1,+\infty)$, and therefore, $0<w<1$. This gives a straightforward continuation to regions $1$ and $2$. When continuing to other regions, we fix $z_3=i$,\, $z_4=-i$ and move $z_1$, $z_2$ along the unit circle, pushing them inward to get around $z_3$, $z_4$. Thus,
\begin{equation}
\arg\lt(\frac{\vp_{32}}{\vp_{14}}\rt)=\begin{cases}
0 & \text{in regions } 1,2,3,4,\\
\pi & \text{in region } 5,\\
-\pi & \text{in region } 6,
\end{cases}
\end{equation}
whereas the continuation of $B_{\lambda,\nu,-\nu}(w)$ from region $2$ to regions $3$ and $6$ is shown in Figure~\ref{fig_anacon}. Considering the other regions and the ``$-$'' case, we arrive at the following equation:
\begin{equation}\label{tpf-disc}
\wideboxed{
\Pi^{\nu}_{\lambda,\pm}(x,x')
=\left|\frac{\vp_{23}}{\vp_{14}}\right|^{\nu}
\Gamma(\lambda+\nu)\,\Gamma(\lambda-\nu)
\begin{dcases}
B_{\lam, \nu, -\nu}(w) & \text{in regions } 1, 2\\
B^{\mp}_{\lam, \nu, -\nu}(w) & \text{in region } 3\\
B^{\pm}_{\lam, \nu, -\nu}(w) & \text{in region } 4\\
e^{\pm i \pi \nu}B^{\mp}_{\lam, \nu, -\nu}(w) & \text{in region } 5\\
e^{\mp i \pi \nu}B^{\mp}_{\lam, \nu, -\nu}(w) & \text{in region } 6
\end{dcases}
}
\end{equation}
(Here $B^{\pm}_{\lam,\nu,-\nu}(w)$ is the analytic continuation of $B_{\lam,\nu,-\nu}(w)$ from $w\in(0,1)$ to the other parts of the real axis through the upper half-plane for the ``$+$'' sign and lower half-plane for the ``$-$'' sign.)

\subsection{The algebra of $\widetilde{\SL}(2,\RR)$-invariant two-point functions}\label{app:opalg}

This subsection is concerned with the functions $\Psi^{\nu}_{\lambda,\mu}[R]$ (see \eqref{tp_def}) for variable $\lambda$ and $\mu$. Here $R$ is an operator acting in the intertwiner space $\calL^{\nu}_{\lambda,\mu}$. In the discrete series case, $R$ is simply a complex number, and $\Psi^{\nu}_{\lambda,\pm}[R]=R\,\Pi^{\nu}_{\lambda,\pm}$.

More generally, let us consider $\widetilde{\SL}(2,\RR)$-invariant $(\nu,-\nu)$-spinors on $\widetilde{\AdS}_2\times\widetilde{\AdS}_2$. Such spinors may be interpreted as integral kernels: the kernel of operator $F$ acting in $\calH^{\nu}$ is $F(x;x')=\bra{x}F\ket{x'}$. Therefore, the product and the Hermitian conjugate are given by these formulas:
\begin{equation}
(FG)(x,x'')
=\int_{\widetilde{\AdS}_2}d^{2}x'\,\sqrt{-g(x')}\,F(x,x')\,G(x',x''),\qquad
F^{\dag}(x,x')=F(x',x)^{*}.
\end{equation}
Using the orthogonality relation \eqref{op_ip} for the basis one-point functions, we obtain the following identity:
\begin{equation} \label{cwfnorm}
\Psi^{\nu}_{1/2+is,\mu}[R]
\cdot \Psi^{\nu}_{1/2+is',\mu'}[R']
=\Psi^{\nu}_{1/2+is,\mu}[RR']\,
\frac{\delta(s-s')\,\delta(\mu-\mu')}{\rcont(s,\mu)},
\end{equation}
where
\begin{equation} \label{rhoc}
\rcont(s,\mu)=\frac{s}{(2\pi)^{2}}\,
\frac{\sinh(2\pi s)}{\cosh(2\pi s)+\cos(2\pi\mu)}.
\end{equation}
Similarly, for the discrete series,
\begin{equation} \label{dwfnorm}
\Psi^{\nu}_{\lambda,\pm}[R]
\cdot\Psi^{\nu}_{\lambda',\pm}[R']
=\Psi^{\nu}_{\lambda,\pm}[RR']\,
\frac{\delta(\lambda-\lambda')}{\rdisc(\lambda)},\qquad
\rdisc(\lambda)=\frac{\lambda-1/2}{(2\pi)^{2}}.
\end{equation}
The product of functions $\Psi^{\nu}_{\lambda,\mu}$, $\Psi^{\nu}_{\lambda,\pm}$ associated with different irreps is always zero.

Now, let $R$ be a function of $s$ and $\mu$ for the continuous series and a function of the $\pm$ sign and $\lambda$ and for the discrete series. Then we may define
\begin{equation} \label{genop}
\Psi^{\nu}[R]=\int_{0}^{\infty}ds \int_{-1/2}^{1/2}d\mu\,\rcont(s,\mu)\,
\Psi^{\nu}_{1/2+is,\mu}[R(s,\mu)]
+\int_{1/2}^{\infty}d\lambda\,\rdisc(\lambda) \sum_{\alpha=\pm}
\Psi^{\nu}_{\lambda,\alpha}[R_{\alpha}(\lambda)].
\end{equation}
All $\widetilde{\SL}(2,\RR)$-invariant $(\nu,-\nu)$-spinors can be cast in this form, and we have the identities
\begin{equation}
\Psi^{\nu}[R]\cdot\Psi^{\nu}[R']=\Psi^{\nu}[RR'],\qquad
\Psi^{\nu}[R^{\dag}]=\Psi^{\nu}[R]^{\dag}.
\end{equation}
Furthermore, one can define a formal trace as follows:
\begin{equation}\label{tr_def}
\tr\bigl(\Psi^{\nu}[R]\bigr)
=\int_{0}^{\infty}ds \int_{-1/2}^{1/2}d\mu\,\rcont(s,\mu)\kern1pt
\Tr(R(s,\mu))
+\int_{1/2}^{\infty}d\lambda\,\rdisc(\lambda)\sum_{\alpha=\pm}
R_{\alpha}(\lambda).
\end{equation}
Note that $\tr\bigl(\Psi^{\nu}[R]\bigr)$ is not the usual trace of the operator $\Psi^{\nu}[R]$ because the latter is infinite. Essentially, equation \eqref{tr_def} is a way to normalize the trace and make it finite while satisfying the cyclic property, $\tr(FG)=\tr(GF)$.

It is also possible to define $\tr(F)$ directly, not using the irreducible decomposition. Indeed, under certain assumptions, the function $F(x;x')$ has the same asymptotic form at $x\to x'$ as $f_j[R](w)$ for $j=1,2,3,4$ and $w\to 0$ (see \eqref{tpf-12} and \eqref{tpf-34}). Specifically,
\begin{equation}\label{F_asymp}
F(x;x')\approx\left|\frac{\vp_{23}}{\vp_{14}}\right|^{\nu}\begin{cases}
-\tr(F)\ln|w|+q-q' & \text{in region }1,\\
-\tr(F)\ln|w|+q+q' & \text{in region }2,\\
-\tr(F)\ln|w|+q+q'' & \text{in region }3,\\
-\tr(F)\ln|w|+q-q'' & \text{in region }4,
\end{cases}
\end{equation}
where $q,q',q''$ are some complex numbers. To see this, let $F=\Psi^{\nu}(R)$ with $R(s,\mu)$ and $R(\lambda)$ decaying sufficiently fast at large values of $s$ and $\lambda$. Then equation \eqref{F_asymp} follows from \eqref{tpf-12}, \eqref{tpf-34} for the continuous series and \eqref{tpf-disc} for the discrete series. Thus, $\tr(F)$ may be defined as the coefficient in front of $\ln\frac{1}{|w|}$ in the asymptotic form of $F(x;x')$.

We will now formulate a somewhat more natural condition that guarantees particular asymptotic behaviors at all region corners and boundaries. It may be viewed as a statement of generalized smoothness. First, let us write $F$ in a form similar to \eqref{tPsi_f}, namely,
\begin{equation}
F(x;x')=\left|\frac{\vp_{23}}{\vp_{14}}\right|^{\nu}f(x;x'),
\end{equation}
where $f$ is an $\widetilde{\SL}(2,\RR)$-invariant scalar function. Essentially, $f$ is a function on the quotient space
\begin{equation} \label{QS_def}
\tGG\backslash\widetilde{\AdS}_2\times\widetilde{\AdS}_2
=H\backslash\widetilde{\AdS}_2=H\backslash\tGG/H,
\end{equation}
where $\tGG=\widetilde{\SL}(2,\RR)$ and $H\subseteq\tGG$ is the subgroup preserving the point $x'=0$. (As mentioned at the beginning of this appendix, $H$ is generated by the Lie algebra element $\Lambda_2$.) Of three equivalent quotient spaces in \eqref{QS_def}, the simplest is $H\backslash\widetilde{\AdS}_2$, that is, the space of orbits under the (left) action of $H$ on $\widetilde{\AdS}_2$. These orbits are shown in Figure~\ref{fig_regions}b by thin lines; each nondegenerate orbit is represented by a unique point of the skeleton subset $S$ composed of thick lines. So one may consider $f$ as a function on $S$, but it is not clear how to define smoothness at the junctions. To resolve this problem, suppose that $f$ is obtained from a sufficiently smooth spinor supported by some neighborhood of $S$ by the integration along the orbits. The spinor being smooth means that its tilde gauge representation $\psi$ is smooth,\footnote{In the neighborhood of the origin, the disk gauge representation is also smooth.} whereas the integrals is defined using the disk gauge:
\begin{equation}
f(x;0)=\int_{-\infty}^{\infty}d\vt\,
\mathring{\psi}\bigl(e^{\vt\Lambda_2}x\bigr).
\end{equation}
It is fairly easy to elaborate these conditions and prove that they imply the asymptotic form \eqref{F_asymp} with $\tr(F)=\psi(0)$.

At last, we consider the inner product between $\widetilde{\SL}(2,\RR)$-invariant $(\nu,-\nu)$-spinors. Let
\begin{equation}
F(x;x')=\left|\frac{\vp_{23}}{\vp_{14}}\right|^{\nu}f_j(w),\qquad
G(x;x')=\left|\frac{\vp_{23}}{\vp_{14}}\right|^{\nu}g_j(w),
\end{equation}
where $j$ ranges over all regions, including translational copies. By definition, the inner product is
\begin{equation}
\braket{F}{G}=\sum_{j}\int \frac{2\,dw}{(1-w)^2}\,f_j(w)^{*}\,g_j(w).
\end{equation}
(The integration measure is the ratio of volume elements for $\widetilde{\AdS}_2$ and $H$.) One can show that the Casimir operator $Q$ is Hermitian with respect to this inner product; more exactly, if $F$ and $G$ satisfy the aforementioned smoothness condition, then $\bra{F}Q\ket{G}=\bra{G}Q\ket{F}^{*}$. Hence, functions of the form $\Psi^{\nu}_{\lambda,\mu}[R]$ or $\Psi^{\nu}_{\lambda,\pm}[R]$ are orthogonal to each other if they correspond to different irreps. In general, their inner product is equal to a $\delta$-function with some coefficient that can be calculated from the asymptotic formula \eqref{asymp_infty} or a similar expression for the discrete series. Thus,
\begin{equation}
\begin{aligned}
\Bigl\langle\Psi^{\nu}_{1/2+is,\mu}[R] \Big|
\Psi^{\nu}_{1/2+is',\mu'}[R']\Bigr\rangle
&=\Tr(R^{\dag}R')\,\frac{\delta(s-s')\,\delta(\mu-\mu')}{\rcont(s,\mu)},
\\[3pt]
\Bigl\langle\Psi^{\nu}_{\lambda,\pm}[R] \Big|
\Psi^{\nu}_{\lambda',\pm}[R']\Bigr\rangle
&=R^*R'\,\frac{\delta(\lambda-\lambda')}{\rdisc(\lambda)},
\end{aligned}
\end{equation}
It follows that
\begin{equation}
\wideboxed{
\braket{F}{G}=\tr(F^{\dag}G)
}
\end{equation}

\bibliography{AdS2particles}
\bibliographystyle{JHEP}
 
\end{document}